\documentclass[10pt,journal,compsoc]{IEEEtran}
\usepackage[hyphens]{url}

\usepackage{mathtools}

\usepackage{ragged2e}
\usepackage{blindtext, graphicx}
\graphicspath{{Pictures/}{Profile/}} 
\usepackage{enumitem} 
\usepackage{hyphenat} 
\usepackage[tracking=true]{microtype} 
\SetTracking[no ligatures = f]{encoding = *,shape = sc}{120}
\usepackage{amsmath, amsthm, amssymb}
\usepackage{amsmath,amssymb,enumitem,mathtools}
\usepackage{xcolor}

\usepackage[most]{tcolorbox}
\newtcolorbox{highlighted}{colback=yellow,coltext=black,breakable}
\usepackage{lipsum}
\usepackage{float}
\usepackage{dblfloatfix}
\usepackage{soul}
\usepackage{xcolor}
\usepackage{graphicx}
\usepackage[export]{adjustbox}

\usepackage{color, soul}

\soulregister\cite7
\soulregister\ref7
\soulregister\pageref7

\usepackage{cite}
\ifCLASSINFOpdf
\else
\fi
\ifCLASSOPTIONcompsoc
  \usepackage[caption=false,font=footnotesize,labelfont=sf,textfont=sf]{subfig}
\else
  \usepackage[caption=false,font=footnotesize]{subfig}
\fi
\begin{document}
	%
	\title{TA-LRW: A Replacement Policy for Error Rate Reduction in STT-MRAM Caches}
	%
	%
	%
	%
	
	\author{Elham~Cheshmikhani,~Hamed~Farbeh,\IEEEmembership{}
%
%
Seyed~Ghassem~Miremadi,~\IEEEmembership{Senior Member,~IEEE,}
		Hossein~Asadi,~\IEEEmembership{Senior Member,~IEEE}
		\IEEEcompsocitemizethanks{\IEEEcompsocthanksitem E. Cheshmikhani and H. Asadi are with the Department of Computer Engineering, Sharif University of Technology, Tehran 11155-11365, Iran.\protect\\
			E-mail: echeshmikhani@ce.sharif.edu; asadi@sharif.edu

\IEEEcompsocthanksitem H. Farbeh is with the Department of Computer Engineering, Amirkabir University of Technology, Tehran 15916-34311, Iran.\protect\\
			E-mail: farbeh@aut.ac.ir

\IEEEcompsocthanksitem S. G. Miremadi was with the Department of Computer Engineering, Sharif University of Technology, Tehran 11155-11365, Iran (deceased on April 13, 2017).\protect\\
			E-mail: miremadi@sharif.edu

			}
		\thanks{Manuscript received December 27, 2017; revised June 13, 2018; revised July 29, 2018.}}
	
	%
	%

	\markboth{IEEE Transactions on Computers,~Vol.~XX, No.~X, August~2018}%
	{Shell \MakeLowercase{\textit{et al.}}: Bare Demo of IEEEtran.cls for Computer Society Journals}
	%



	\IEEEtitleabstractindextext{%
		{\justify
			\begin{abstract}
				As technology process node scales down, on-chip SRAM caches lose their efficiency because of their low scalability, high leakage power, and increasing rate of soft errors.
Among emerging memory technologies, \textit{$Spin$}-\textit{$Transfer~Torque~Magnetic~RAM$} (STT-MRAM) is known as the most promising replacement for SRAM-based cache memories.
The main advantages of STT-MRAM are its non-volatility, near-zero leakage power, higher density, soft-error immunity, and higher scalability.
Despite these advantages, high error rate in STT-MRAM cells due to \textit{$retention~failure$}, \textit{$write~failure$}, and \textit{$read~disturbance$} threatens the reliability of cache memories built upon STT-MRAM technology.
The error rate is significantly increased in higher temperature, which further affects the reliability of STT-MRAM-based cache memories.
The major source of heat generation and temperature increase in STT-MRAM cache memories is write operations, which are managed by cache \textit{$replacement~policy$}.
To the best of our knowledge, none of previous studies have attempted to mitigate heat generation and high temperature of STT-MRAM cache memories using replacement policy.
In this paper, we first analyze the cache behavior in conventional \textit{$Least$}-\textit{$Recently~Used$} (LRU) replacement policy and demonstrate that the majority of consecutive write operations (more than 66\%) are committed to adjacent cache blocks.
These adjacent write operations cause accumulated heat and increased temperature, which significantly increase the cache error rate. 
To eliminate heat accumulation and the adjacency of consecutive writes, we propose a cache replacement policy, named \textit{$Thermal$}-\textit{$Aware~Least$}-\textit{$Recently~Written$} (TA-LRW), to smoothly distribute the generated heat by conducting consecutive write operations in distant cache blocks. 
TA-LRW guarantees the distance of at least three blocks for each two consecutive write operations in an 8-way associative cache.
This distant write scheme reduces the temperature-induced error rate by 94.8\%, on average, compared with the conventional LRU policy, which results in 6.9x reduction in cache error rate.
The implementation cost and complexity of TA-LRW is as low as \textit{$First$}-\textit{$In,~First$}-\textit{$Out$} (FIFO) policy while providing a near-LRU performance, having the advantages of both replacement policies. 
The significantly reduced error rate is achieved by imposing only 2.3\% performance overhead compared with the LRU policy.
			\end{abstract}}
			
			}

			\maketitle

			\section{Introduction}
			
			\IEEEPARstart{C}{ache}~memories occupy a large part of the processor chip area and play an important role in performance, reliability, and power consumption of the whole computer systems~\cite{Iyer2010cacheintro},~\cite{guo2017sanitizer}. 
			Conventional SRAM caches face several challenges by downscaling technology process node, e.g., high leakage power, increased error rate, low density, and high sensitivity to process variations. 
			Recent developments have suggested emerging memory technologies, e.g., \textit{$Phase~Change~Memory$} (PCM), \textit{$Resistive~RAM$} (ReRAM), \textit{$Ferroelectric~RAM$} (FeRAM), and \textit{$Spin$}-\textit{$Transfer~Torque~Random~Access~Memories$} (STT-MRAM), as replacements for currently used memories~\cite{vatajelu2017challenges, EDCC, ZAZADTE}. Among these memories, STT-MRAM is known as the most promising alternative for SRAMs in on-chip caches according to recent reports~\cite{itrs, imani2016approximate, kim2016exploration}. 
				Several recent studies have evaluated STT-MRAM and SRAM as \textit{Last-Level Caches} (LLCs) and illustrated that STT-MRAM outperforms SRAM in terms of energy consumption and performance. STT-MRAM improves the performance by increasing the cache capacity (due to its higher density) and reduces the total energy consumption due to its near-zero leakage power \cite{park2012future, wang2013oap, jog2012cache, Jiang2012cons, Smullen2011relax, lee2012perf}.
			
			Unlike SRAM technology, STT-MRAM cells are non-volatile and their leakage power is negligible.
			 In addition, their density is higher than that of SRAMs, and they are not vulnerable to radiation-induced soft errors. 
			 However, the high error rate of STT-MRAM cells should be addressed to make it applicable in on-chip caches.
			 STT-MRAM caches are error-prone to three types of failures in write and read accesses as well as memory idle intervals as follows:
			 1) Due to stochastic behavior of magnetization process of STT-MRAM cells, a cache cell may not switch on a write operation, resulting to a \textit{$write~failure$}~\cite{choi2017nvm},~\cite{farbeh2016floating};
			 2) A cache cell may unintentionally flip by the current applied for reading a cache block, which leads to a \textit{$read~disturbance~error$}~\cite{khvalkovskiy2013basic},~\cite{wang2015selective};
			 3) Without applying any current, an idle cache cell may flip stochastically, leading to a \textit{$retention~failure$}~\cite{Chen2016}, \cite{Chintaluri2015}.
			The rates of these three types of errors are significantly increased by raising the temperature. 
			Increasing the temperature reduces the~\textit{$thermal~stability~factor$} ($\Delta$) of STT-MRAM cells, which results in exponentially increase in read disturbance and retention failure rates. 
			On the other hand, write current decreases in higher temperature, which results to a higher write failure rate~\cite{bi2012analysis}.
			
			The major source of heat generation and high temperature in STT-MRAM caches is  read and write accesses~
			\cite{beigi2016tesla, bi2012stt, bishnoi2014asynchronous, sun2009novel, dong2008circuit}.
			The energy dissipated per write access in the cache is by one order of magnitude higher than that per read access. Therefore, the increase in cache temperature is mainly due to write operations. 	
			The write operations affect the temperature and consequently the cache error rate in two aspects: 1) the number of write requests to the cache per unit of time for an arbitrary cache configuration and 2) the distribution pattern of the writes over the cache blocks. 
			Although the former is beyond the control of the cache, the latter is handled by the cache controller. 
			One of the main factors in write distribution non-uniformity is the decisions made by cache replacement policy, as it determines into which cache block the incoming data should be written.
			To the best of our knowledge, none of the existing cache replacement policies are aware of the temperature and the heat generated by write operations. Hence, it is highly probable that consecutive write accesses are performed in adjacent cache blocks, which leads to heat accumulation in some cache regions and the increased error rate.
			
			In this paper, we first conduct a set of experiments using finite-volume based \textit{$Computational~Fluid~Dynamics$} (CFD) tools (ANSYS FLUENT 14.5 and GAMBIT FLUENT~\cite{ansys}) and show that each write operation into a cache block increases the temperature of the written STT-MRAM cells by 9$^{\circ}$K. 
			Then, we investigate the sequence of L2 cache write operations in conventional \textit{$Least$}-\textit{$Recently~Used$} (LRU) replacement policy and demonstrate that a large fraction of consecutive incoming data are written into the same or adjacent cache blocks. 
			These write operations, originated by a cache miss or writeback from L1 caches, accumulate the generated heat locally and increase the error rate in hot blocks. 
			Using these observations, we propose a cache replacement policy, so-called \textit{$Thermal$}-\textit{$Aware~Least$}-\textit{$Recently~Written$}~(TA-LRW), to uniformly distribute the temperature and prevent the heat accumulation.
			Unlike LRU policy, TA-LRW policy sorts the blocks based on their last write access instead of their last read/write access and evicts the least-recently written block on a cache miss.
			On a writeback hit, the incoming data is written in the least-recently written block instead of overwriting the block containing the previous version of data.
			To prevent the heat accumulation, the target block to eviction is selected \textit{physically far enough}\footnote{
			Two blocks are physically far enough if the effect of generated heat during the write operation in one of them is negligible on the temperature of the other block.}
			from the previously written block in the cache set.

			TA-LRW policy uses a pointer to indicate the next block to be written in a set. To prevent the heat accumulation, we present a permutation of write sequence in which the distance of two consecutive writes is at least three blocks and the pointer is updated after each write based on this permutation. TA-LRW policy does not require the complicated LRU policy controller and peripherals and its implementation is as simple as the low-cost \textit{$First$}-\textit{$In,~First$}-\textit{$Out$} (FIFO) policy. Meanwhile, the performance of TA-LRW is very close to that of LRU and significantly higher than that of FIFO, since it decides based on write access history, which we show that is an effective approximation for the total access history.

			We evaluate the proposed TA-LRW policy by running a set of workloads from SPEC CPU2006 benchmark suite~\cite{spec2006} using gem5 cycle-accurate simulator~\cite{gem5}. 
			The STT-MRAM cache parameters are extracted from NVSim tool~\cite{nvsim} and the cache temperature is measured using ANSYS FLUENT 14.5 and GAMBIT FLUENT tools~\cite{ansys}.
			The L2 cache configuration is adjusted according to STT-MRAM parameters and the cache is modified to operate based on TA-LRW policy. 
			The simulation results show that the TA-LRW policy reduces the temperature-induced error rate by 6.9x compared with LRU policy. 
			Furthermore, TA-LRW increases the miss rate and  \textit{$Cycles~Per~Instruction$} (CPI) by only 0.5\% and 2.3\%, respectively, compared with LRU, while these values for FIFO are 9.5\% and 10.3\%, respectively.

			Briefly, the \textbf{main contributions} of this work are as follows:
			
			\begin{enumerate}
 				 \item This is the first study that investigates the effect of temperature on STT-MRAM cache error rate and demonstrates that heat accumulation increases the error rate by 110.9\%. We also illustrate that this heat accumulation is mainly due to locality of committed write operations in the cache. 
			 	 \item As a first effort, this work models the STT-MRAM cells using CFD tools to analyze its temperature behavior. Using these finite-volume based tools, we show that each write operation increases the cell temperature by 9$^{\circ}$K. This significant increase in STT-MRAM cell temperature implies that write operations are the main source of heat generation in STT-MRAM-based caches. 
			 	 \item We reveal that using conventional LRU replacement policy, write operations cause a substantial increase in error rate due to heat accumulation in L2 cache. Our evaluations show that more than 66\% of write operations are performed in the hottest block or in its adjacent blocks in LRU replacement policy.
			 	 \item 
			 	 We propose a simple yet effective replacement policy, called TA-LRW, to prevent the heat accumulation in the cache. Unlike conventional replacement policies, TA-LRW decides to write into the blocks that are far enough apart to minimize the accumulated heat and reduce the cache error rate. We explore all possible permutations for write sequences and select the most suitable permutation that guarantees the minimum heat accumulation in TA-LRW policy.
			 	\item Detailed evaluations and comparisons are conducted to show the efficiency of TA-LRW. Our evaluations show that the conventional LRU replacement policy increases the STT-MRAM cache error rate by 110.9\%, which is reduced to 16.1\% by TA-LRW. These values indicate 6.9x reduction in temperature-induced error rate. 
			\end{enumerate}

			The rest of this paper is organized as follows. Section 2 describes the basics of STT-MRAM and its reliability challenges. In Section 3, the observations and motivations for this work are discussed. The details of the proposed TA-LRW policy are presented in Section 4. Section 5 gives the simulation setup and evaluation results. Related work and discussions are investigated in Section 6. Finally, we conclude the paper in Section 7.
			\begin{figure}[t]
				\centering
				\subfloat[]{\includegraphics[width=0.3\linewidth]{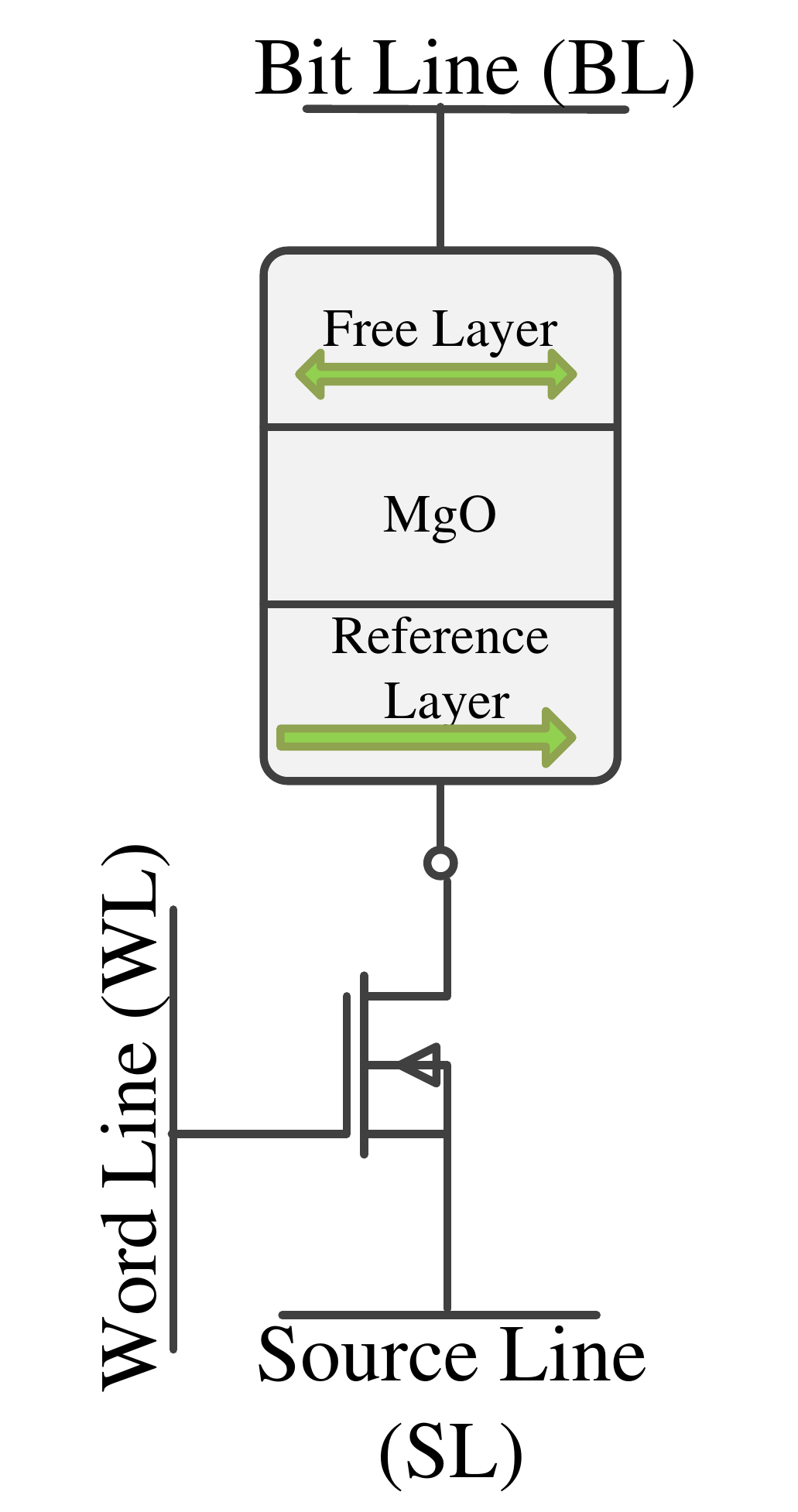}}
				\hspace{9pt}
				\subfloat[]{\includegraphics[width=0.5\linewidth]{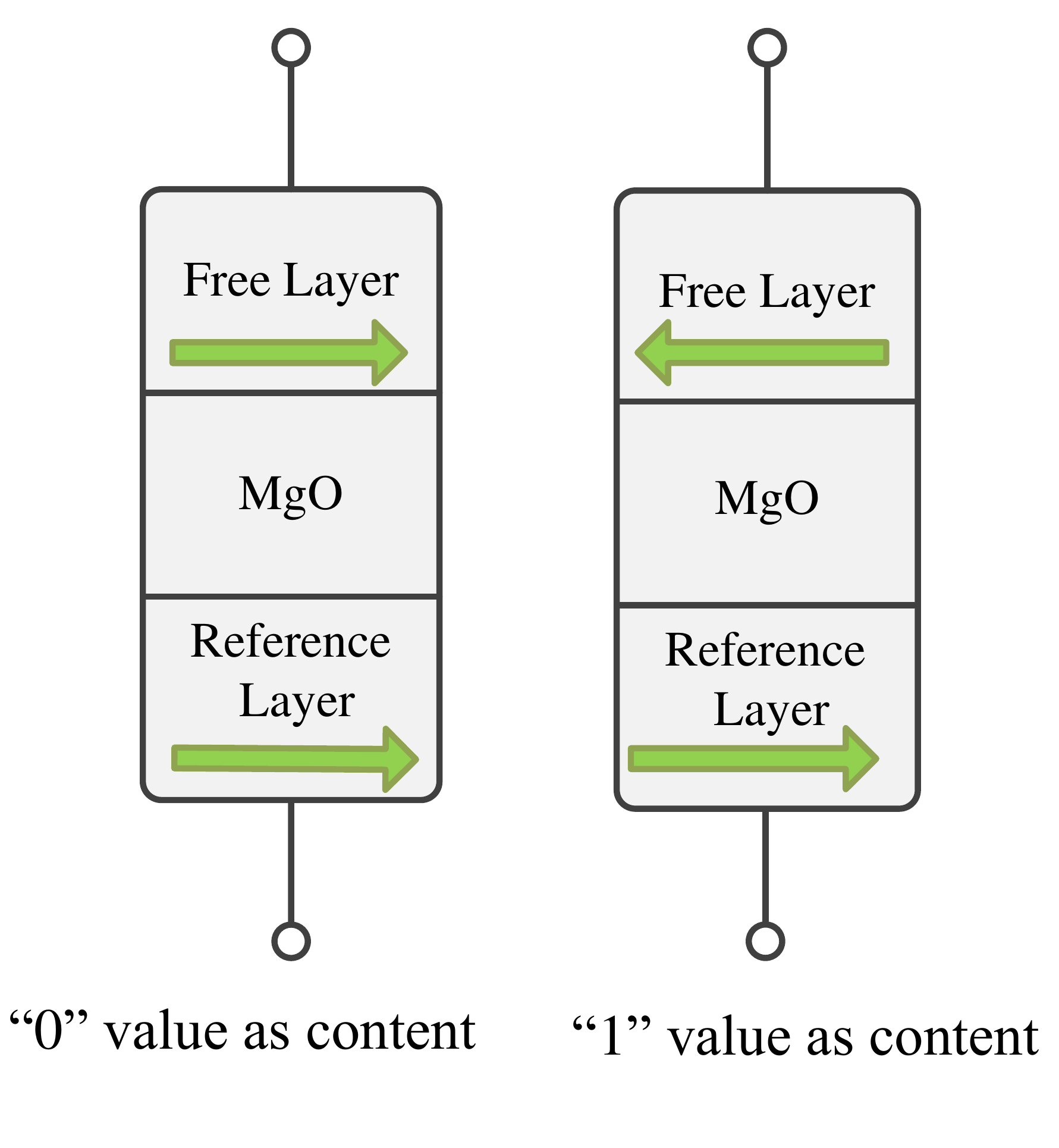}}
				\caption{STT-MRAM schematic~\cite{EDCC}: (a) 1T1J STT-MRAM structure and (b) MTJ states.}
				\label{fig:basics}
			\end{figure}
				
			\section{Preliminaries}
			\label{sec:PRELIMINARIES}
				\subsection{STT-MRAM Basics}
			Magnetoresistive RAM (MRAM) cells store data bits using magnetic charge instead of electrical charge. These magnetic storage elements, known as \textit{$Magnetic~Tunnel~Junction$} (MTJ), store the charge and are formed from two ferromagnetic layers, separated by a thin oxide barrier layer~\cite{12-EDCC-sun2012process},~\cite{14-EDCC-apalkov2013spin}. The barrier layer is made up of crystallized MgO. There are some methods to torque the MRAM cell. The preferred technique is \textit{$Spin$}-\textit{$Transfer~Torque$} (STT) that uses spin-aligned electrons to torque the domain~\cite{dieny2016introduction}. STT-MRAM cell is comprised of a MTJ and an access transistor, which is named 1T1J STT-MRAM, as shown in Fig.~\ref{fig:basics}(a). One of the ferromagnetic layers in MTJ, named \textit{$reference~layer$}, has a fixed magnetic field direction and the field direction of the other layer, named \textit{$free~layer$}, can be changed through an applied current~\cite{15-EDCC-zhao2011design, mittal2016survey}.

			The resistance of the MTJ alternates between a low and high values based on the magnetic field direction of the free layer. If it is parallel with the magnetic field direction of the fixed layer, the MTJ is in low resistance state and if is antiparallel with that, the resistance of MTJ is high. The two states of the MTJ unit are shown in Fig.~\ref{fig:basics}(b). These high and low resistances are interpreted as binary logic `1' and `0', respectively. High resistance stores logic `1' in the MTJ and low resistance is assumed as logic `0', as shown in Fig.~\ref{fig:basics}(b).

			\subsection{Read and Write Operations}
			The resistance of the MTJ shows the cell value. To read the data stored in a cell, word line is set to V$_{DD}$ to turn on the access transistor. Then, a small read current/voltage is applied to the bit line~\cite{8-EDCC-zhao2012failure, EDCC}. This current/voltage generates a current in the cell, which can be compared to a reference value. If the sensed voltage/current is higher than the reference, the MTJ resistance is high and the cell contains `1'. Otherwise, the resistance is low and the cell contains `0'.
			
			To write a value in a STT-MRAM cell, the magnetic field direction of the free layer should be changed. To this end, a write voltage/current is applied to the bit line or source line. If the current flows from source line to bit line (voltage is applied to source line), electron charges flow from the free layer to the reference layer. Due to the strong magnetic field in reference layer, the electrons with the opposite direction of the reference layer are reflected to the free layer. This reflection creates a torque in the free layer, antiparallelizes the magnetic of MTJ, and leads to write `1'~\cite{ZAZADTPDS}.  To write `0' in the cell, the spin-polarized current flows from bit line to source line and causes the electron charges to flow from the reference layer to the free layer. Electrons with the spin direction same as that of electrons spin in the reference layer pass through the free layer and generate a torque that parallelize the two MTJ ferromagnetic layers and leads to write `0'~\cite{ZAZADTPDS},~\cite{14-zazad-eken2014novel}.

		\begin{figure}[t]
				\centering
				\includegraphics[width=0.67\linewidth]{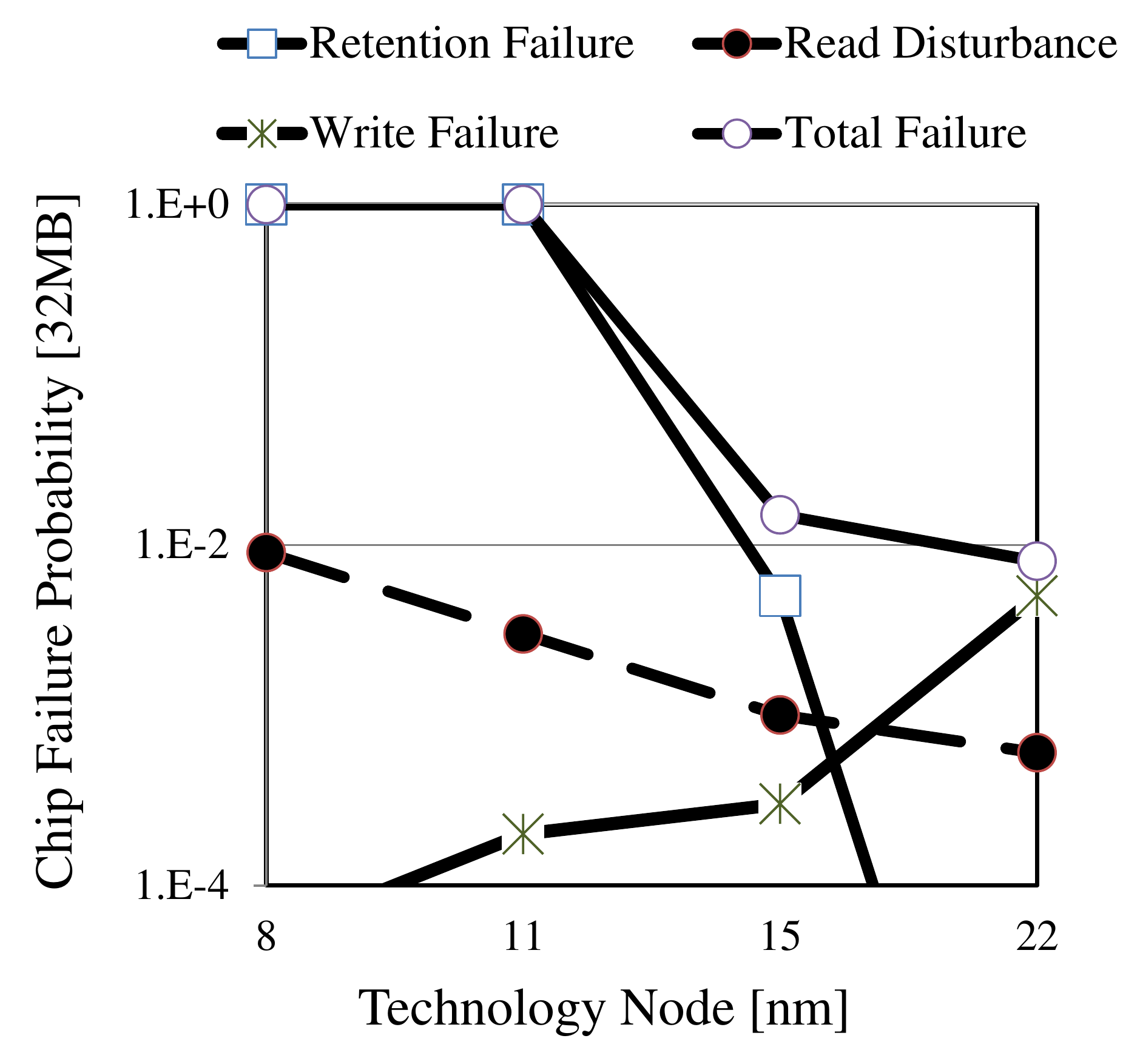}\vspace{-6pt}
				\caption{STT-MRAM failure rates in technology node scaling~\cite{naeimi2013intel}.}
				\label{fig:naeimi}
			\end{figure}

			\subsection{Reliability of STT-MRAM} 
			The major reliability challenges in STT-MRAM cells are \textit{$retention~failure$}, \textit{$read~disturbance$}, and \textit{$write~failure$}~\cite{EDCC, naeimi2013intel, Chintaluri2015}. A  retention failure occurs when the content of an idle cell unintentionally flips without applying any voltage/current. A read disturbance occurs when the content of a cell is flipped during a read operation. Due to the stochastic switching behavior of a cell, it is probable that its content flips even by a small read current. The error remains until the next write operation in the cell~\cite{naeimi2013intel, Chintaluri2015}. Write operation in a STT-MRAM cell is also stochastic. Write failure is inability of a cell to switch when the write current is applied~\cite{Chintaluri-ESTCS2016}. This means that the value of the cell is not changed during the write pulse interval. In the subsequent requests to the cell, an erroneous value is read until the next write operation.

			Fig.~\ref{fig:naeimi} shows a prediction for the rates of three mentioned errors in different technology process nodes for a 32MB STT-MRAM cache memory~\cite{naeimi2013intel}. As depicted, write failure and read disturbance are the main reliability threat in 22nm technology process node, while the rate of retention failure is negligible. However, the retention failure rate increases significantly by technology downscaling and is higher than both other two errors in 15nm technology process node and beyond. On the other hand, write failure rate decreases while read disturbance rate increases in smaller technology process nodes. The total error rate considering all three sources of errors is significantly increased by technology downscaling, in which the retention failure is the dominant contributor.
				
			\begin{figure}[b]
				\centering
				\includegraphics[width=0.95\linewidth]{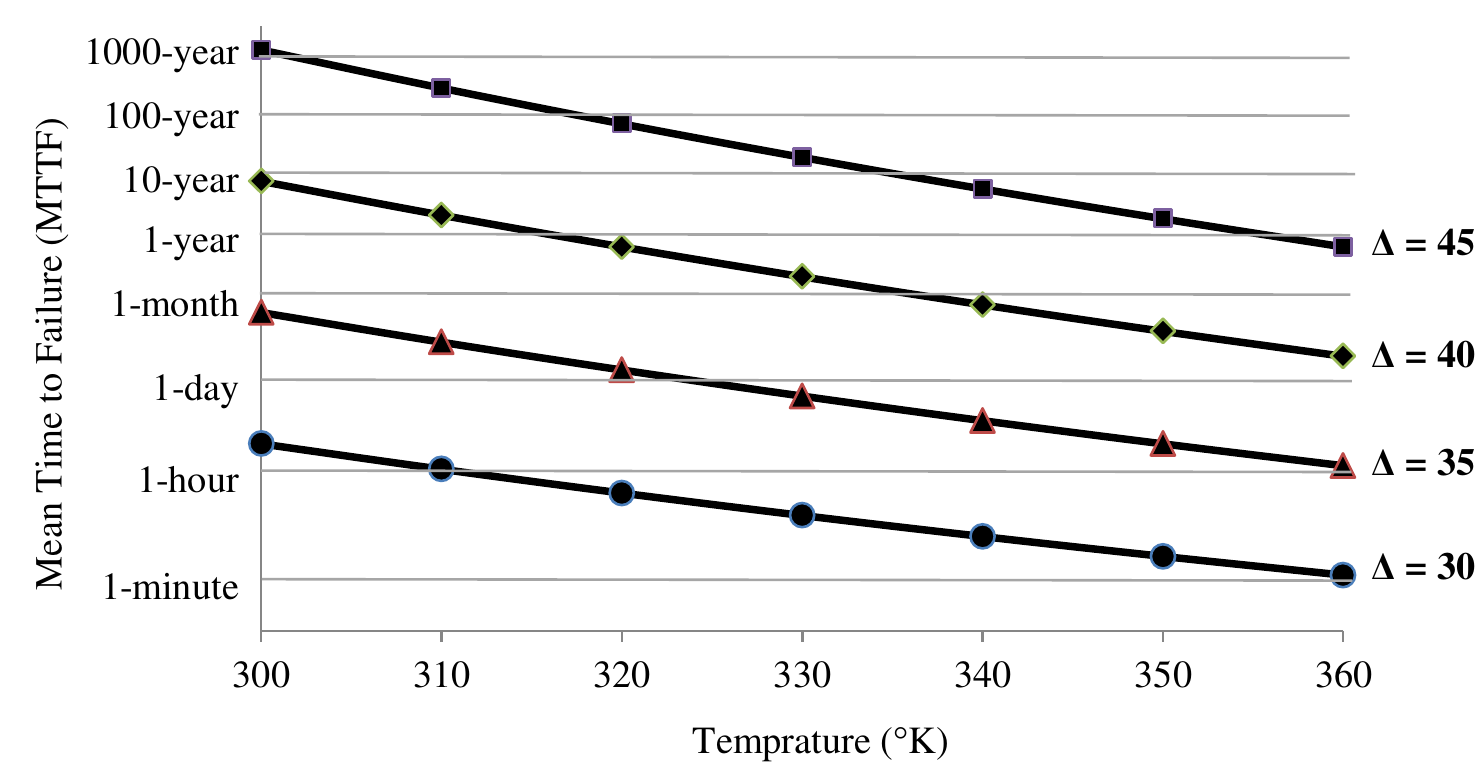}\vspace{-10pt}
				\caption{Mean time to retention failure versus temperature for four different values of thermal stability factor ($\Delta$).}
				\label{fig:fig3}\vspace*{-10pt}
			\end{figure}

			\begin{figure}[b]\vspace{-8pt}
				\centering
				\includegraphics[width=1\linewidth]{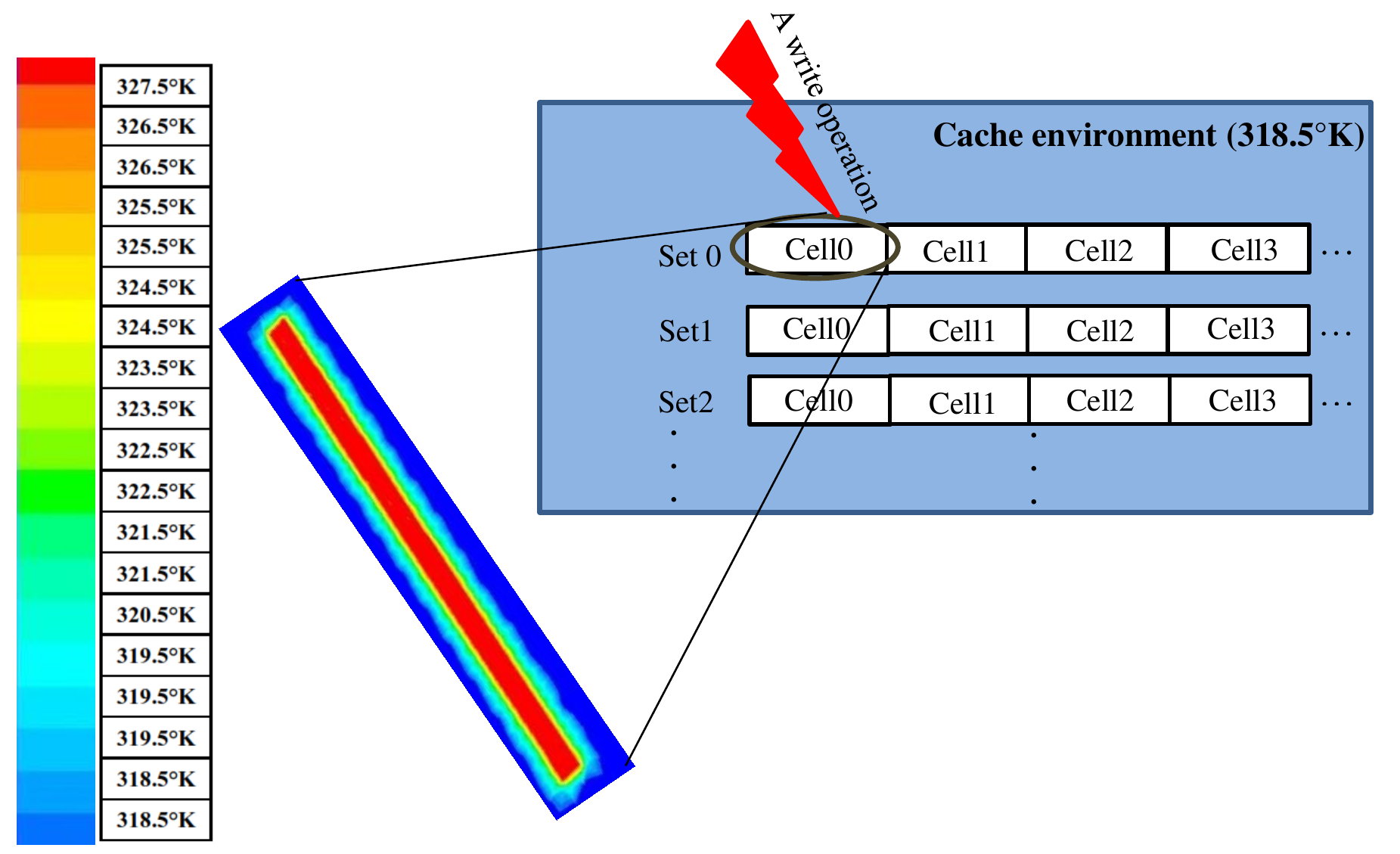}\vspace{-6pt}
				\caption{Temperature increase of a cell due to a write operation.}\vspace*{-10pt}
				\label{fig:figg3}
			\end{figure}

			\section{Motivation} 
			Retention failure occurs stochastically and increases in higher temperature. Thermal effects cause a flip in a STT-MRAM cell that follows Poisson distribution with timing characteristics of \textit{$\tau$$e^{\Delta}$}~\cite{naeimi2013intel}. Equation (\ref{eq:1}) shows the probability of \textit{n} times bit flip in the unit of time \textit{t}.
			\begin{flalign}
			\label{eq:1}
			\resizebox{.36\linewidth}{!}{$ P_{bit-flip} = \frac{\lambda^n e^{-\lambda} }{ n! } $}     \phantom{\hspace{1.2cm}}
			\end{flalign}
			 where \textit{$\lambda$=t/($\tau$$e^{\Delta}$)} is failure rate and \textit{$\tau$} is attempt period, which is assumed to be 1ns~\cite{naeimi2013intel}. If the number of probable bit-flips goes to infinity and it occurs for odd numbers to cause an error, the retention failure probability during time \textit{t} is calculated according to (\ref{eq:2})~\cite{naeimi2013intel}.
			\begin{flalign}
			\label{eq:2}
			\resizebox{.89\linewidth}{!}{$ P_{Ret-Failure} =\displaystyle\sum_{k=0, n=2k+1}^{\infty}P_{flip}(n) = 1- exp(\frac{-t}{e^\Delta}) $}
			\end{flalign}
			The approximated retention failure probability equation shows that this failure rate is exponentially dependent on the thermal scalability factor ($\Delta$). Thermal stability factor is according to (\ref{eq:3})~\cite{naeimi2013intel}.
			\begin{flalign}
			\label{eq:3}
			\resizebox{.18\linewidth}{!}{$\Delta = \frac{E_{b}}{ K T}$}
			\end{flalign}
			where, \textit{E$_b$} is barrier energy, \textit{K} is Boltzman constant, and \textit{T} is temperature. Thermal stability factor ($\Delta$) is inversely proportional to temperature and reduces by increasing the temperature. As observed in (\ref{eq:2}), reduction in $\Delta$ exponentially increases the retention failure rate. Fig. \ref{fig:fig3} depicts the \textit{$Mean~Time~To~Failure$} (MTTF) of a STT-MRAM cell  for some known values of $\Delta$ and shows how MTTF is reduced in higher temperature. For example, MTTF is about 10 years in 300$^{\circ}$K for a STT-MRAM with nominal $\Delta$ of 40 and is reduced to about two days when the temperature increases to 360$^{\circ}$K.

			Next, we explore the temperature effects on read disturbance and write failure rate. The occurrence probability of a read disturbance in a STT-MRAM cell is according to (\ref{eq:4})~\cite{Ran2016JSA}.

			\begin{figure*}[t]
				\centering
				\subfloat[bzip2]{\includegraphics[width=0.88\linewidth]{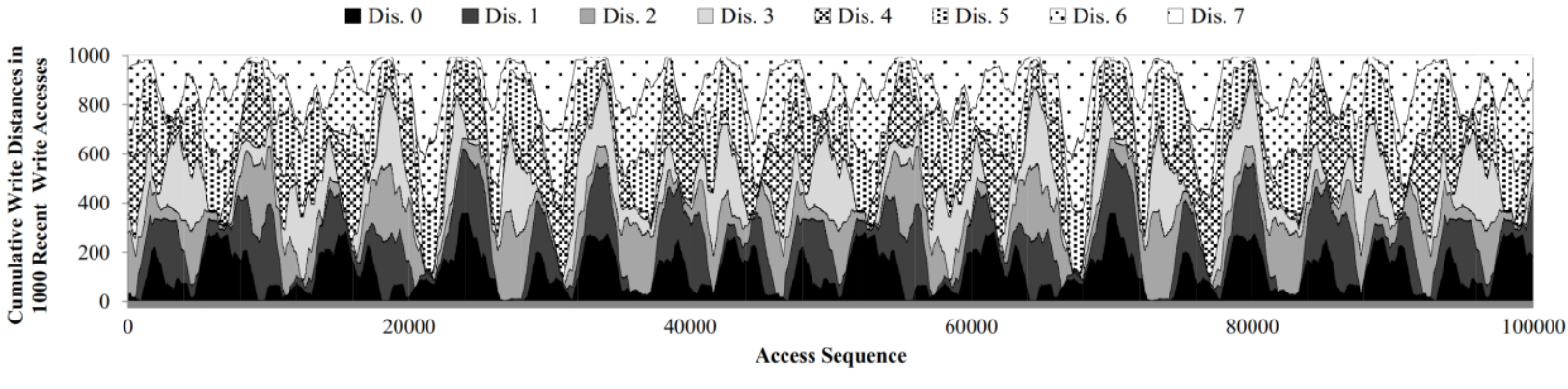}}
				\hfill
				\subfloat[gcc]{\includegraphics[width=0.88\linewidth]{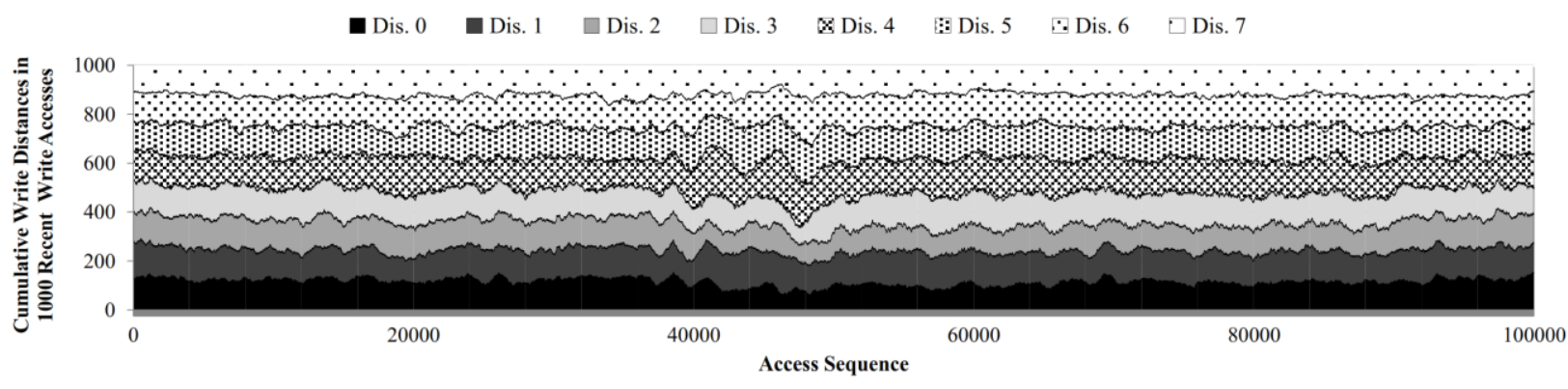}}
				\hfill
				\subfloat[cactusADM]{\includegraphics[width=0.88\linewidth]{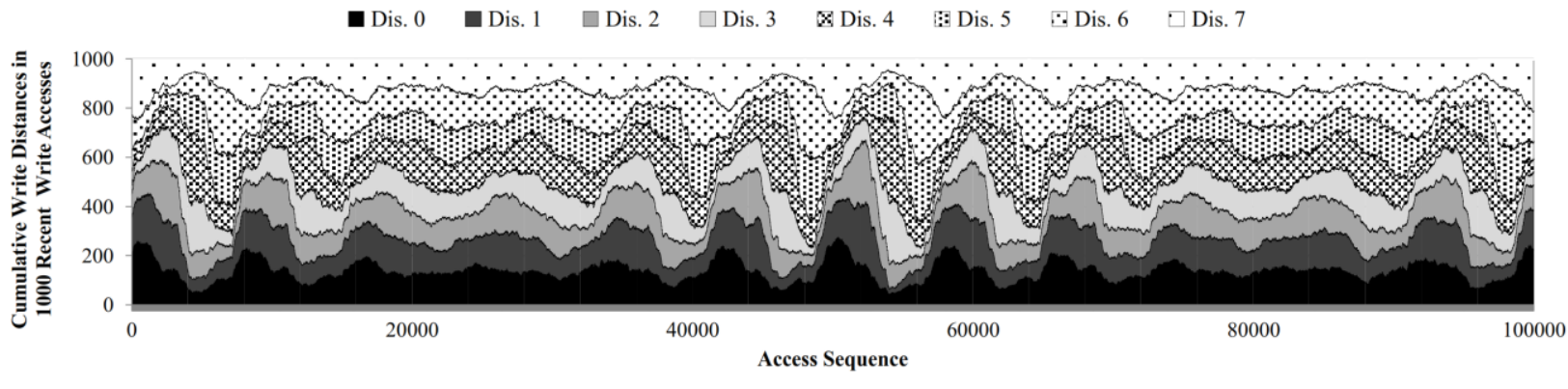}}
				\hfill
				\subfloat[lbm]{\includegraphics[width=0.89\linewidth]{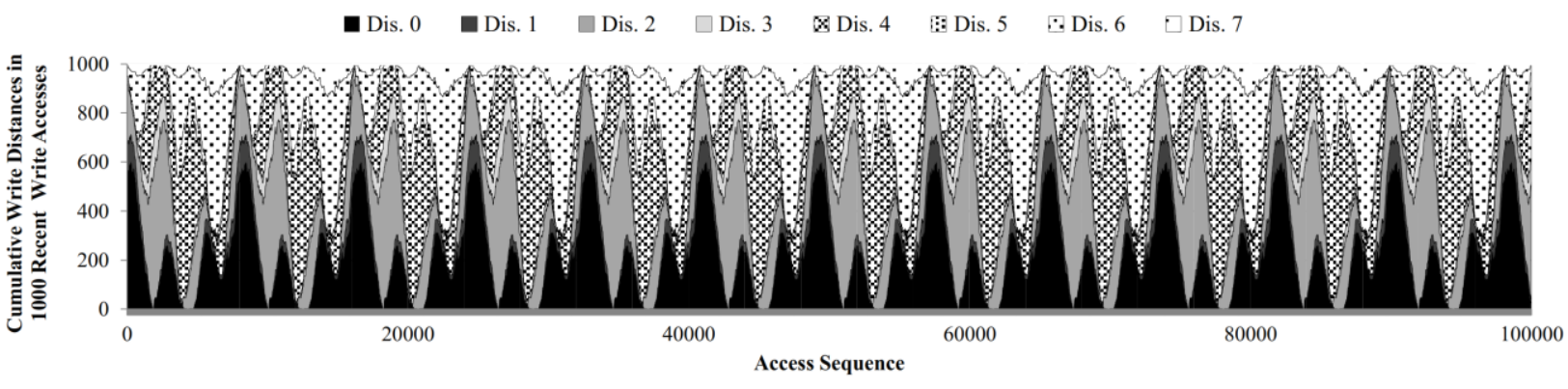}}
				\caption{Accumulated write distances for every 1000 recent writes in (a) bzip2, (b) gcc, (c) cactusADM, and (d) lbm workloads in LRU policy.}\vspace{-8pt}
				\label{fig:4}
			\end{figure*}

			\begin{equation}
			\begin{multlined}
			\label{eq:4}
			 P_{Read-Disturbance} = 1- exp(\frac{-t_{read}}{\tau}\times \\
			\shoveleft[1cm]{exp(\frac{-\Delta(I_{read}-I_{C_0})}{I_{C_0}}))} 
			\end{multlined}
			\end{equation}
where, \textit{$\tau$} is attempt period and assumed to be 1ns, \textit{I$_{read}$} is read current, \textit{I$_{C0}$} is the current needed to write in 0$^{\circ}$K and \textit{t$_{read}$} is the read pulse. As depicted, read disturbance rate exponentially increases by $\Delta$ reduction; and, $\Delta$ reduces by increasing the temperature, as mentioned earlier. The occurrence probability of a write failure for a STT-MRAM cell is shown in (\ref{eq:5})~\cite{Pajouhi2016JETC, AMir2016TDMR}.

			\begin{equation}
			\begin{multlined}
			\label{eq:5}
			 P_{Write-Failure} = exp( -t_{write}\times \\
			\shoveleft[1cm]{\frac{2 \times \mu_{\beta}\times p\times(I_{write}-I_{C_0})}{c+\log_{e}(\pi^2\times\Delta/4)\times (e\times m\times (1+p^2))})} 
			\end{multlined}
			\end{equation}		
where, \textit{I$_{write}$} is write current, \textit{c} is Euler constant, \textit{e} is electron charge, \textit{m}  is magnetic momentum of the free layer, \textit{p} is tunneling spin polarization, \textit{$\mu$$_{\beta}$} is Bohr magneton, and \textit{t$_{write}$} is write pulse width.

			Unlike retention failure and read disturbance, reducing $\Delta$ decreases the rate of write failure. However, by increasing the temperature, which results in $\Delta$ reduction,  \textit{I$_{write}$} is also reduced due to drivability degradation of the access transistor. As demonstrated in~\cite{bi2012analysis}, the adverse effect of temperature on \textit{I$_{write}$} is larger than its positive effect on $\Delta$ and the write failure rate increases in higher temperature.
			As discussed, increasing in temperature, which is mainly due to the heat generated by write operations, significantly increases the total error rate in STT-MRAM cells.

			To quantify the effects of write operations on the cache temperature, we conduct a set of experiments using \textit{Computational}~\textit{Fluid}~\textit{Dynamics} (CFD) tools and calculate the temperature of a STT-MRAM cell for a write operation. Simulation parameters which are obtained from NVSim simulator~\cite{nvsim} are depicted in Table~\ref{table:2}. According to Fig. \ref{fig:figg3}, simulation results show that the STT-MRAM cell temperature increases from about 318.5$^{\circ}$K to 327.5$^{\circ}$K per write operation. 
			This 9$^{\circ}$K increase in the cell temperature can significantly increase the error rates in the cache blocks.

			\begin{table}[t]
				\centering
				\caption{Details of STT-MRAM cell configuration in CFD tools simulation}\vspace{-10pt}
				\includegraphics[width=1\linewidth]{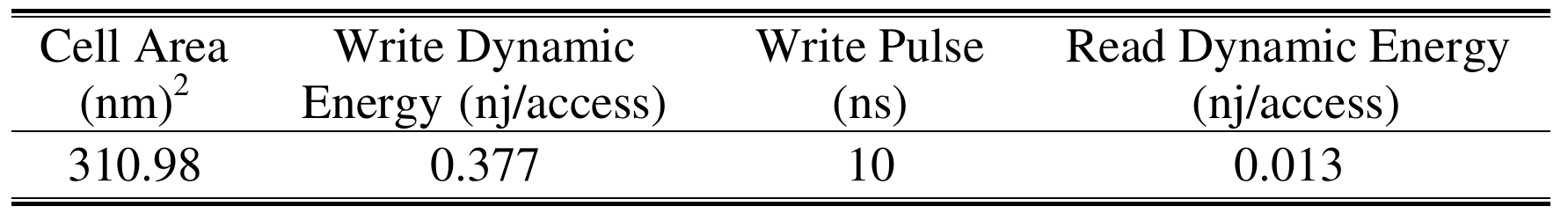}
				\label{table:2}\vspace{-25pt}
			\end{table}

			Although a write operation increases the temperature of the cache for a short interval and the block can cool down, the locality in write operations not only expands the high-temperature region, but also slows down the cooling process of hot blocks. 
			Write operations in L2 caches are not well distributed. It is highly probable that temporally adjacent requests are written in neighboring cache blocks. The locality in write operations leads to heat accumulation in some cache regions and increases the error rates for long intervals.

			To explore non-uniformity in a sequence of write operations in L2 cache, we extract the write locations in the cache sets using gem5 cycle-accurate simulator~\cite{gem5} running the workloads of SPEC CPU2006 benchmark suite~\cite{spec2006}. The details of system configuration are given in Table \ref{table:3} in Section 5. 
			Considering an 8-way set-associative cache, the distance of two consecutive writes can be any value between zero and seven. For example, if the most recent data is written in block 2 and the current data is written in block 7, the distance of these two consecutive writes is five. Fig. \ref {fig:4} shows the breakdown of write distances for four workloads, i.e., \textit{bzip2}, \textit{gcc}, \textit{cactusADM}, and \textit{lbm} as examples of write locality\footnote{The results for the other benchmarks are given in Fig. \ref{fig:app1} in the Appendix.}. For the sake of visibility, the distances of most recent 1000 write operations is depicted in each access for the workloads and the Y-axis is the cumulative write distances for these recent writes.

			Considering \textit{bzip} workload in Fig. \ref{fig:4}(a), there are several intervals in which the write distance of more than 20\% of accesses is zero, indicating that data is written into the hottest block. In addition, the zero and one write distances contribute in more than 40\% of writes for considerable intervals. According to Fig. \ref{fig:4}(b), more than 20\% of incoming blocks are written into the most-recently written block (write distance 0) or into its adjacent block (write distance 1) throughout the \textit{gcc} workload execution interval. Fig. \ref{fig:4}(c) shows that for \textit{cactusADM} workload, in majority of time intervals, write distances for more than 30\% of writes are zero or one. Finally, there are several intervals in which more than 60\% of incoming blocks are written into the hottest block in \textit{lbm} workloads, as shown in Fig. \ref{fig:4}(d). 
			
			These observations confirm the existence of a high temporal locality in writing into adjacent cache blocks. 
			The above observations also indicate a high spatial locality in consecutive write operations in which a considerable number of incoming data is written into the recently written block or one of its adjacent blocks. These localities in write operations can cause heat accumulation in recently written cache regions, which in turn can increase the error rate.

			\section{Proposed TA-LRW Policy}

			Write operations in an L2 cache are from two sources, i.e., writebacks from L1 caches and read misses. On a read miss, the cache replacement policy selects a victim block to replace its content with the new incoming data. On a writeback, if the older version of the data is already available in the cache, the block will be overwritten; otherwise, the cache replacement policy determines a block that the incoming data block should be written into.
		
			In any case, the location of the write operation in a cache set either is determined by the cache replacement policy, in the case of a cache miss, or is predetermined, in the case of a writeback hit. A large fraction of accumulated heat in cache sets is due to decisions of cache replacement policy on cache misses and the inevitable decision of cache controller for overwriting the cache blocks on writeback hits.

The goal of TA-LRW is evenly distributing the heat generated by consecutive write operations in the blocks. SLC- or MLC-based cells as well as in-plane or perpendicular cell structure do not affect the functionality of TA-LRW, since it operates on blocks of a cache set. Furthermore, as the write operation in MLC STT-MRAM caches requires larger write current and consumes more energy~\cite{chen2017energy}, it imposes more heat accumulation. Therefore, MLC structure in STT-MRAM memories further exacerbates the heat accumulation challenge. In addition, MLC cell faces \textit{$write~disturbance$} as another error type besides the three well-known errors in SLC STT-MRAM, which introduces a new challenge to the STT-MRAM reliability~\cite{chen2017energy, seoking2014iccd}.

			Meanwhile, it is predicted that 3D stacked design is the future structure of STT-MRAM caches~\cite{beigi2016tesla}. Temperature-induced error rate is more severe for 3D STT-MRAM caches than its 2D counterpart due to the more difficult heat transferring in 3D STT-MRAMs. Hence, employing TA-LRW in 3D and/or MLC STT-MRAM caches is even more crucial than in their 2D and/or SLC counterparts.
	
			\subsection{Basics of TA-LRW Policy}
			We have already observed a high probability of heat accumulation in LRU policy, as the most conventional and widely-used replacement policy, in Fig. \ref{fig:4}. To reduce the heat accumulation in cache sets, our approach is to write the consecutive incoming data in non-adjacent blocks. This approach requires a different policy to select victim blocks than that of the existing replacement polices as well as redirecting a writeback hit from an already available block to another block.

			The primary goal of the replacement policies is to minimize the cache miss rate based on localities in accessing cache blocks. Replacing the cache blocks based on the location of the recent writes should not violate the locality principle in the cache. To this aim, our proposed cache replacement policy not only selects a suitable block to be written on a cache miss, but also redirects the overwrite operation of already available data to a suitable block. Meanwhile, as the primary goal of the existing cache replacement policies, it provides a low miss rate and high performance.

			Our proposed cache replacement policy, named \textit{$Thermal$}-\textit{$Aware~Least$}-\textit{$Recently~Written$} (TA-LRW), offers four main features: 1) all consecutive incoming data are written in non-adjacent blocks, 2) the write operations are evenly distributed over all blocks in a set, 3) it replaces the cache blocks based on a semi-LRU decision to provide a near-LRU performance, 4) its implementation cost is as low as FIFO replacement policy and does not require the LRU age bits per block to keep track of access history as well as the complicated LRU controller and operations for each access. In the following, we explain the TA-LRW replacement policy in more details.

			In TA-LRW policy, there is a pointer, named \textit{$write~pointer$}, in each cache set to indicate the next block that should be written. This pointer traverses over all blocks in a set band returns back to its original location. This round-robin writing strategy guarantees that no recent \textit{N} incoming data blocks are written in the same cache block.
			To guarantee that no sequence of incoming data is written in adjacent blocks, TA-LRW updates the pointer after each write in a way that the next write is performed in a far enough block.

			\subsection{Write Access Manipulation}
			
			As we will show later in Section 5, the majority of writes in LRU policy are performed in the same block as that of the previous write or in their adjacent blocks. Using TA-LRW policy, it is guaranteed that the distance between two consecutive writes in a set is at least three blocks and the written blocks have enough time to cool down until their next write. To this aim, we need to find a permutation of sequence of writes into blocks in which physically adjacent blocks are located in distant places of permutation.

\begin{figure}[t]
				\centering
				\includegraphics[width=1\linewidth]{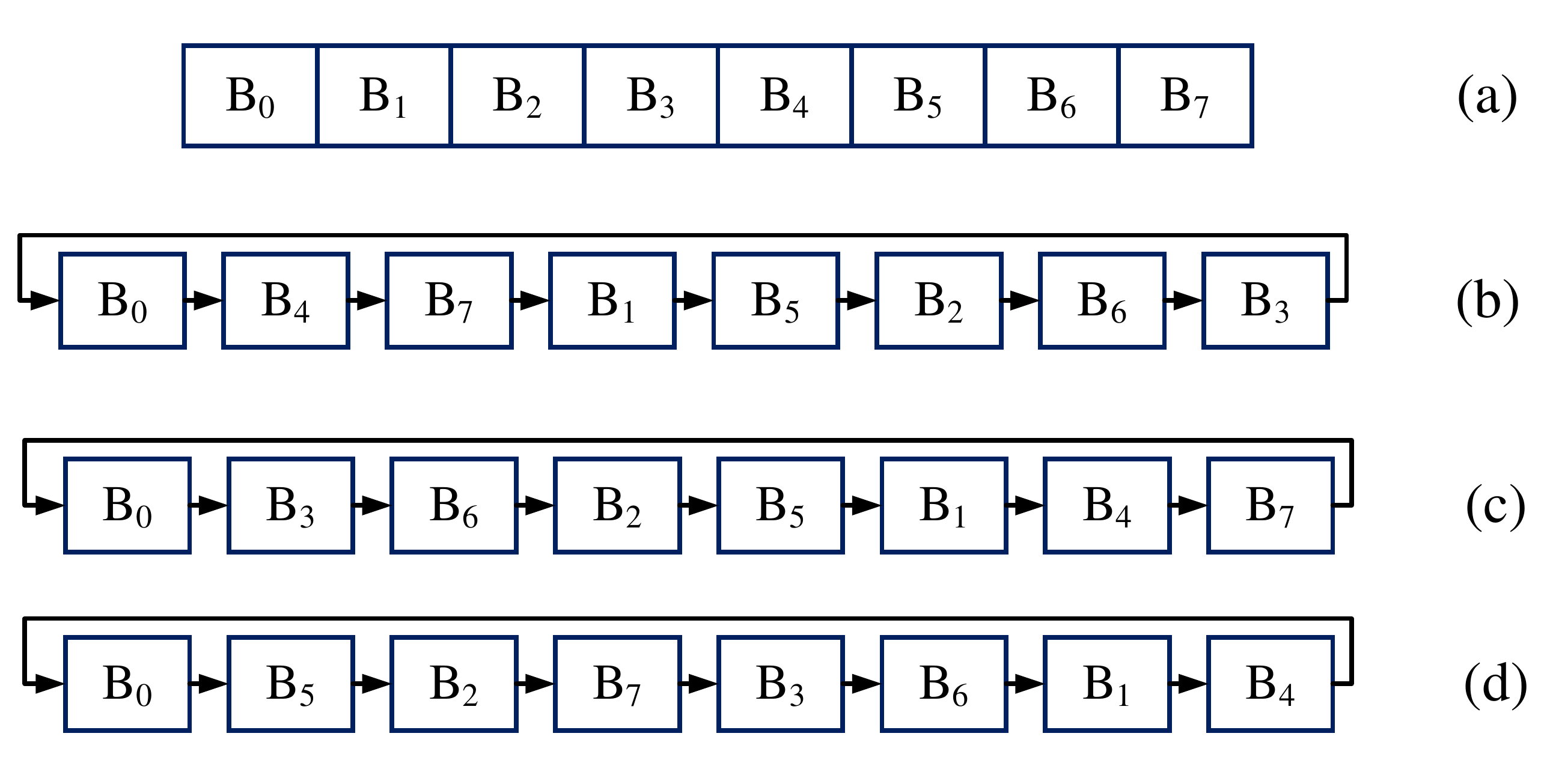}\vspace{-10pt}
				\caption{Three samples of suitable permutations for TA-LRW policy: (a) cache blocks in a set, (b) first permutation for sequence of writes, (c) second permutation, (d) third permutation used for this work.}\vspace{-10pt}
				\label{fig:5}
		\end{figure}

			A set of 8-way set-associative cache is shown in Fig. \ref{fig:5}(a). Three permutations that satisfy the TA-LRW requirements are shown in Fig. \ref{fig:5}(b)-(d). In these permutations, the distance between each two consecutive writes is at least three blocks. For the permutation in Fig. \ref{fig:5}(b), after one round in which all blocks are written once, the distance of four out of eight consecutive writes is three blocks and the distance of three and one out of eight consecutive writes is four and 	six blocks, respectively. In each round for the permutation in Fig. \ref{fig:5}(c), the distance of five, two, and one out of eight writes is three, four, and seven blocks, respectively.

			The third permutation (Fig. \ref{fig:5}(d)) includes three different distances for the sequence of writes. The distance of three and two out of eight writes in each round is three and four, respectively, and the distance of the other three writes is five. There are several other permutations in which the write distances are three or more blocks. It also can theoretically be proven that there is no permutation with write distance of at least four. Therefore, the best permutations are those with minimum write distance of three. 
			
			We have generated all permutations (8! = 40,320) for 8-way cache out of which the minimum distance between every two adjacent number is three for 176 permutations. All of these 176 permutations guarantee that every two consecutive incoming data in a cache set are written in blocks with the distance of at least three.
			The implementation complexity and performance of these permutations are the same and their heat distributions are not much different.		
			To find the best permutation between the existing 176 suitable permutations, we first analyze how a write operation in each block increases the temperature of blocks in a set based on their distance from the write location. Then, we calculate the total heat accumulation after one round of writing into all blocks for these permutations. The best choice is a permutation with minimum heat accumulation. We consider the permutation in Fig. \ref{fig:5}(d) in this work based on the assumed values of heat accumulation for each write distance.

		\subsection{Architectural Aspects of TA-LRW}
			TA-LRW policy selects the victim block to be replaced according to its write pointer value for a read miss or writeback miss. The block indicated by the pointer is the least-recently written block in the set and is the most suitable block to be written for preventing the heat accumulation.
			For a writeback hit, existing replacement policies, e.g., LRU policy, have no choice other than overwriting the found block. However, it is probable that this block is among the most-recently written blocks and the current write operation leads to the accumulation of heat in the block region. TA-LRW policy overwrites the found block only if the block is selected by the write pointer. Otherwise, the block is invalidated and the incoming data is written in a block indicated by the pointer.

	\begin{figure}[t]
				\centering
				\includegraphics[width=0.95\linewidth]{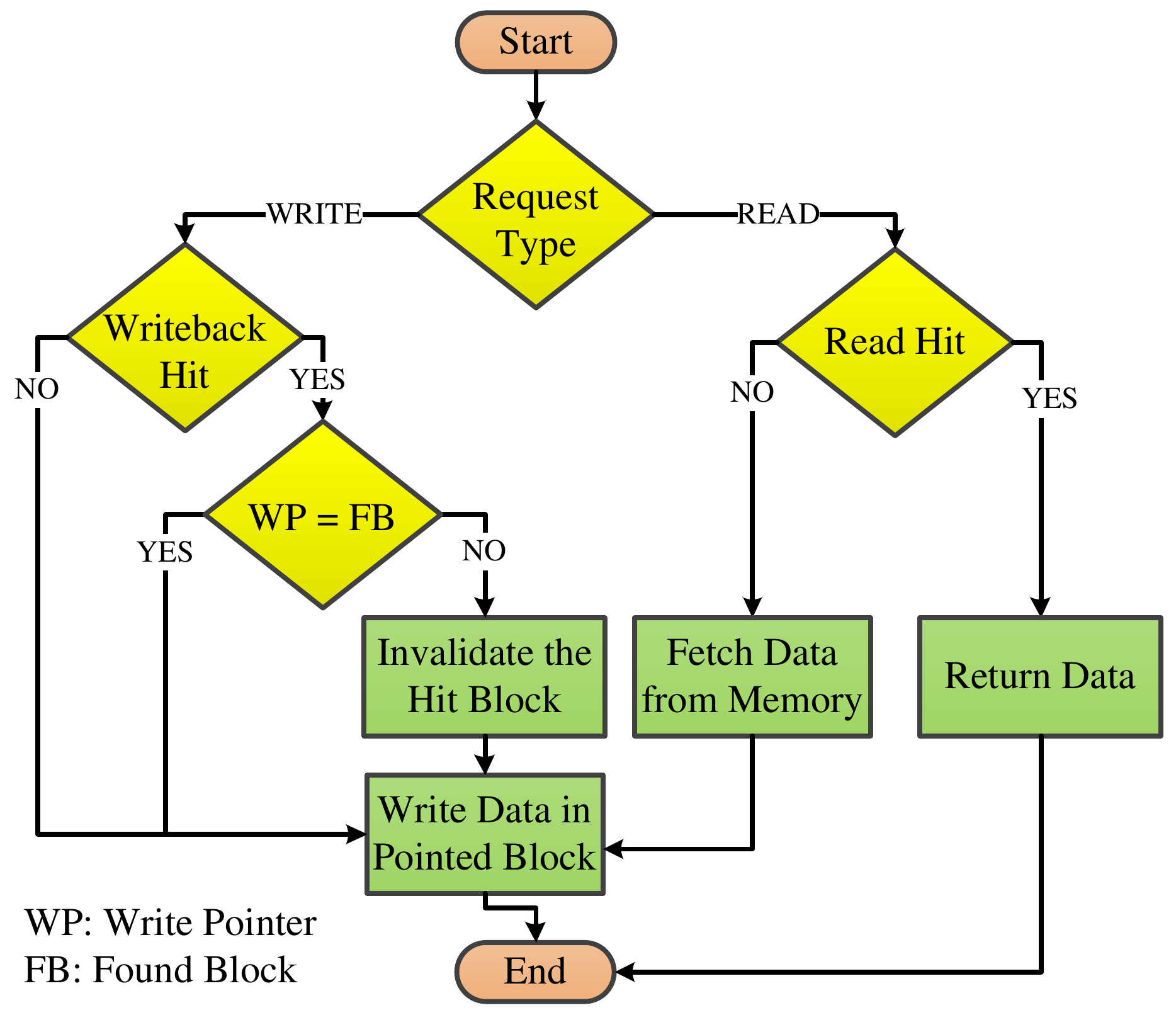}\vspace{-5pt}
				\caption{
				Cache operation on an access request based on TA-LRW policy.}
				\label{fig:flowchart}\vspace{-5pt}
	\end{figure}	

		Beside its uniform write distribution and capability in preventing heat accumulation, another main feature of TA-LRW policy is its simple yet effective victim block selection process. Despite the popularity and widespread use of the LRU policy due to a high performance provided, its complicated controlling operations and the required peripherals limit its applicability in high associative caches~\cite{sudarshan2004highly}, \cite{liu2008cache}. TA-LRW policy provides a near-LRU performance with a very simple and low-cost controller.

		TA-LRW sorts the blocks in a set based on their write time and evicts the oldest block on a cache miss. Both LRU and TA-LRW policies select the victim block based on the age of the blocks. While LRU policy updates the ages for every read/write accesses, TA-LRW policy updates them only for every write accesses. The oldest block in LRU policy is the least-recently read/written block and in TA-LRW policy is the least-recently written one.

Since the accesses to cache blocks are a mix of read and write operations, sorting the blocks based on their write sequence is a suitable representative for sorting them based on their read/write sequence. More clearly, the least-recently written block is highly probable to be among the elder blocks, and therefore TA-LRW policy likely evicts an old block, if not the oldest one, on a cache miss. On the other hand, recently written blocks (younger blocks in TA-LRW policy) are highly probable to be among the recently accessed blocks (younger blocks in LRU policy). Therefore, similar to LRU policy, they are not evicted from the cache on a cache miss but gotten older in TA-LRW policy. Fig. \ref{fig:flowchart} depicts the TA-LRW decision according to the type of the cache access request.

\begin{figure}[t]\vspace{-8pt}
				\centering
				\subfloat[First method]{\includegraphics[width=0.48\linewidth]{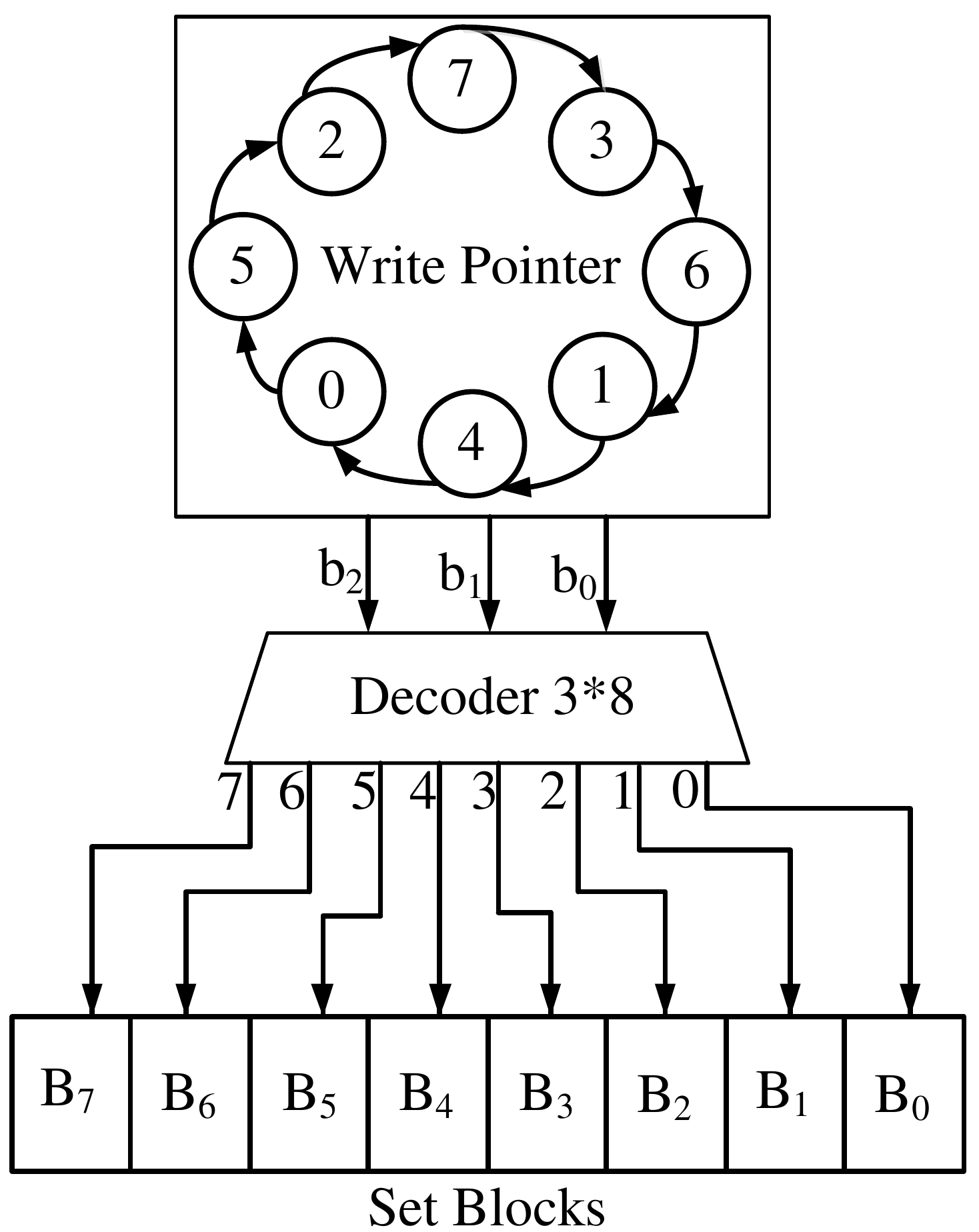}}
				\hfill
				\subfloat[Second method]{\includegraphics[width=0.48\linewidth]{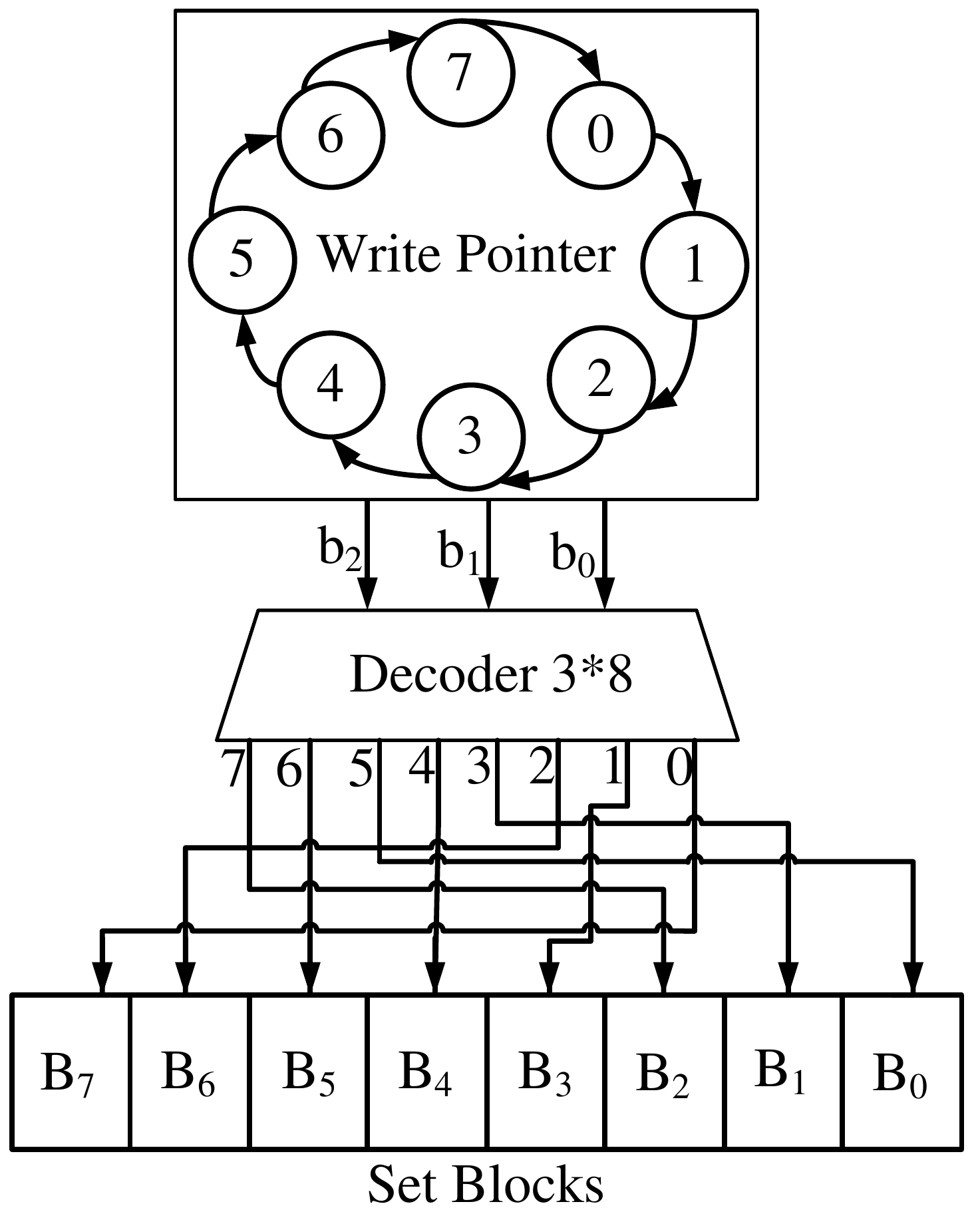}}
				\caption{Two implementations of TA-LRW pointer to select the suitable block to be written: (a) updating pointer according to TA-LRW permutation, (b) connecting decoder output to blocks according to TA-LRW permutation.}
				\label{fig:6}\vspace{-5pt}
			\end{figure}

		\begin{figure*}[t]
				\centering
				\includegraphics[width=1\linewidth]{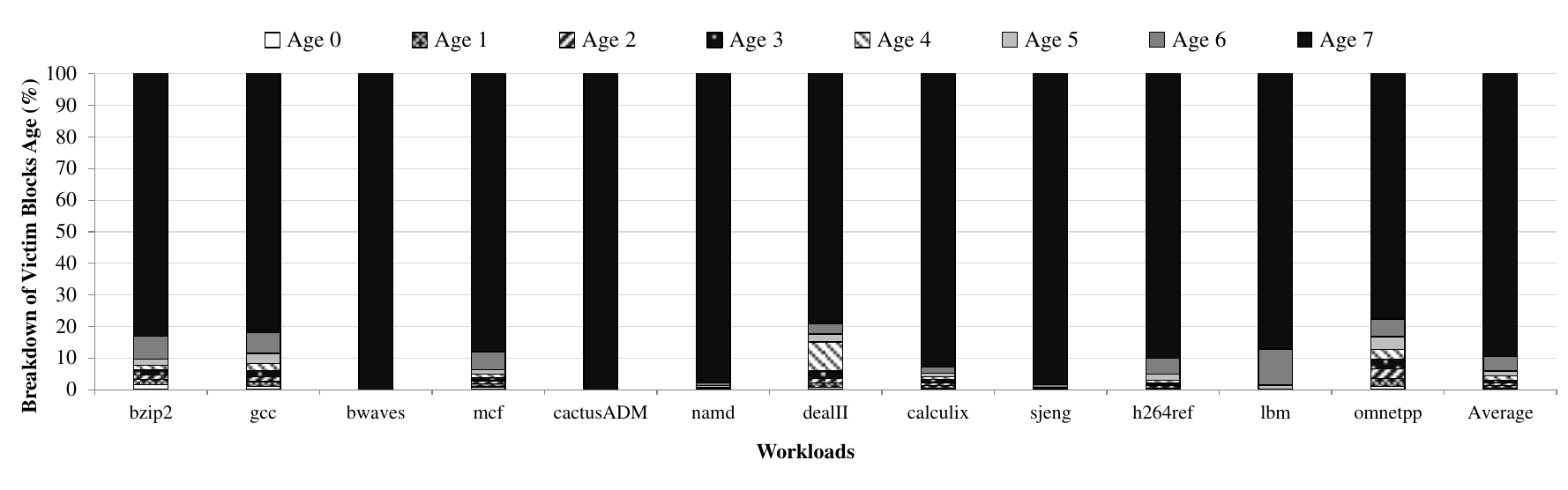}\vspace{-20pt}
				\caption{Age of evicted blocks in TA-LRW replacement policy.}\vspace{-12pt}
				\label{fig:9}
			\end{figure*}

			In summary, the age of blocks in TA-LRW policy is determined the same as that in LRU policy except that the read accesses have no effect on the ages. The detailed results will be given in Section 4.4 to show that the age of blocks in TA-LRW policy is very close to that in LRU policy. Therefore, TA-LRW policy provides semi-LRU functionality with much simpler operations. Unlike LRU policy, which requires an age field per cache block and a complicated logic to manage these ages per access, the main component in TA-LRW policy is write pointer per cache set, which can be implemented simply in two ways shown in Fig. \ref{fig:6}. The first implementation is shown in Fig. \ref{fig:6}(a) in which the write pointer is updated according to TA-LRW write sequence permutation. The second implementation is to shuffle the output of way selection decoder according to the TA-LRW write permutation, as depicted in Fig. \ref{fig:6}(b).



		\subsection{Selection of Proper Victim Blocks}	

		The primary goal of cache replacement policies is to provide higher performance by evicting the block that is predicted not to be needed for a longer time in the future. This prediction in both LRU and TA-LRW policies is based on the recent history of accessing to blocks.			
			
	LRU policy always evicts the oldest block (the least-recently read/written) on a cache miss. TA-LRW policy has an approximation on the access history for the block, as it only keeps track of write access history. The performance overhead of TA-LRW depends on the accuracy of this approximation. TA-LRW will be closer to LRU in term of performance by discarding the elder blocks on a cache miss. The ages of discarded blocks, based on their recent accesses, in TA-LRW policy demonstrate how close this policy is to LRU policy. Fig. \ref{fig:9} shows the ages of total blocks discarded by TA-LRW policy for all workloads. The age of a block can be any value from zero to seven in our 8-way associative cache.

			On average, 89.5\% of discarded blocks are the oldest one in a set and the block selected by TA-LRW policy is the least-recently used one. Therefore, for an average of 89.5\%, TA-LRW policy decision is the same as that of LRU policy. TA-LRW policy discards the second oldest blocks (blocks with age 6) for 4.6\% of cache misses, on average, and 1.5\% of evictions are on blocks with age 5. Therefore, an average of 96.6\% of all discarded blocks in TA-LRW policy is either the oldest block or among these elder blocks in a set. 

			TA-LRW policy discards the youngest block for only 0.5\%, on average, for all workloads and only 0.8\% of discarded blocks are the second youngest block, on average. For 3.1\% of cache misses, the age of discarded blocks is two, three, or four. As observed, the majority of victim blocks selected by TA-LRW policy are old enough and suitable for eviction. Therefore, it can be expected that the impact of the proposed replacement policy on the cache miss rate and performance is minimal.

			\begin{table}[tb]
				\centering\vspace{-1pt}
				\caption{Details of processor and caches configuration}\vspace{-10pt}
				\includegraphics[width=1\linewidth]{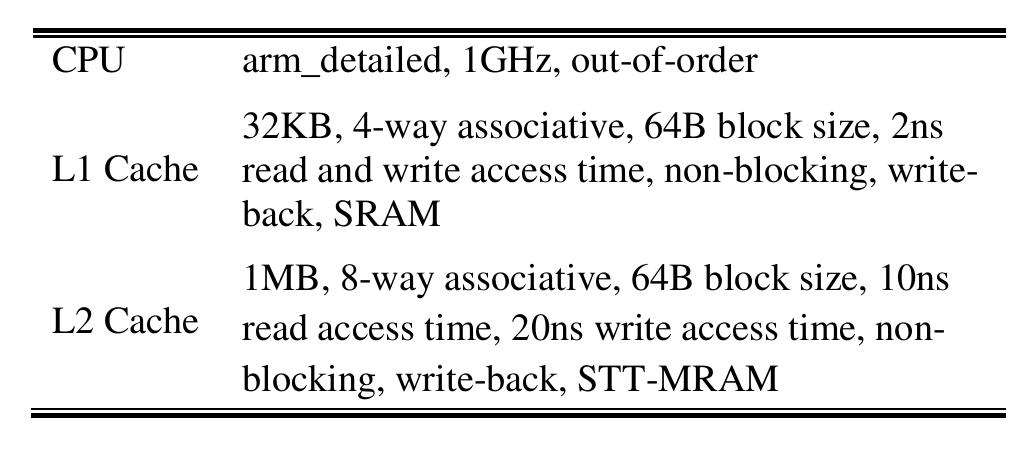}
				\label{table:3}\vspace*{-25pt}
			\end{table}
		
			\section{Simulation Setup and Results}	

			To evaluate the proposed TA-LRW policy, we implement it in gem5 cycle-accurate simulator~\cite{gem5} and use SPEC CPU2006 benchmark suite~\cite{spec2006} as the workloads. TA-LRW policy is employed in the L2-cache of our modeled processor. The processor includes a dedicated 4-way instruction and data L1-cache and a shared 8-way L2-cache. The configuration details are given in Table \ref{table:3}. For each workload, the first one billion instructions are skipped as warm-up phase and the results are extracted from the next one billion instructions.

			TA-LRW policy is compared with the well-known LRU policy in terms of write distribution, heat accumulation, error rate, and performance. To better illustrate how the performance of the simple TA-LRW policy approximates that of the complicated LRU policy, we also reports the performance of the FIFO policy, as a replacement policy similar to TA-LRW in terms of implementation cost and complexity. 
			
			\begin{figure*}[t]\vspace{5pt}
				\centering
				\includegraphics[width=1\linewidth]{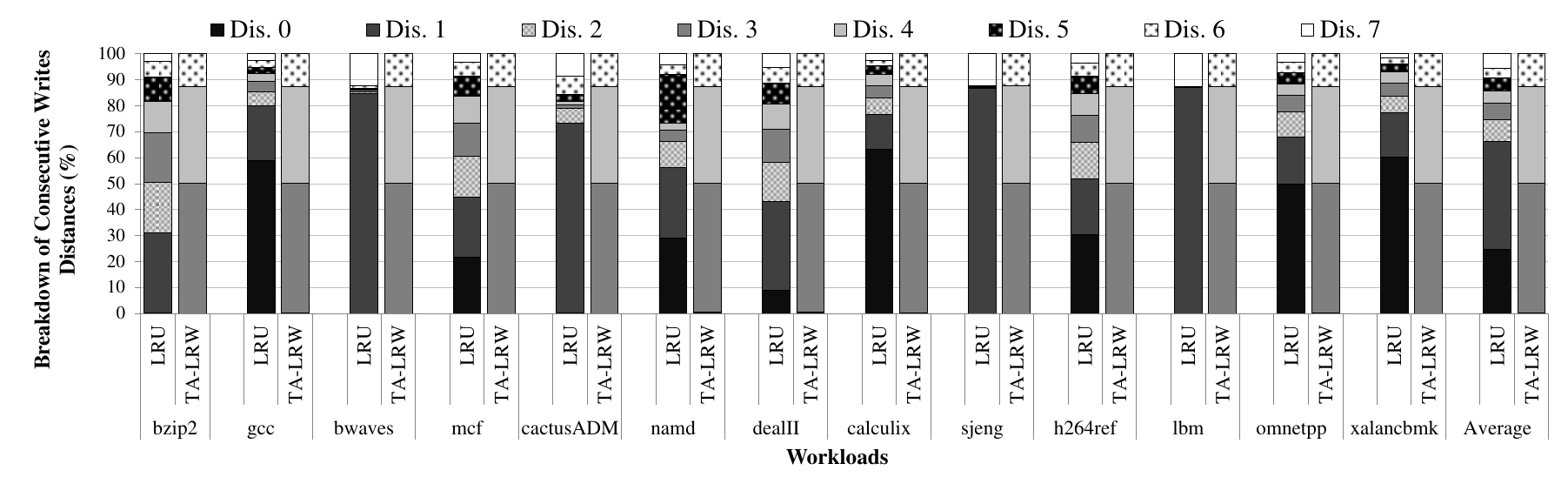}\vspace{-10pt}
				\caption{Contribution of each write distance in total write operations for LRU and TA-LRW replacement policies.}\vspace{-15pt}
				\label{fig:7}\vspace{5pt}
			\end{figure*}

			\subsection{Write Distribution}
			
		 	The distance between each two consecutive write operations in TA-LRW policy is three, four, or five blocks. As mentioned in Section 4, these distances are for three, two, and three out of eight consecutive write operations, respectively, for each round of writing all eight blocks in a set. Therefore, it can be predicated that 37.5\% of incoming data blocks are written by three blocks away from the previous write in a set. The write distance for another 25.0\% of incoming data is four blocks and the remaining 37.5\% are written in blocks that are away from the previously written block by five blocks. 

			These values are the same for all workloads and are independent of the cache access pattern. On the other hand, the distance between two consecutive writes in LRU can be any value from zero to seven. In the worst case, this distance is zero and the most-recently written (hottest) block is written again, leading to high accumulation of heat.

			In the case of one-block distance, the currently written block is adjacent to the hottest block and again leads to high heat accumulation. On the contrary, for large values of the write distances, e.g., five, six, and seven, two consecutive writes are so far to each other that almost no heat is accumulated. The heat accumulation in LRU policy depends on the contribution of write distances in total write accesses in the cache. The higher contribution of large distances, the lower heat is accumulated. However, the results for LRU policy show that the majority of incoming data blocks are written in neighboring cache blocks.

			Fig. \ref{fig:7} depicts the contribution of each write distance in total write operations for LRU and TA-LRW polices in all workloads. As shown, only distances of three, four, and five blocks contribute in TA-LRW policy by 37.5\%, 25.0\%, and 37.5\%, respectively. While TA-LRW policy guarantees the minimum distance of three blocks, the write distance for all workloads in LRU policy is less than three blocks for more than 50.0\% of write operations. On average, 24.8\% of write operations are performed in a block similar to the previous write and the distance of 41.3\% of write operations is only one block. Therefore, more than 66\% of write operations in LRU policy are performed in the hottest block or its adjacent blocks.

		\subsection{ Heat Accumulation}

\begin{figure}[tb]
				\centering
				\includegraphics[width=.81\linewidth]{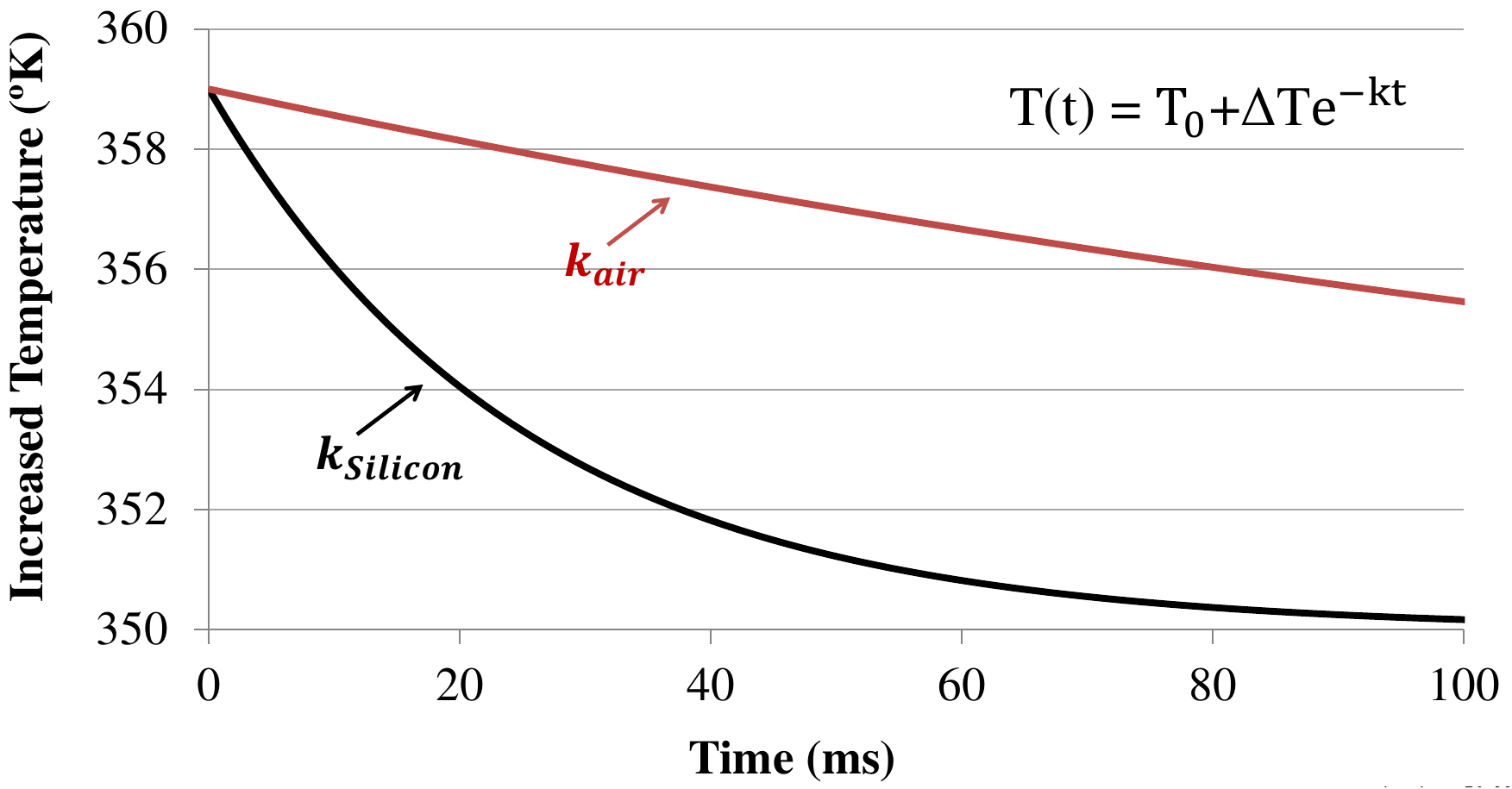}\vspace{-10pt}
				\caption{
				Cooling down of a heat accumulated block to base temperature of the cache over the time.}
				\label{fig:tempInTime}\vspace{5pt}
			\end{figure}

			To calculate the accumulated heat for write operations, we need to indicate the temperature increase of the target block and its neighboring blocks based on the distances from the write location. Our experiments using CFD tool show that the temperature of the written block increases by 9$^{\circ}$K just after committing the write operation.
Afterward, the temperature of the neighboring blocks in the target set and its adjacent sets is increased according to Table \ref{table:4}, considering the equation presented in~\cite{sellitto2014flux} for radial heat transfer in silicon nanolayers.  
				On the other hand, the temperature of these blocks cools down toward their surrounding area afterward according to Fig. \ref{fig:tempInTime} based on Newton's law of cooling.

\begin{figure*}[t]
				\centering\vspace*{-20pt}
				\subfloat[dealII in LRU policy]{\includegraphics[width=0.43\linewidth]{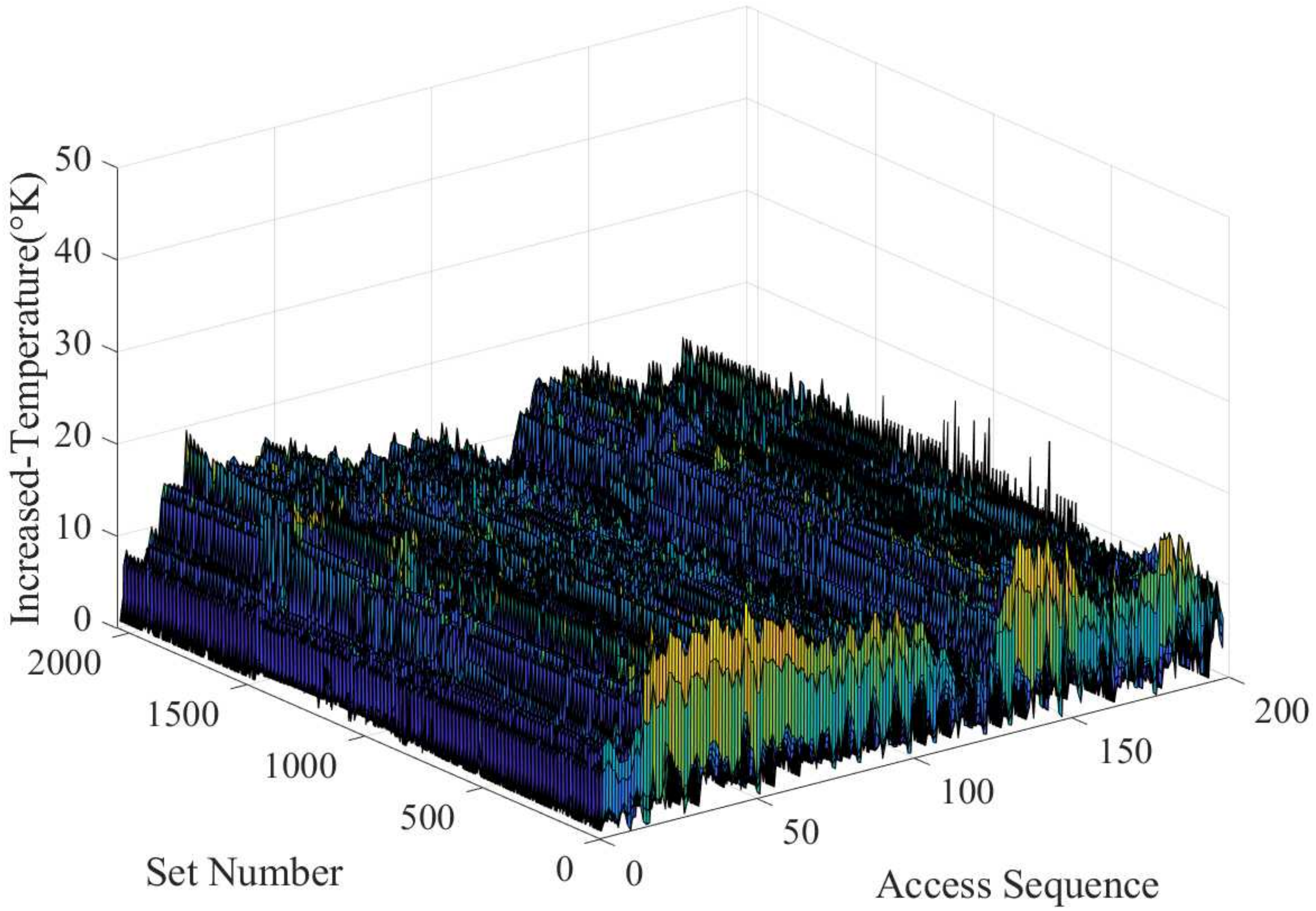}}
				\hfill
				\subfloat[dealII in TA-LRW policy]{\includegraphics[width=0.43\linewidth]{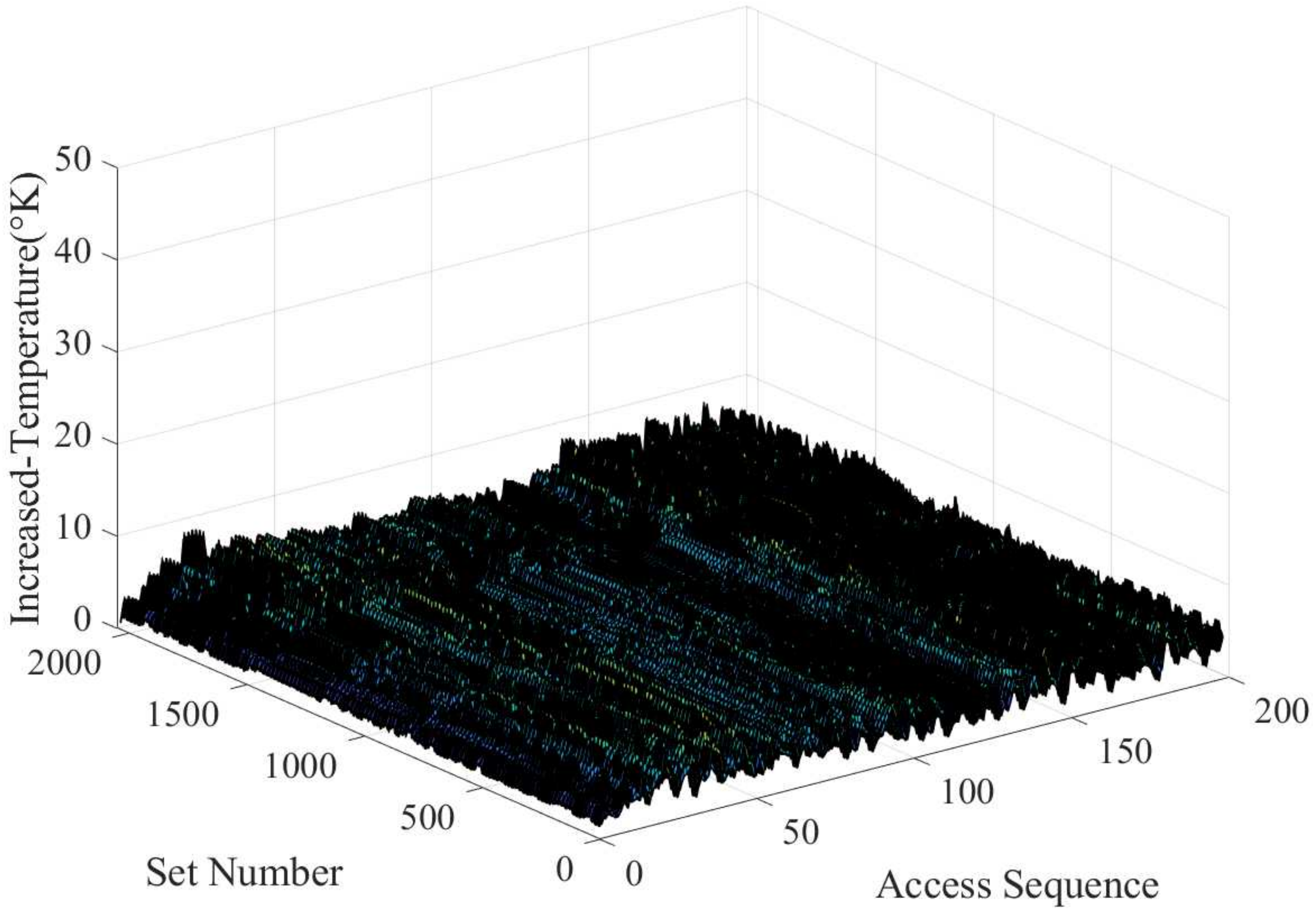}}\\
				\subfloat[namd in LRU policy]{\includegraphics[width=0.43\linewidth]{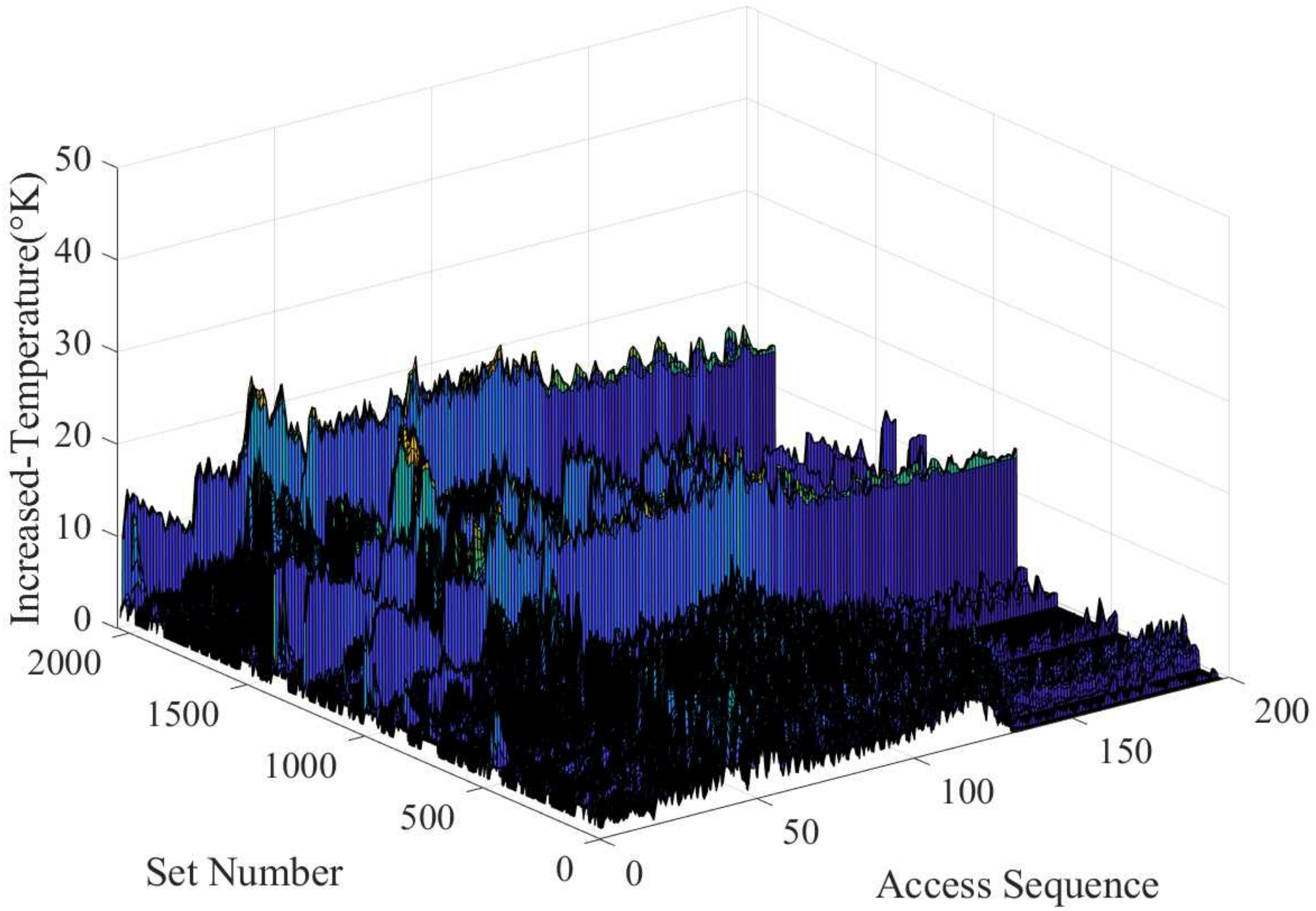}}
				\hfill
     				\subfloat[namd  in TA-LRW policy]{\includegraphics[width=0.43\linewidth]{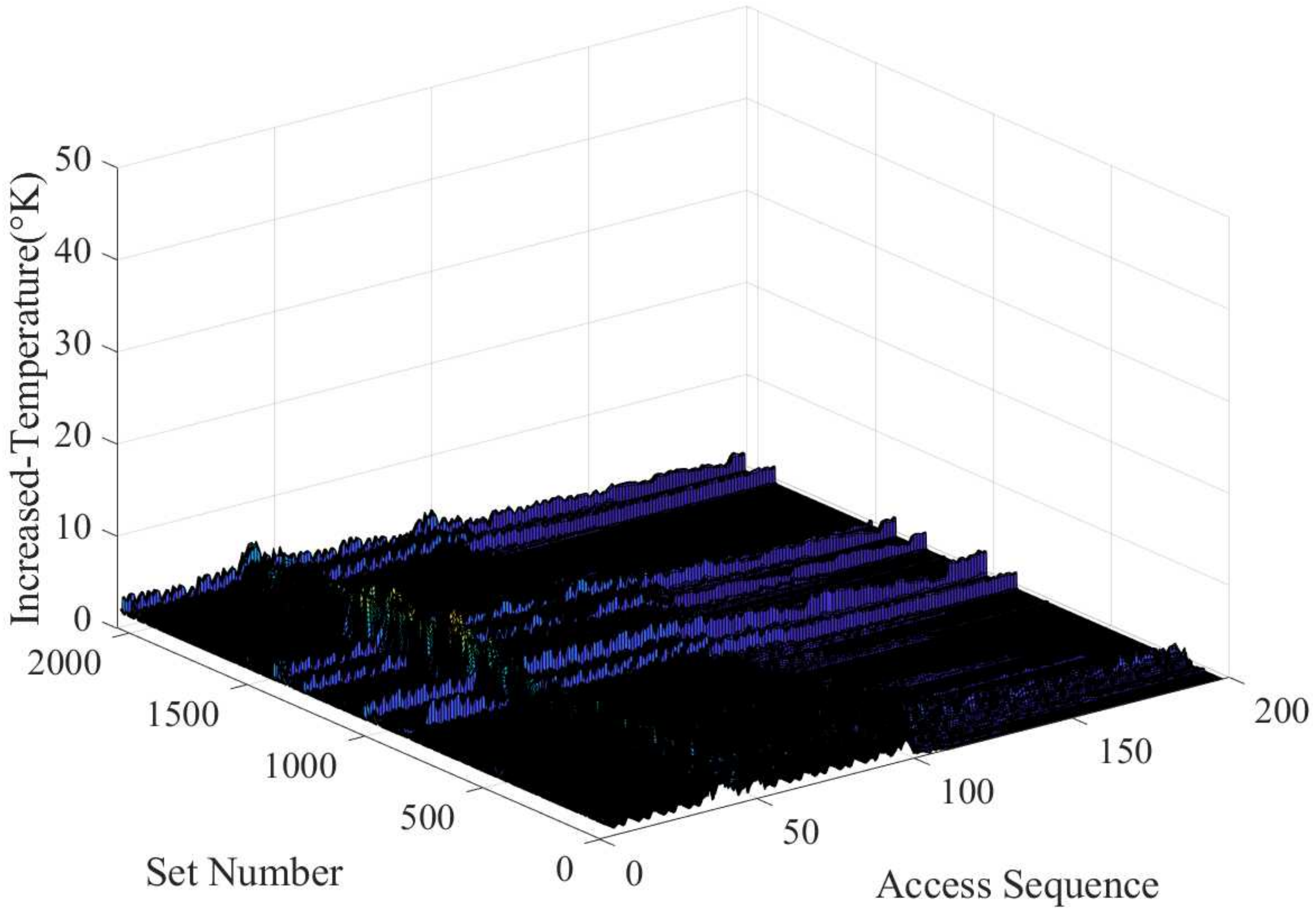}}\\
				\subfloat[h264ref in LRU policy]{\includegraphics[width=0.43\linewidth]{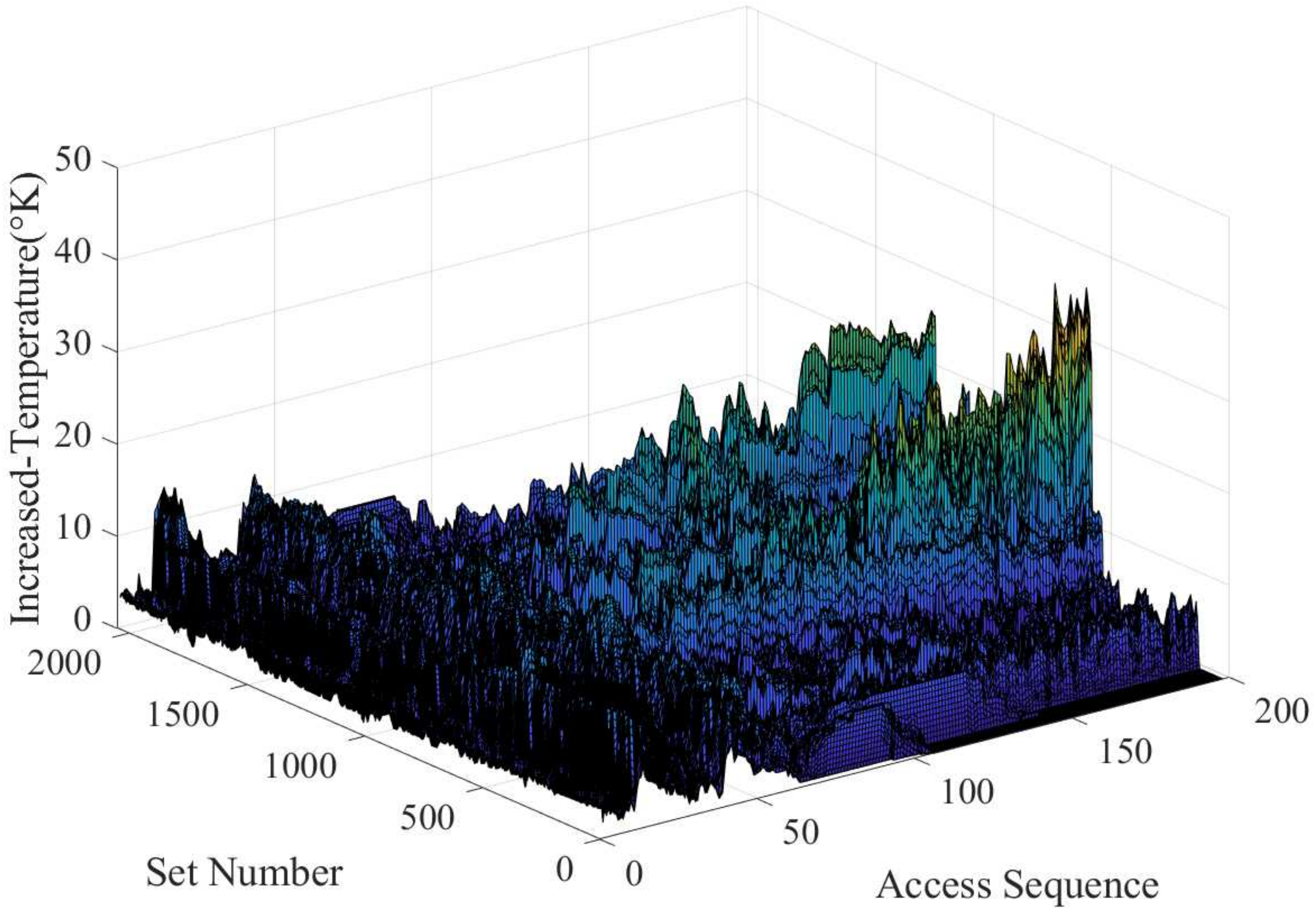}}
				\hfill
				\subfloat[h264ref in TA-LRW policy]{\includegraphics[width=0.43\linewidth]{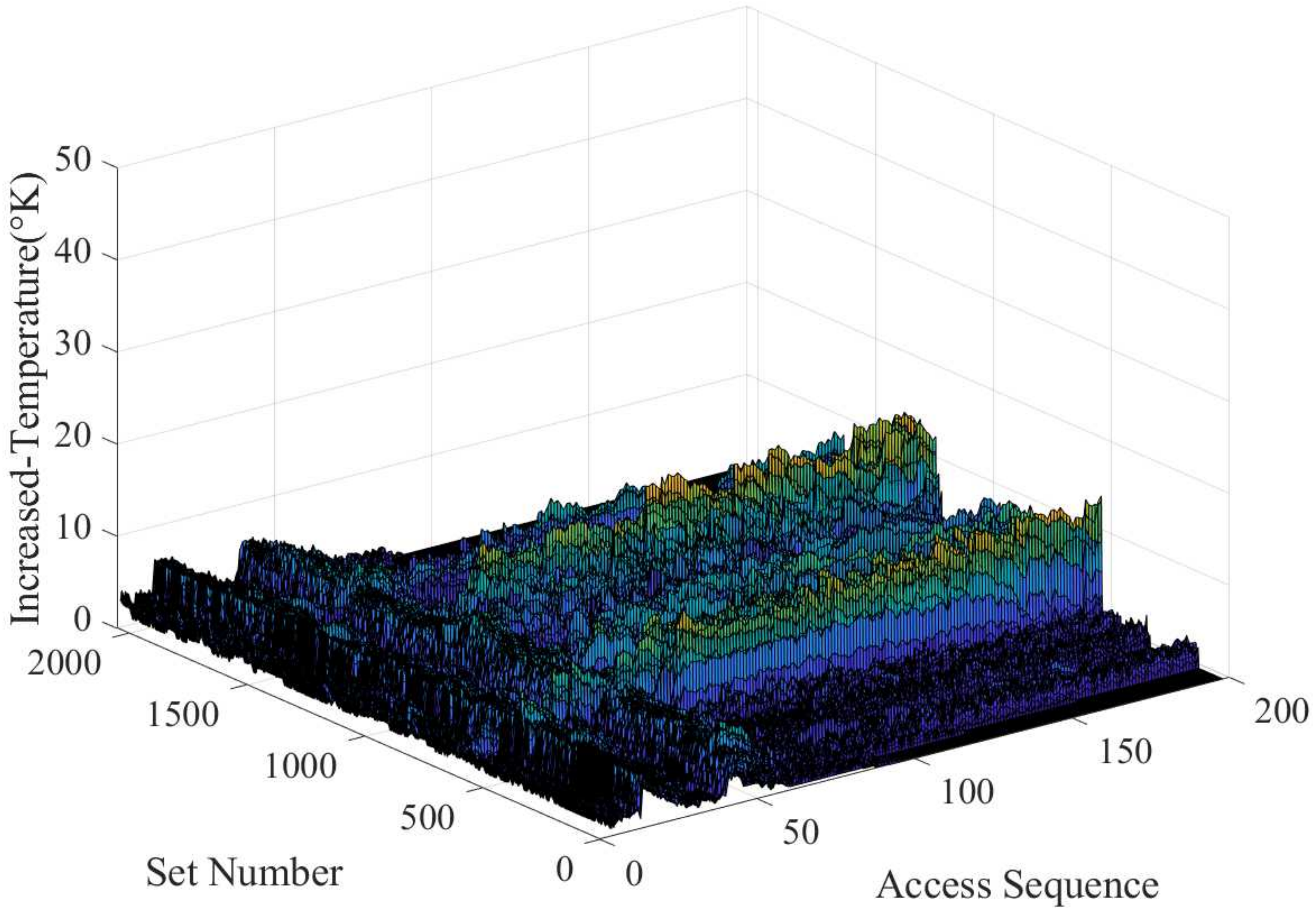}}
				\hfill
				\caption{
				Temperature increase in blocks of STT-MRAM cache in (a, b) dealII, (c, d) namd, and (e, f) h264ref workloads in comparison with LRU and TA-LRW replacement policy, respectively.}
				\label{fig:8}

			\end{figure*}

			Fig. \ref{fig:8} shows the amount of temperature increase due to heat accumulation in cache sets. 
			For the sake of visibility, we depicted the temperature increase for an interval of 200 successive writes in each cache set, and the results of only three workloads in LRU and TA-LRW policies are given in Fig. \ref{fig:8}(a)-(f)
			\footnote{The results for the other benchmarks are given in Fig. \ref{fig:app2} in the Appendix Section.}.
		
	\begin{table}[tb]\vspace{5pt}
				\centering\vspace{-1pt}
				\caption{
				Temperature increase in cache blocks based on their distance from write location}\vspace{-10pt}
				\includegraphics[width=0.9\linewidth]{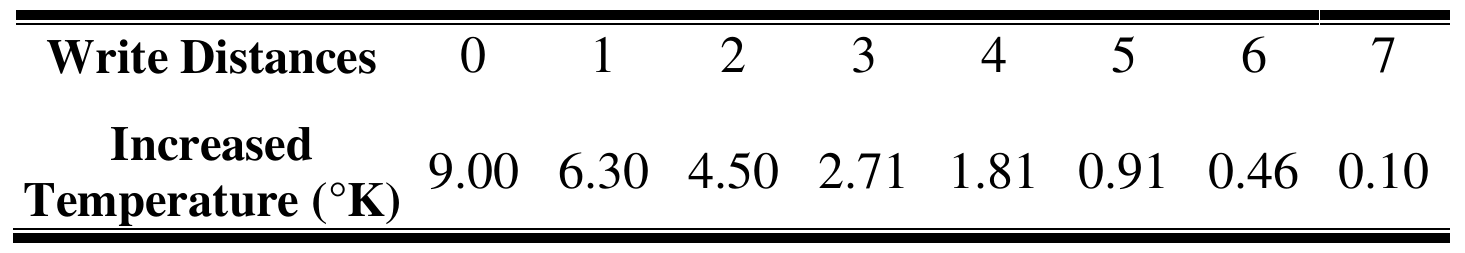}
				\label{table:4}
			\end{table}

			\begin{figure*}[t]
				\centering
				\includegraphics[width=1\linewidth]{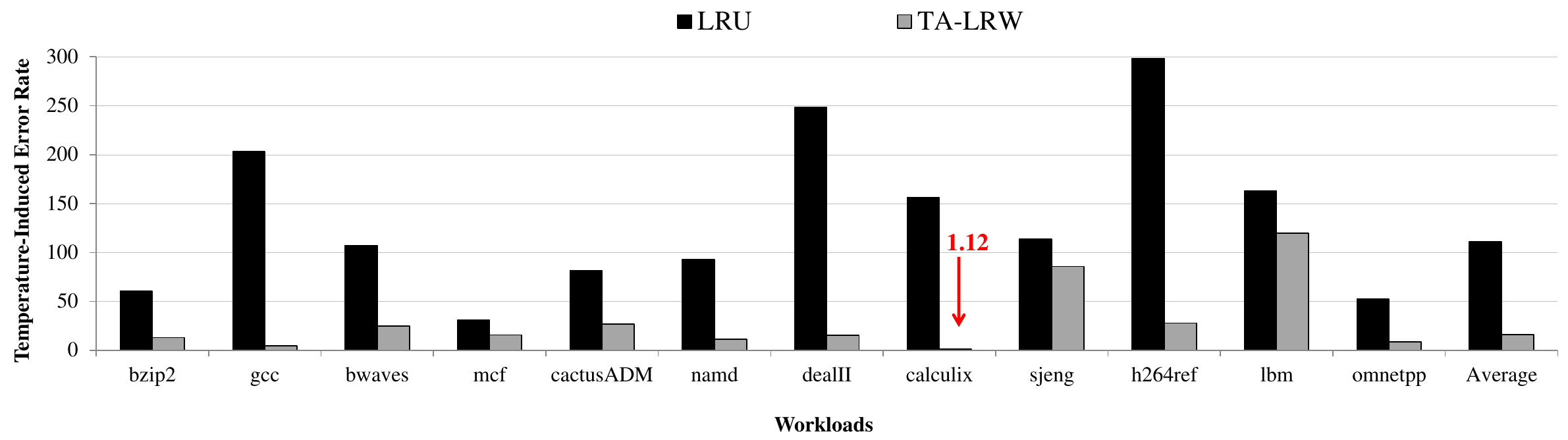}\vspace{-10pt}
				\caption{Temperature-induced error rate in LRU and TA-LRW replacement policies. The values are normalized to intrinsic error rate in each workload.}\vspace{-12pt}
				\label{fig:9-rate}
			\end{figure*}

			As can be seen, there is a large diversity in the pattern of heat accumulation in LRU policy for the given workloads and time intervals. 
			TA-LRW policy not only reduces the peak temperature increase to less than half of its value in LRU policy, but also significantly mitigates the variations in cache regions temperature.
			Fig. \ref{fig:8}(a) shows the increased temperature for \textit{dealII} workload. 
			For the majority of cache sets, the temperature is increased by 20$^{\circ}$K in a large fraction of accesses.
			TA-LRW significantly decreases this value to less than 10$^{\circ}$K, as shown in Fig. \ref{fig:8}(b).
			As depicted in Fig. \ref{fig:8}(c) for~\textit{namd} workload in LRU policy, a large diversity of temperature increase is observed in different cache sets. 
			While the increased temperature of some sets is permanently about 20$^{\circ}$K, this value fluctuates around 10$^{\circ}$K for a large number of sets and is less than 5$^{\circ}$K in some other sets.
			By employing TA-LRW, Fig. \ref{fig:8}(d) illustrates that the worst-case temperature increase is less than 5$^{\circ}$K and the variations among different sets and access sequences are not considerable.
			According to Fig. \ref{fig:8}(e), the temperature of almost all cache sets is increased by around 10$^{\circ}$K in the initial accesses of \textit{h264ref} workload. 
			This value gradually increases to 20$^{\circ}$K for some sets and reaches to 35$^{\circ}$K by the end of access sequences.
			By distributing the write accesses in cache sets, TA-LRW reduces the increased temperature to less than 10$^{\circ}$K for almost all accesses in all cache sets (Fig. \ref{fig:8}(f)). For the majority of cache sets, this value is reduced to less than 5$^{\circ}$K.

			\subsection{Error Rate Evaluation}

			As mentioned, heat accumulation in the cache increases the cache error rate. A cache operating in a given temperature with no heat accumulation has an \textit{intrinsic error rate}. Heat accumulation increases the cache error rate to a value higher than its intrinsic error rate. We define the difference between these two error rates as \textit{temperature-induced error rate}. 
			Intrinsic error rate is the summation of retention failure, write failure, and read disturbance for a cache operating in the base temperature. The cache is in its base temperature when no heat accumulation due to consecutive writes is considered. \textit{Total failure}~\textit{rate} is the summation of retention failure, write failure, and read disturbance rates when the heat accumulation during the execution time in different parts of the cache is considered.
The equations in Section 3 show the dependency of the three STT-MRAM error types to cells temperature. The transient temperature increase in the cache due to heat accumulation leads to variable rate of each error in different cache parts as well as different time intervals. The minimum values for these variable error rates are observed in the cache parts operating in the base temperature with no heat accumulation. However, the heat accumulation is observed in most parts whose temperature increases according to the amount of heat. In this case, the temperature-induced error rate is the increased error rate in the cache due to the heat accumulation effects.
Therefore, the total error rate of the cache is the summation of intrinsic and temperature-induced error rates, as given in (\ref{eq:6}).

			\begin{equation}
			\begin{multlined}
			\label{eq:6}
			 Total~error~rate = (intrinsic~error~rate) + \\
			\shoveleft[1cm]{(temperature$-$induced~error~rate)} 
			\end{multlined}
			\end{equation}

			Intrinsic error rate exists in the cache regardless of the replacement policy used in the cache. The goal of TA-LRW replacement policy is to minimize the temperature-induced error rate, which is a considerable fraction of the total error rate in LRU policy. Fig. \ref{fig:9-rate} depicts the temperature-induced error rate in the cache for all workloads normalized to the intrinsic error rate. On average, the temperature-induced error rate in LRU policy is about 110.9\% and is reduced to about 16.1\% in TA-LRW. This observation indicates that TA-LRW reduces the temperature-induced error rate by 6.9x.

			Temperature-induced error rate in some workloads, e.g., \textit{gcc}, \textit{dealII}, and \textit{h264ref}, is more than 200\% in LRU policy. This value indicates that heat accumulation can increase the total error rate in the cache by more than twice. The minimum value of the temperature-induced error rate in LRU is about 30.5\% for \textit{mcf}. For TA-LRW, the temperature-induced error rate is less than 30\% for all workloads except in \textit{sjeng} and \textit{lbm} workloads. These values are 85.6\% and 119.9\% for \textit{sjeng} and \textit{lbm} workloads, respectively, which are still much lower than that in LRU.


The proposed TA-LRW policy aims to minimize the heat accumulation and its functionality is independent of external parameters, e.g., cores activity, transactions on buses, and cooling system. In this regard, we assume that external parameters affect the cache temperature evenly and provide a uniform base temperature for all cache regions. The heat accumulation challenge is the difference between cache momentary temperature and its base temperature. The main aim of TA-LRW policy is to minimize the gap between the base temperature and the cache momentary temperature. Heat accumulation due to consecutive writes remains a major challenge without any dependency on STT-MRAM cache base temperature, its diversity during time, cooling system functionality, and the uniformity or non-uniformity of heat transition. The consequence of heat accumulation is the increase in heat-sensitive cache error rate, which is mitigated by TA-LRW.

			\begin{figure*}[tp]
				\centering
				\includegraphics[width=1\linewidth]{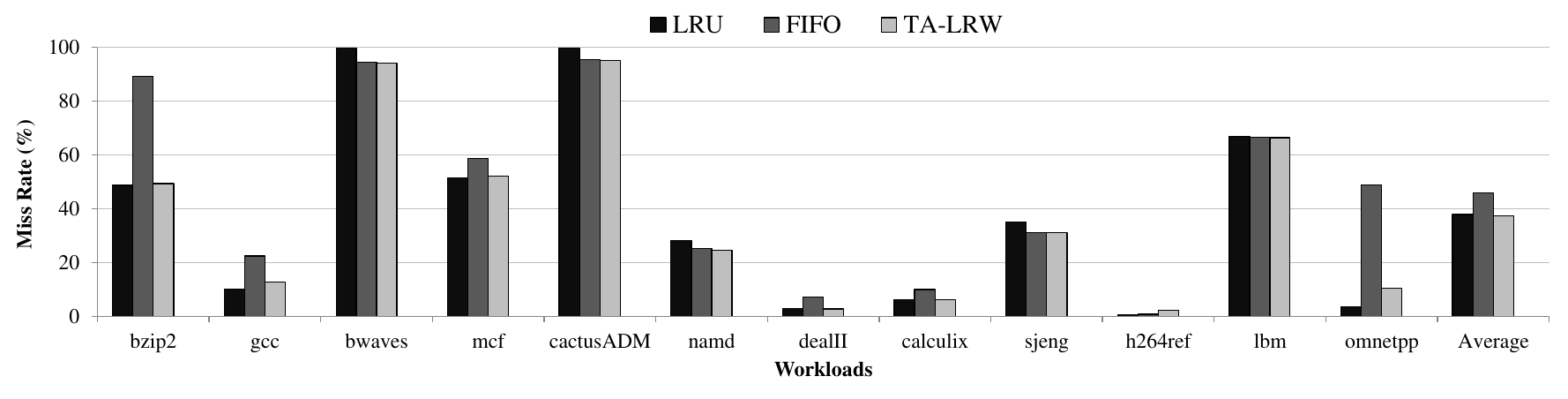}\vspace{-9pt}
				\caption{Miss Rate of TA-LRW in comparison with LRU and FIFO replacement policies.}
				\label{fig:10}
			\end{figure*}

			\subsection{Performance Evaluation}

			\begin{figure*}[t]\vspace{5pt}
				\centering
				\includegraphics[width=1\linewidth]{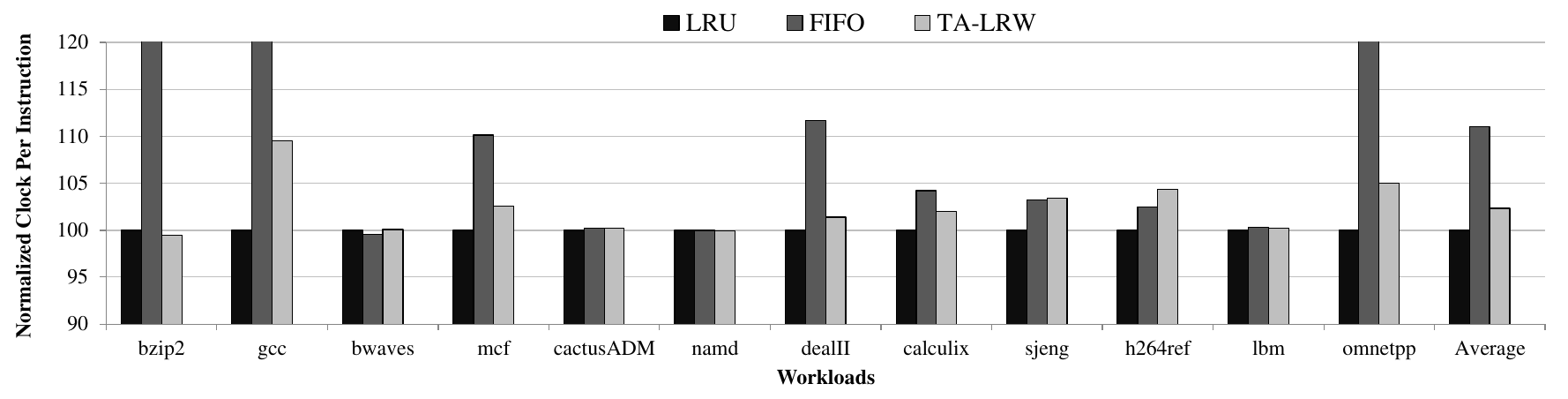}\vspace{-9pt}
				\caption{Clock Per Instruction (CPI) in FIFO and TA-LRW replacement policies normalized to CPI in LRU replacement policy.}
				\label{fig:11}
			\end{figure*}

			Fig. \ref{fig:10} depicts the L2 cache miss rate in both LRU and TA-LRW policies for all workloads. To better illustrate how TA-LRW policy is close to LRU policy in term of performance, we have also included the miss rate of FIFO policy in Fig. \ref{fig:10}. As mentioned, the complexity and implementation cost of TA-LRW is almost the same as that of FIFO. Hence, it may be misinterpreted that their strategy on victim block selection is almost the same, resulting in similar performance. However, this is not the case and the proposed FIFO-like replacement policy in term of complexity, is close to LRU in term of performance. Including the results of FIFO policy provides a better insight into the efficiency of TA-LRW policy by demonstrating how far is FIFO policy from LRU and TA-LRW policies in term of performance.

			According to Fig. \ref{fig:10}, TA-LRW increases the miss rate by an average of 2.9\%, compared with LRU policy. This value for FIFO policy is as high as 9.5\%. For some workloads, e.g.,  \textit{bwaves}, \textit{cactusADM}, \textit{namd}, and \textit{sjeng}, the miss rate in both FIFO and TA-LRW is slightly lower than that in LRU. The miss rate of TA-LRW in \textit{bzip2} workload is almost the same as that in LRU, whereas FIFO significantly increases the miss rate. In \textit{omnetpp} workload, TA-LRW increases the miss rate by 6.2\% and this increase in FIFO is 45.1\%. FIFO policy increases the miss rate in \textit{mcf}, \textit{dealII}, and \textit{calculix} workloads, while this rate in TA-LRW is almost the same as that in LRU. The difference in L2 cache miss rate by the evaluated replacement policies affects the performance of the system.
 
			We consider \textit{$Cycles~Per~Instruction$} (CPI) as the performance metric and depict the CPI for all three policies in Fig. \ref{fig:11}. On average, the performance overhead in FIFO is 10.3\%, compared with LRU policy. However, TA-LRW increases the CPI by only 2.3\%, on average. FIFO increases the CPI by more than 20\% in \textit{bzip2}, \textit{gcc}, and \textit{omnetpp} workloads, whereas the worst case overhead of TA-LRW is 9.3\%, which is observed in \textit{gcc} workload. The performance of all evaluated replacement policies are almost the same in some workloads, e.g., \textit{bwaves}, \textit{cactusADM}, \textit{namd}, and \textit{lbm}. There is only two workloads, i.e., \textit{sjeng} and \textit{h264ref}, in which the CPI of TA-LRW is slightly higher than that of FIFO. TA-LRW outperforms LRU in \textit{bzip2} workload, in which the performance overhead of FIFO is more than 50\%.  

			In summary, TA-LRW offers a near-LRU performance without requiring the complicated LRU components and circuitry. The implementation of TA-LRW is also as simple as FIFO, while providing a much better performance. Considering an N-way associative cache, LRU policy needs to add \textit{$\log_2 N$} extra bits to each cache block for sorting the age values. Therefore, the number of extra bits in each cache set is \textit{$N\log_2 N$}. The extra bits in each cache set required by TA-LRW is only \textit{$\log_2 N$} (same as in FIFO), which is $N$ times lower than that in LRU. Moreover, LRU conducts a set of operations to update the age bits of blocks for each cache access and to find the suitable victim for each cache miss. TA-LRW only updates the pointer of accessed set for each write access. Due to its smaller components and simpler operations, replacing LRU policy with TA-LRW policy significantly reduces the area and energy consumption of the cache replacement policy, in addition to uniformly distributing the generated heat in the cache.

		It is noteworthy that TA-LRW is the first, but not necessary the optimal thermal-aware replacement policy for STT-MRAM caches. Utilizing a more precise thermal model for the cache on which the replacement policy makes decisions can help for a better heat distribution. Besides, including a trade-off between miss rate and temperature can provide a higher performance. This study will spark further research in designing more advanced thermal-aware replacement policies for STT-MRAM caches.

			\section{Related Work and Discussion} 

			There are several studies addressing the reliability of STT-MRAM caches and presenting techniques to overcome either write failure, read disturbance, or retention failure.  
			These techniques, in the best case, reduce the rate of one error type without affecting the other error types. 
			However, most of them adversely affect the rate of other error types to alleviate the target error type.

			To reduce the rate of write failure, some studies increase the amplitude and/or width of write pulse~\cite{sun-TMAG-12, Emre-SiPS-12, lakys2012self}.
			In addition to imposing high latency and energy overhead, these techniques increase the probability of oxide barrier breakdown in STT-MRAM cells. 
			In studies presented in~\cite{sun2011design, 15-AZAD-TETC}, the content of memory is read after each write operation and the write operation is repeated until the data is correctly written. Extra read operations increase the occurrence probability of read disturbance in these techniques.

			\textit{$Error$}-\textit{$Correcting~Codes$} (ECCs) have been widely used to tolerate write failures~\cite{14-AZAD-TETC, 15-AZAD-TETC, wen2013cd, ZAZADTE, ZAZADTPDS, ahn2013selectively, seyedzadeh2016leveraging,azad2018orient}.
			These codes increase the read disturbance rate as well as the cache area and energy consumption.
			Some studies try to reduce the number of bit switching in write operations by encoding the incoming data or writing into a cache block with the minimum hamming distance~\cite{wen2013cd, maddah2015cafo, AMir2016TDMR}. 
			These techniques complicate the design and increase the read disturbance rate by imposing extra read operations.

			To reduce the read disturbance rate, read current is decreased in studies presented in~\cite{15-EDCC-zhao2011design},~\cite{Na-TCAS-II-16}. 
			To correctly read the contents of memory cells at reduced read current, these studies present more accurate sense amplifiers.
			\textit{Destructive~read~and~restoring} technique performs an extra write operation after each read operation to correct possible read disturbance error~\cite{Takemura-IMW-10},~\cite{Jiang-ASPDAC-16}. 
			This technique not only imposes energy consumption overhead, but also significantly increases the write failure rate.
			The number of restore operations is reduced in~\cite{wang2015selective} by selectively restore the erroneous blocks.
			Using ECC is another approach to correct read disturbance errors, which imposes area, performance, and energy overhead~\cite{seyedzadeh2016leveraging}.
			
			To reduce the retention failure rate, some studies suggested to increase the STT-MRAM cell retention time by manipulating the MTJ layers characteristics or the cell thermal stability factor~\cite{bi2012analysis, sun2011design, li2010design, 15-EDCC-zhao2011design, driskill2010latest}. These changes in STT-MRAM characteristics increase the write failure rate.
			Employing ECC is another approach for overcoming the retention failure rate~\cite{naeimi2013intel}. 
			Same as the side effects of ECC in tolerating read disturbance and write failure, ECC increases the occurrence probability of these two error types in addition to imposing energy consumption and area overhead.

			Refreshing the STT-MRAM cells is another technique to reduce the retention failure rate~\cite{smullen2011relaxing}. However, refreshing is beneficial only if assuming a deterministic retention time for all of the memory cells, which is not the case as the retention failure occurs stochastically. 
			Considering some error correction mechanisms, e.g., using ECCs, memory scrubbing is suggested in~\cite{mittal2017survey} to periodically check/correct the cache contents and prevent the error accumulation.

			Some studies reduce the write energy and latency by decreasing the write current and/or pulse width, which leads to increase in write failure rate~\cite{arezoomand2017energy, sun2011multi, jog2012cache, sun2011multi, LARS2018}. To moderate this increase, these studies decrease STT-MRAM thermal stability factor ($\Delta$). As its side effect, read disturbance and retention failure rates are exponentially increased. Reduction in write energy decreases the temperature-induced rates of all three error types on the one hand and significantly increases the intrinsic rates of read disturbance and retention failure on the other hand. The total error rate is highly increased in this case. As an example, our investigations show that temperature-induced error rate due to heat accumulation increases the total error rate to 2x, which fades out by using smaller write current or write pulse width. However, reduction of $\Delta$ from 45 to 35 will increase the total error rate by four orders of magnitude. Therefore, the total error rate of STT-MRAM cache significantly increases by reducing the write current and thermal stability factor, which makes this approach useless for error rate reduction.

			Besides the mentioned schemes in emerging STT-MRAM caches, an extensive effort is conducted for decades to improve the reliability of SRAM caches~\cite{zhang2012reliable, farbeh2014psp, manoochehri2011cppc}. Despite the similar functionality of SRAM and STT-MRAM cache, reliability improvement schemes of SRAM caches are generally inapplicable or inefficient in STT-MRAM caches. This is due to the difference in source and pattern of errors in these two technologies. 

			 The main source of errors in SRAM caches is radiation-induced particle strike, which causes bit-flip in a single or multiple adjacent memory cells with exponential probability distribution~\cite{kim2007multi, farbeh2016cache, neale2015adjacent}. Recent studies have shown that STT-MRAM cells are immune to this source of errors. On the other hand, retention failure, write failure, and read disturbance errors in STT-MRAM caches are originated from thermal instability and stochastic behavior of the cells, which is not a concern in SRAM cells. Beside STT-MRAM physical parameters, the occurrence probability and pattern of these errors depends on the cache access behavior and the data content. Because of these differences, not only the reliability improvement schemes in SRAM and STT-MRAM caches are incompatible, but also it is not easy to provide a fair comparison between their reliability. The benefits of replacing SRAM with STT-MRAM are mainly due to non-volatility, near-zero leakage power, and higher density of STT-MRAM, whereas its reliability is among the major challenges for its commercialization. This study is one step ahead to overcoming the drawbacks of STT-MRAM caches.

			To the best of our knowledge, TA-LRW is the first technique that simultaneously mitigates the rates of retention failure, read disturbance, and write failure in STT-MRAM caches.
			As shown in Section 5, TA-LRW is an effective approximation of conventional LRU replacement policy, with significantly simpler design.
			As observed in Fig. \ref{fig:11}, the performance of TA-LRW policy is almost the same as that in LRU policy.
			The only challenge in TA-LRW policy is its performance overhead in a rarely-occurred scenario in which 
			a high fraction of victim blocks in cache misses are not among the old blocks.
			This scenario is probable when the write history of blocks is not an accurate approximation for their access history.
			This inaccuracy increases only when a large fraction of cache accesses is due to read in elder blocks in the cache, which moves the old blocks to the head of LRU queue. 
			However, our evaluations indicate that the mentioned scenario is a rare event.

			\section{Conclusions}

			Emerging STT-MRAM on-chip caches are highly error-prone due to stochastic characteristics of this memory technology. The error rate in cache memory increases exponentially in higher temperature. Our investigation shows that the temporal locality in writing to adjacent cache blocks causes regional heat accumulation inside the cache. To prevent this heat accumulation, we proposed \textit{Thermal-Aware Least-Recently Written} (TA-LRW) replacement policy. TA-LRW policy keeps track of write operations into cache blocks and replaces the blocks based on their write history on a cache miss. The proposed replacement policy guarantees that the distance of two consecutive write operations in a set is at least three blocks. 
				Regarding the advantage of TA-LRW over conventional replacement policies in STT-MRAM LLCs, our evaluation shows that the proposed TA-LRW policy reduces the temperature-induced error rate by 94.8\% compared with the conventional LRU-based LLC. In addition, the hardware complexity, energy consumption, and area of TA-LRW are much lower than those of LRU policy and its performance overhead is only 2.3\%.
			The proposed TA-LRW replacement policy is not only an effective solution for heat distribution in STT-MRAM caches, but also an effective approximation and promising alternative for LRU policy in highly associative cache for minimizing its complexity.

			 \section*{Acknowledgments}
			%
			%
			The authors would like to thank Ali Bijarchi, Dr. Amir Mahdi Hosseini Monazzah, and Dr. Nezam Rohbani for their contribution and discussion on setting up the STT-MRAM cell temperature simulation.	
			\ifCLASSOPTIONcaptionsoff
			\newpage
			\fi

			
			
			\bibliographystyle{IEEEtran}
			\bibliography{IEEEabrv,references}
			%
			%
			%
			
			%
			\begin{IEEEbiography}[{\includegraphics[width=1in,height=1.25in,clip,keepaspectratio]{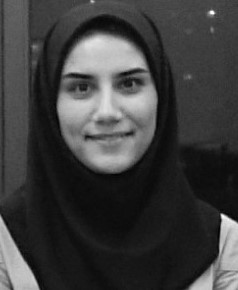}}]{Elham Cheshmikhani}
				received the B.Sc. degree in computer engineering from Iran University of Science and Technology (IUST) and the M.Sc. degree in computer engineering from Amirkabir University of Technology (Tehran Polytechnic), Tehran, Iran, in 2011 and 2013, respectively. She is currently a PhD candidate in computer engineering at Sharif University of Technology (SUT), Tehran, Iran.
She was a member of Design and Analysis of Dependable Systems (DADS) at AUT from 2011 to 2015, and has been a member of the Dependable Systems Laboratory (DSL) and Data Storage, Networks \& Processing Laboratory (DSN) since 2015 and 2017, respectively. Her research interests include emerging nonvolatile memory technologies, dependability analysis, fault tolerance, and storage systems.
		\end{IEEEbiography}
						
			\begin{IEEEbiography}[{\includegraphics[width=1in,height=1.25in,clip,keepaspectratio]{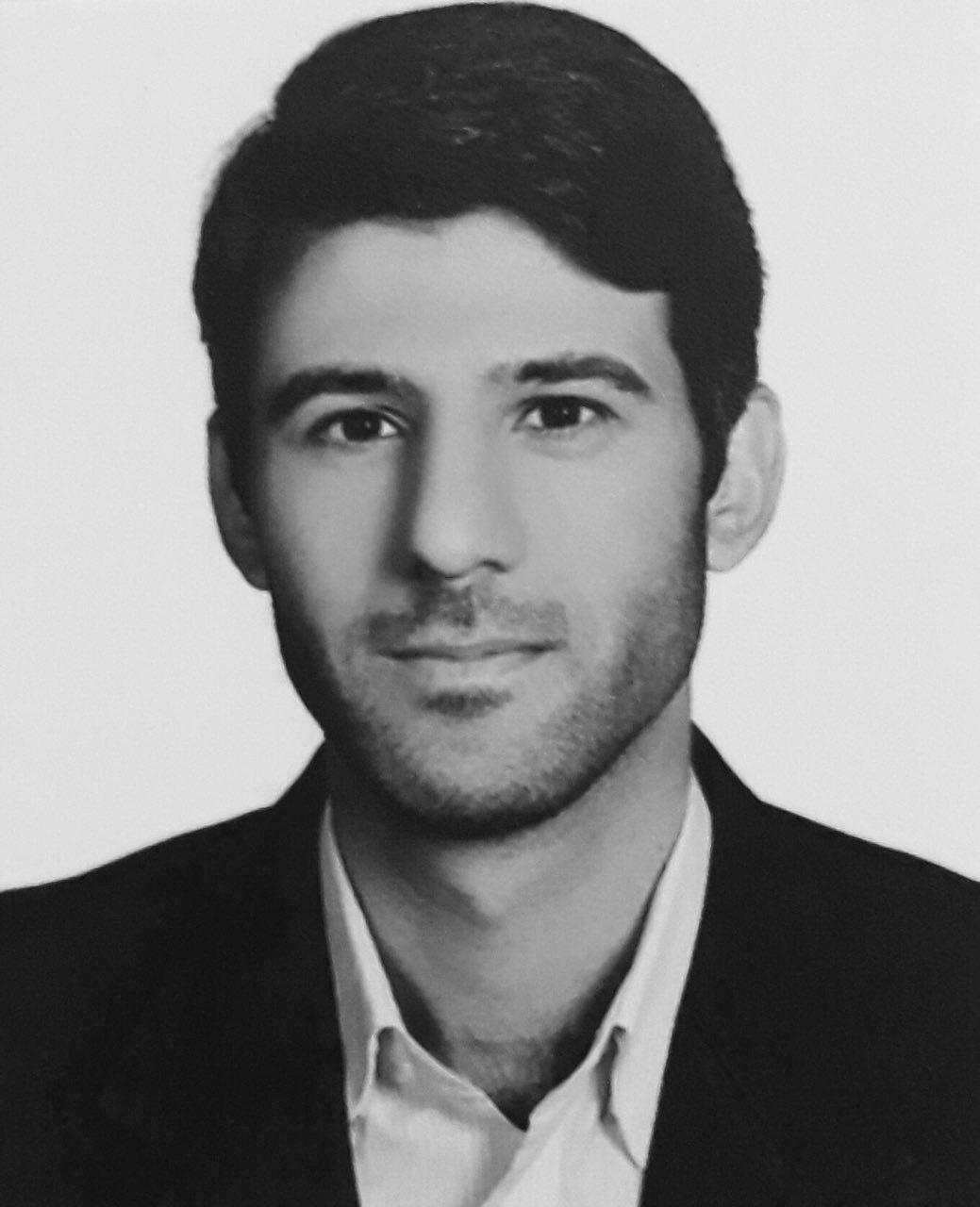}}]{Hamed Farbeh}
			(S'12) received the B.Sc., M.Sc., and PhD degrees in computer engineering from Sharif University of Technology (SUT), Tehran, Iran, in 2009, 2011, and 2017, respectively.
				He was been a member of the Dependable Systems Laboratory (DSL) at SUT from 2007 to 2017 and the head of DSL from April 2017 to February 2018. 
				He is currently the faculty member of the Department of Computer Engineering and Information Technology, Amirkabir University of Technology (Tehran Polytechnic), Tehran, Iran. 
				He was with the Embedded Computing Laboratory (ECL), KAIST, Daejeon, South Korea, as a Visiting Researcher from October 2014 to May 2015 and collaborated with the Institute of Research for Fundamental Sciences (IPM), Tehran, Iran, as Postdoc fellow from May 2017 to January 2018.
				His current research interests include reliable memory hierarchy, reliability challenges in emerging memory technologies, cyber-physical systems. He was the IEEE student member from 2012 to 2017.
%
			\end{IEEEbiography}
%
%
%
%

		\begin{IEEEbiography}[{\includegraphics[width=1in,height=1.25in,clip,keepaspectratio]{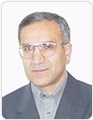}}]{Seyed Ghassem Miremadi}
				(SM'07) received the M.Sc. degree in applied physics and electrical engineering from the Linköping Institute of Technology, Linköping, Sweden, and the Ph.D. degree in computer engineering from the Chalmers University of Technology, Gothenburg, Sweden. 

		He initiated the Dependable Systems Laboratory with the Sharif University of Technology (SUT), Tehran, Iran, in 1996, as fault-tolerant computing is his specialty, and has chaired the laboratory since then. He and his group have done research in physical, simulation-based and software-implemented fault injection, dependability evaluation using HDL models, fault-tolerant embedded systems, fault-tolerant networks-on-chip, fault-tolerant real-time systems, and fault-tolerant storage systems. He was the Education Director from 1997 to 1998, the Head from 1998 to 2002, the Research Director from 2002 to 2006, and the Director of the Hardware Group with the Computer Engineering Department, SUT, from 2009 to 2010. From 2003 to 2010, he was the Director of the Information Technology Program with SUT International Campus-Kish Island, Kish, Iran. From 2010 to 2012 and since 2014, he has been the Vice President of Academic Affairs at SUT, where he is currently a Professor of Computer Engineering. 
		Dr. Miremadi is a Senior Member of the IEEE Computer Society and the IEEE Reliability Society. He served as the General Co-Chair of the 13th International CSI Computer Conference in 2008, the Executive Chair of the International Conference on Engineering Education in 2013, and the General Co-Chair of the International CSI Symposium on Real-Time and Embedded Systems and Technologies in 2015. He is the Editor of the \textit{Scientia Transactions on Computer Science and Engineering}.
			\end{IEEEbiography}
			
			\begin{IEEEbiography}[{\includegraphics[width=1in,height=1.25in,clip,keepaspectratio]{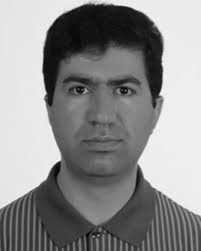}}]{Hossein Asadi}
				(M'08, SM'14) received the B.Sc. and M.Sc. degrees in computer engineering from the SUT, Tehran, Iran, in 2000 and 2002, respectively, and the Ph.D. degree in electrical and computer engineering from Northeastern University, Boston, MA, USA, in 2007. He was with EMC Corporation, Hopkinton, MA, USA, as a Research Scientist and Senior Hardware Engineer, from 2006 to 2009. From 2002 to 2003, he was a member of the Dependable Systems Laboratory, SUT, where he researched hardware verification techniques. From 2001 to 2002, he was a member of the Sharif Rescue Robots Group. He has been with the Department of Computer Engineering, SUT, since 2009, where he is currently a tenured Associate Professor. He is the Founder and Director of the Data Storage, Networks, and Processing (DSN) Laboratory, Director of Sharif High-Performance Computing Center, the Director of Sharif Information Technology Service Center (ITC), and the President of Sharif ICT Innovation Center. He spent three months in the summer 2015 as a Visiting Professor at the School of Computer and Communication Sciences at the Ecole Poly-technique Federele de Lausanne (EPFL). He is also the co-founder of HPDS corp., designing and fabricating midrange and high-end data storage systems. He has authored and co-authored more than seventy technical papers in reputed journals and conference proceedings. His current research interests include data storage systems and networks, solid-state drives, operating system support for I/O and memory management, and reconfigurable and dependable computing. Dr. Asadi was a recipient of the Technical Award for the Best Robot Design from the International RoboCup Rescue Competition, organized by AAAI and RoboCup, a recipient of Best Paper Award at the 15th CSI Internation Symposium on Computer Architecture and Digital Systems (CADS), the Distinguished Lecturer Award from SUT in 2010, the Distinguished Researcher Award and the Distinguished Research Institute Award from SUT in 2016, and the Distinguished Technology Award from SUT in 2017. He is also recipient of Extraordinary Ability in Science visa from US Citizenship and Immigration Services in 2008. He has also served as the publication chair of several national and international conferences including CNDS2013, AISP2013, and CSSE2013 during the past four years. Most recently, he has served as a Guest Editor of IEEE Transactions on Computers, a Program Co-Chair of the 18th International Symposium on Computer Architecture \& Digital Systems (CADS2015), and the Program Chair of CSI National Computer Conference (CSICC2017).

			\end{IEEEbiography}
			
			%
			
			
			


			\begin{figure*}[b]\vspace{-20pt}
				\centering
				\subfloat[bwaves]{\includegraphics[width=0.48\linewidth]{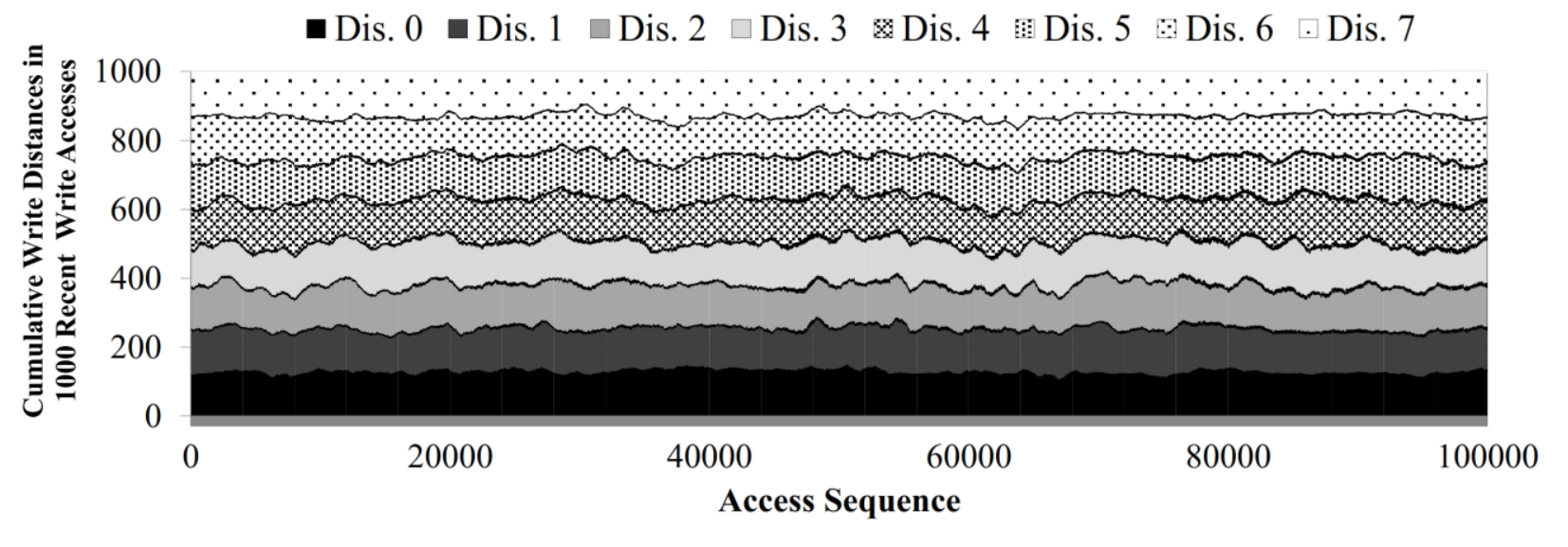}}
				\hfill
				\subfloat[h264ref]{\includegraphics[width=0.48\linewidth]{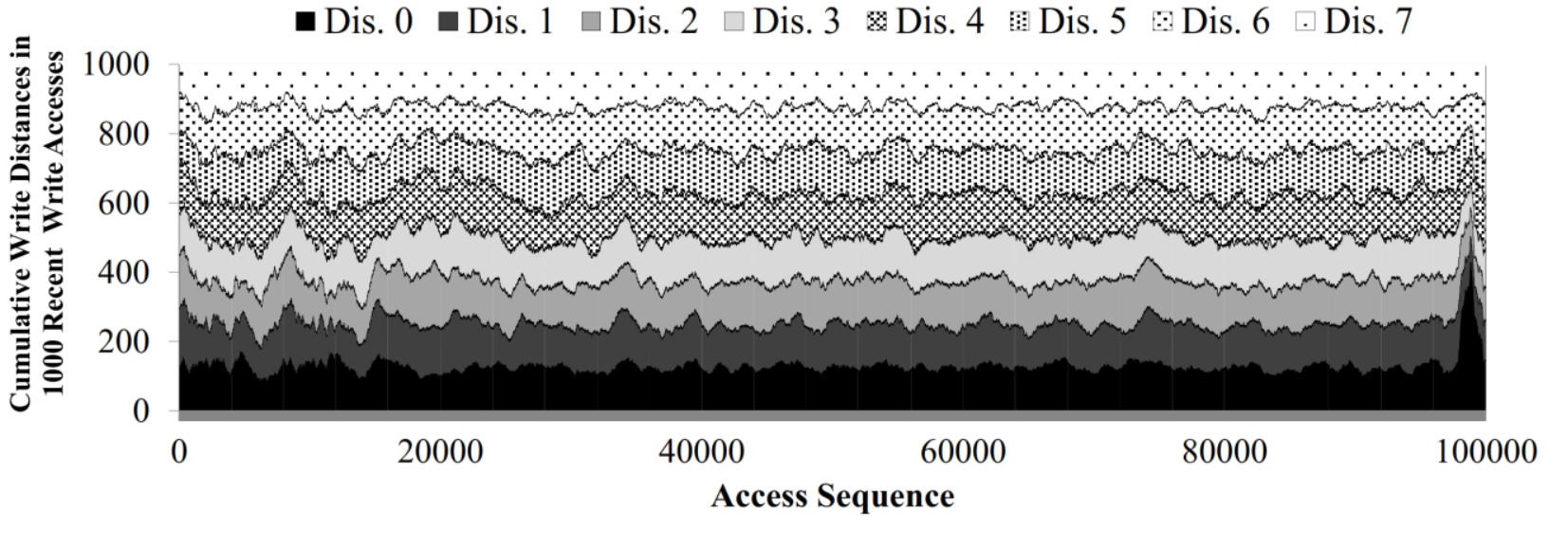}}
				\hfill
				\subfloat[sjeng]{\includegraphics[width=0.48\linewidth]{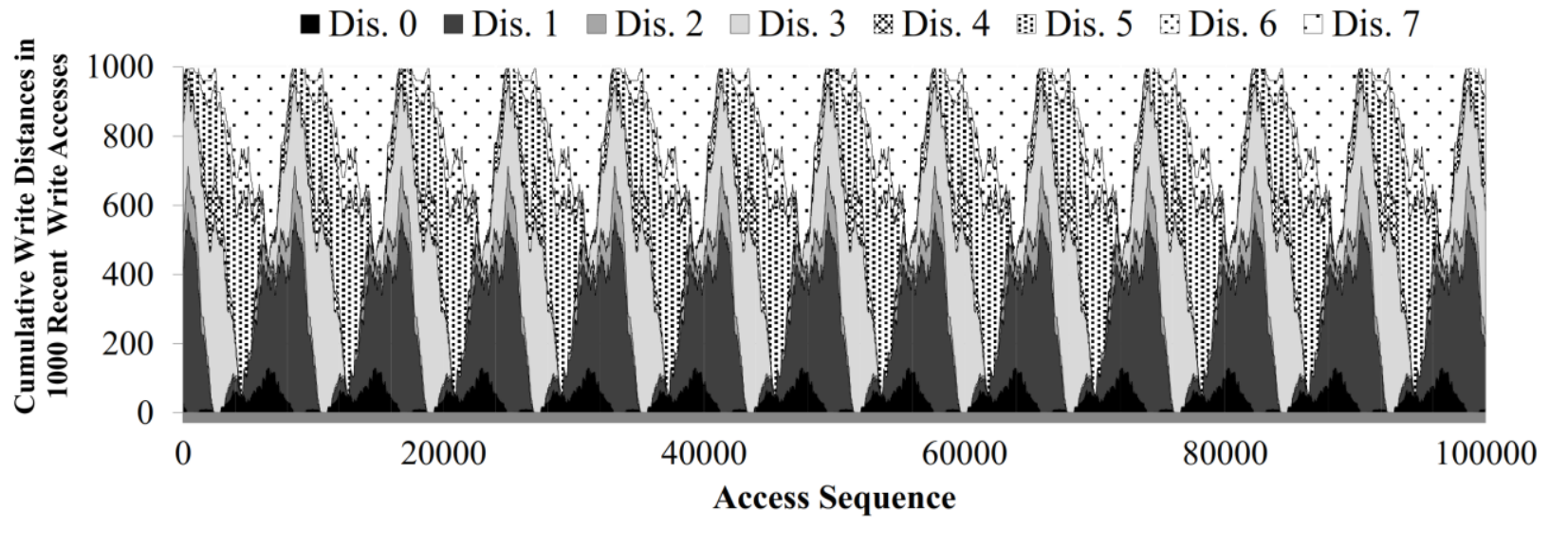}}
				\hfill
				\subfloat[dealII]{\includegraphics[width=0.48\linewidth]{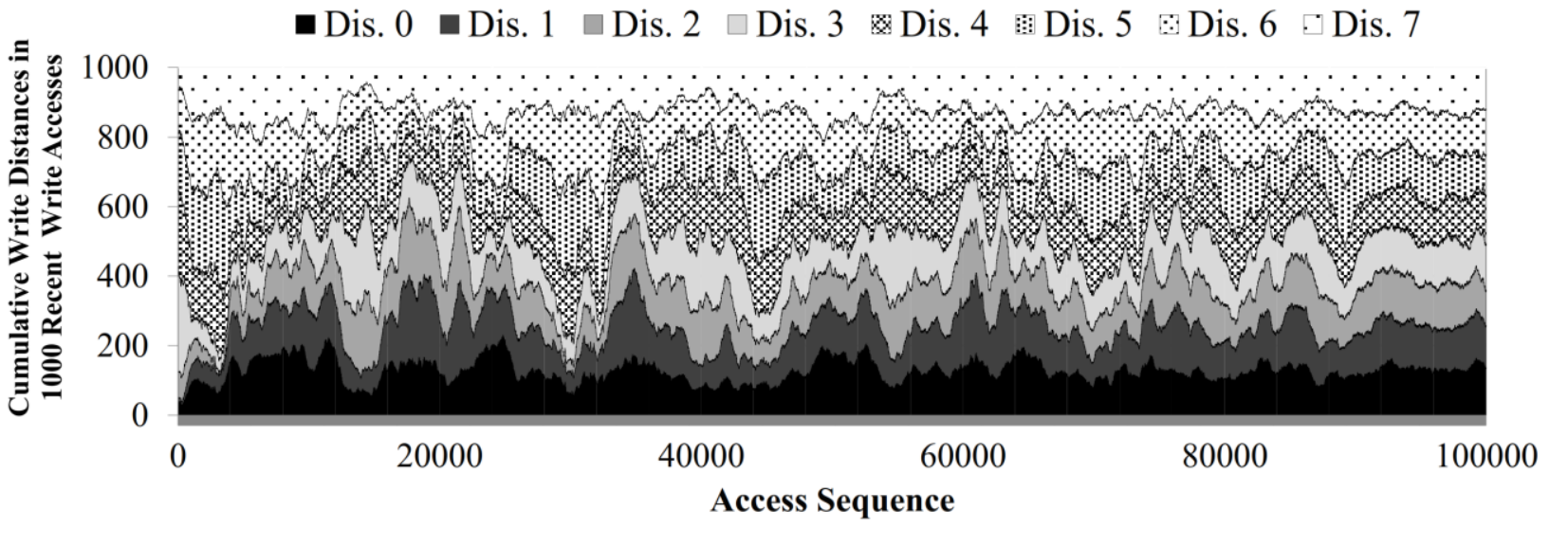}}
				\hfill
				\subfloat[omnetpp]{\includegraphics[width=0.48\linewidth]{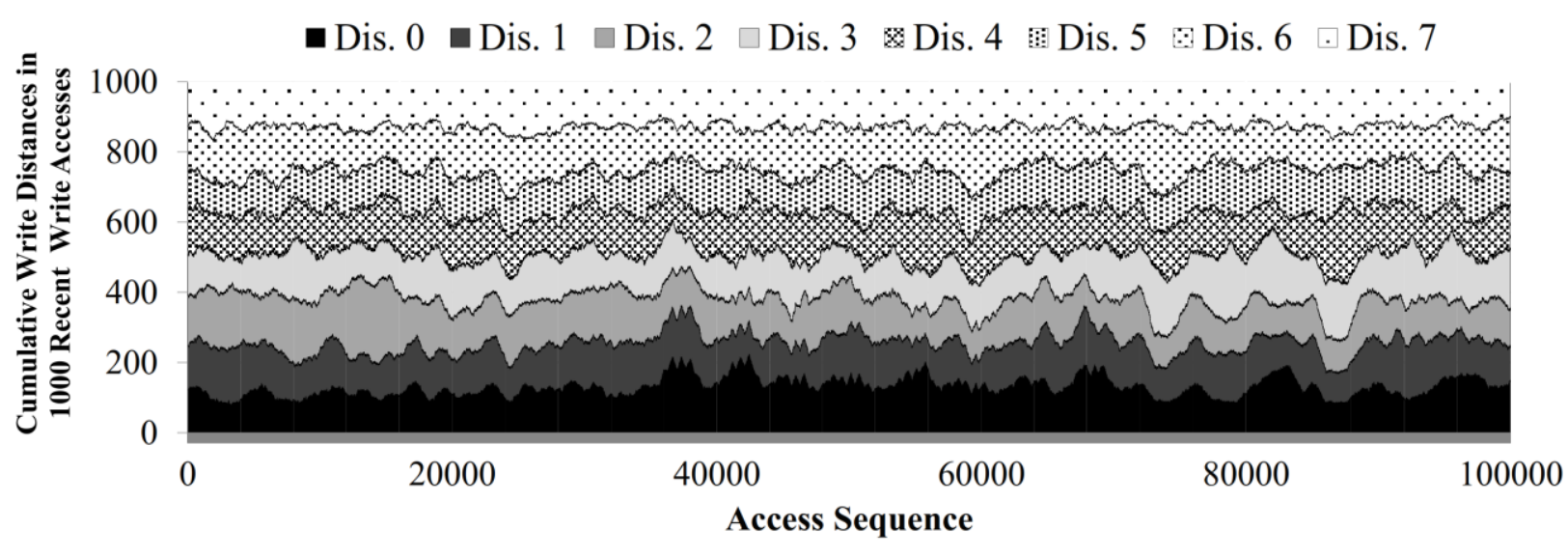}}
				\hfill
				\subfloat[mcf]{\includegraphics[width=0.48\linewidth]{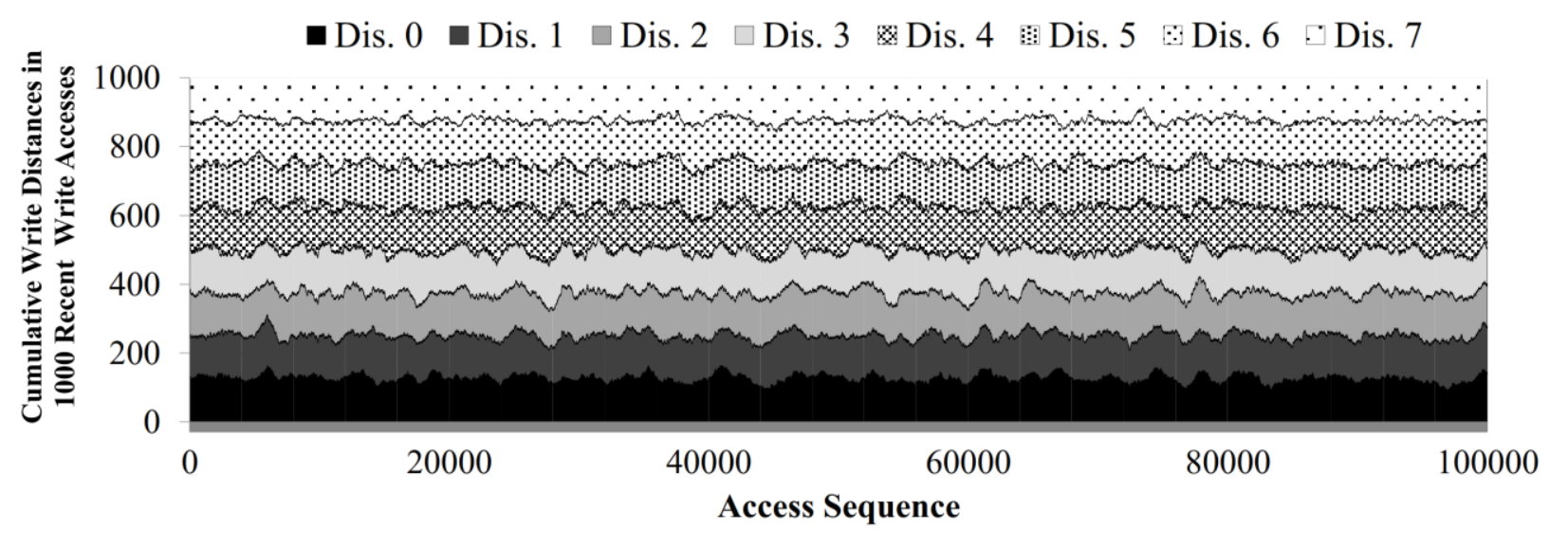}}
				\hfill
				\subfloat[namd]{\includegraphics[width=0.48\linewidth]{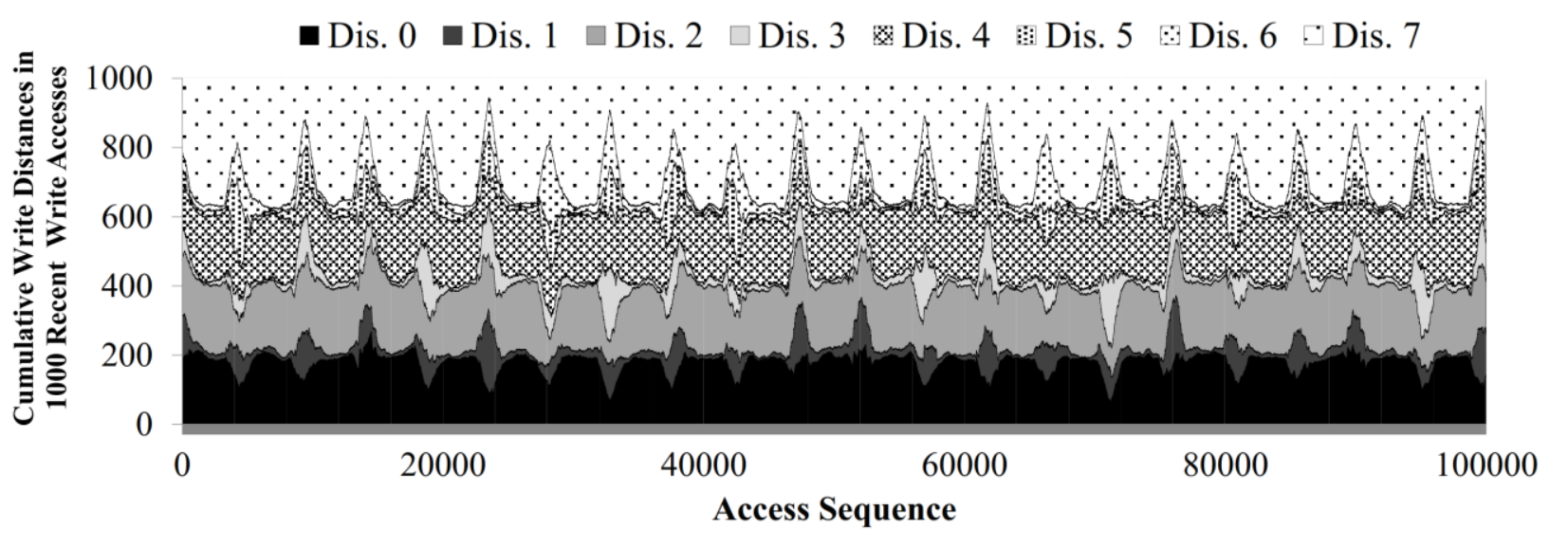}}
				\hfill
				\subfloat[calculix]{\includegraphics[width=0.48\linewidth]{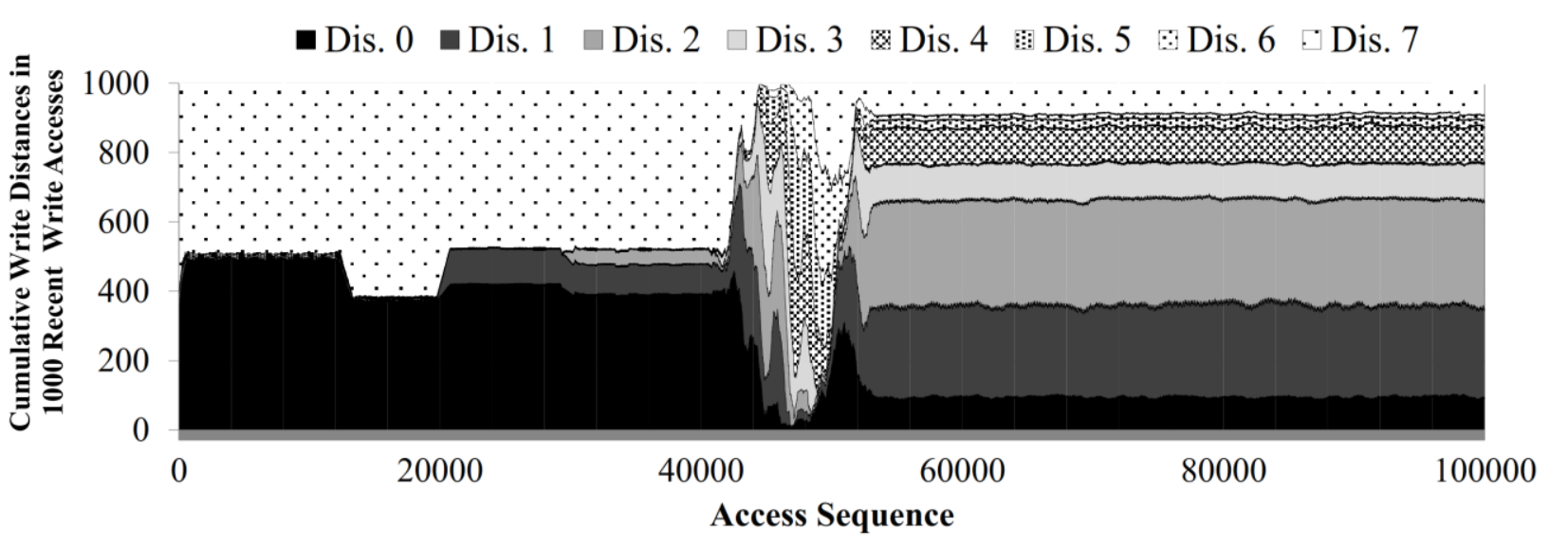}}
				\caption{Accumulated write distances for every 1000 recent writes in (a) bwaves, (b) h264ref, (c) sjeng, (d) dealII, (e) omnetpp, (f) mcf, (g) namd, and (h) calculix workloads in LRU policy.}\vspace{-8pt}
				\label{fig:app1}
			\end{figure*}

				\appendix 
				\section*{A. Accumulated Write Distances in LRU Policy}
				
				To show the breakdown of write distances for all workloads, the results for the remaining workloads are shown in  Fig. \ref{fig:app1}. Write locality in \textit{bwaves} workload shows that less than 30\% of data is written in the blocks with more than 3 distances, as depicted in Fig. \ref{fig:app1}(a). The same behavior is observed in \textit{h264ref}, \textit{omnetpp}, and \textit{mcf} workloads (Fig \ref{fig:app1}(b), (e) and (f), respectively). Fig. \ref{fig:app1}(c) indicates that for some time intervals, a large fraction of consecutive writes in \textit{sjeng} workload is conducted in adjacent blocks. The write distance for about 40\% of accesses in \textit{namd} workload is either zero, one, or two, according to Fig. \ref{fig:app1}(g), which results in a large amount of heat accumulation. Fig. \ref{fig:app1}(h) shows that the heat accumulation in \textit{calculix} benchmark is even higher than that in \textit{namd} workload since the write distance is less than three for more than 50\% of write accesses in the majority of time intervals.

		\begin{figure*}[b]\vspace{-25pt}
				\centering
				\subfloat[bwaves in LRU policy]{\includegraphics[width=0.25\linewidth]{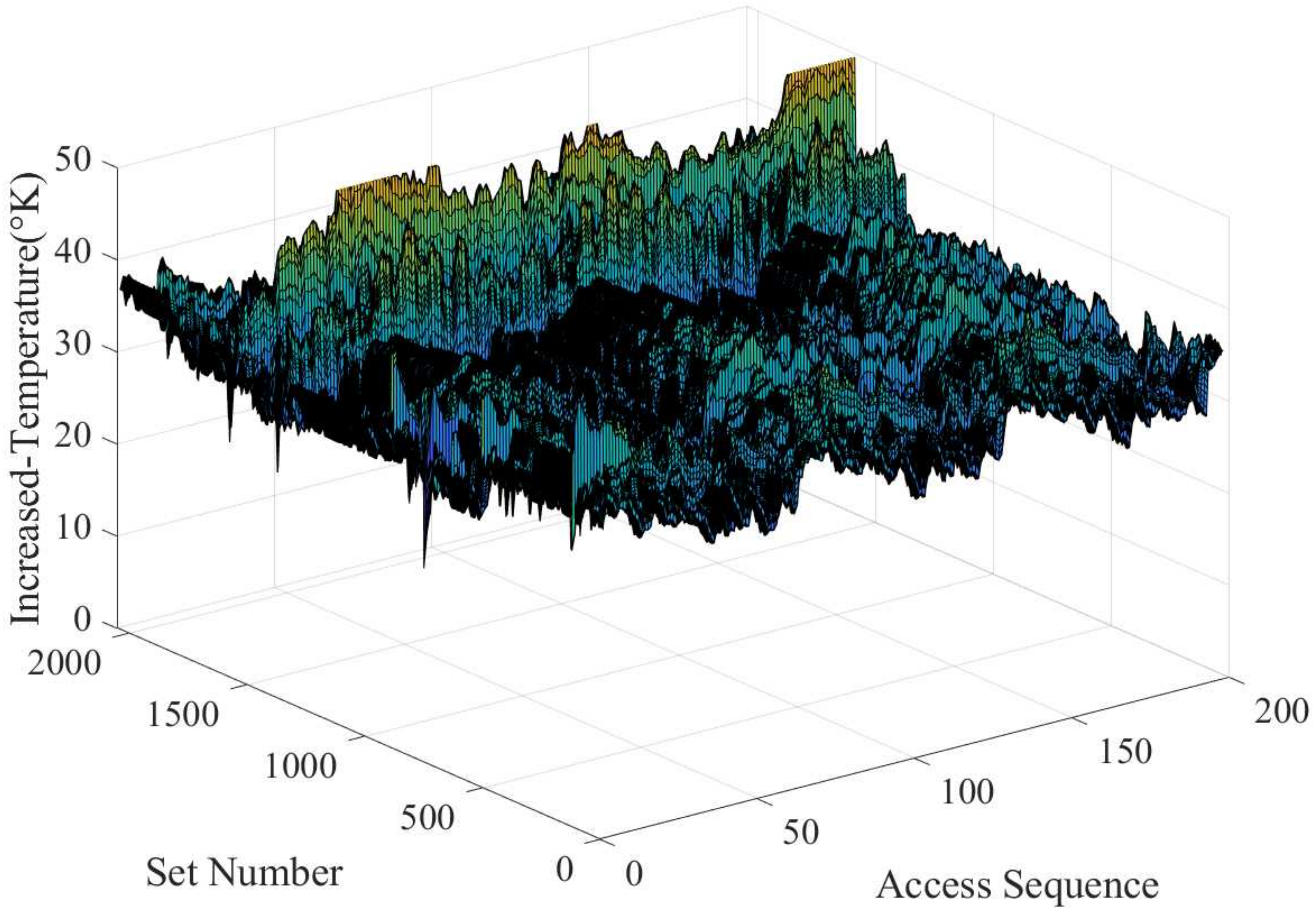}}
				\hfill
				\subfloat[bwaves in TA-LRW policy]{\includegraphics[width=0.25\linewidth]{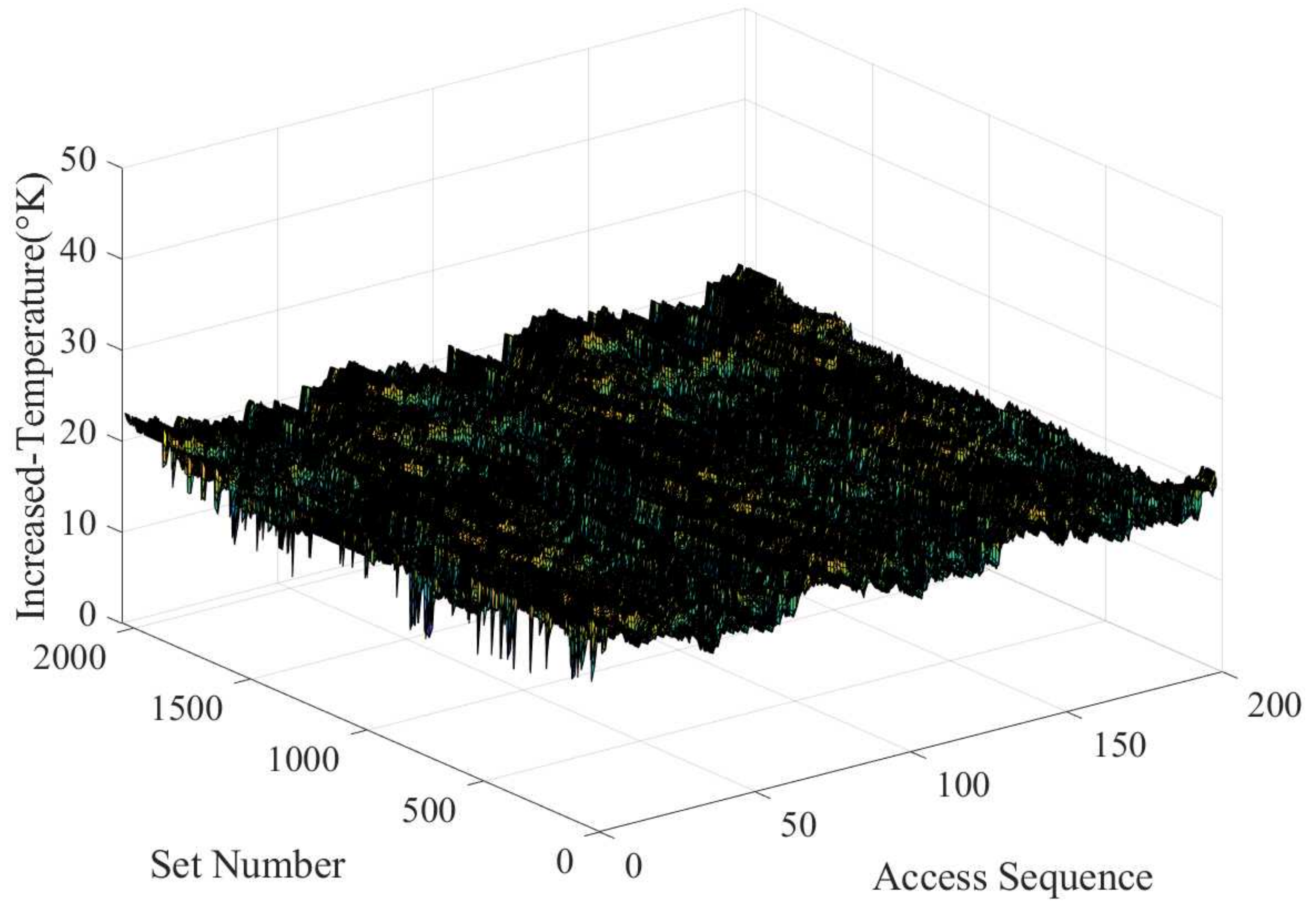}}
				\hfill
				\subfloat[calculix in LRU poliy]{\includegraphics[width=0.25\linewidth]{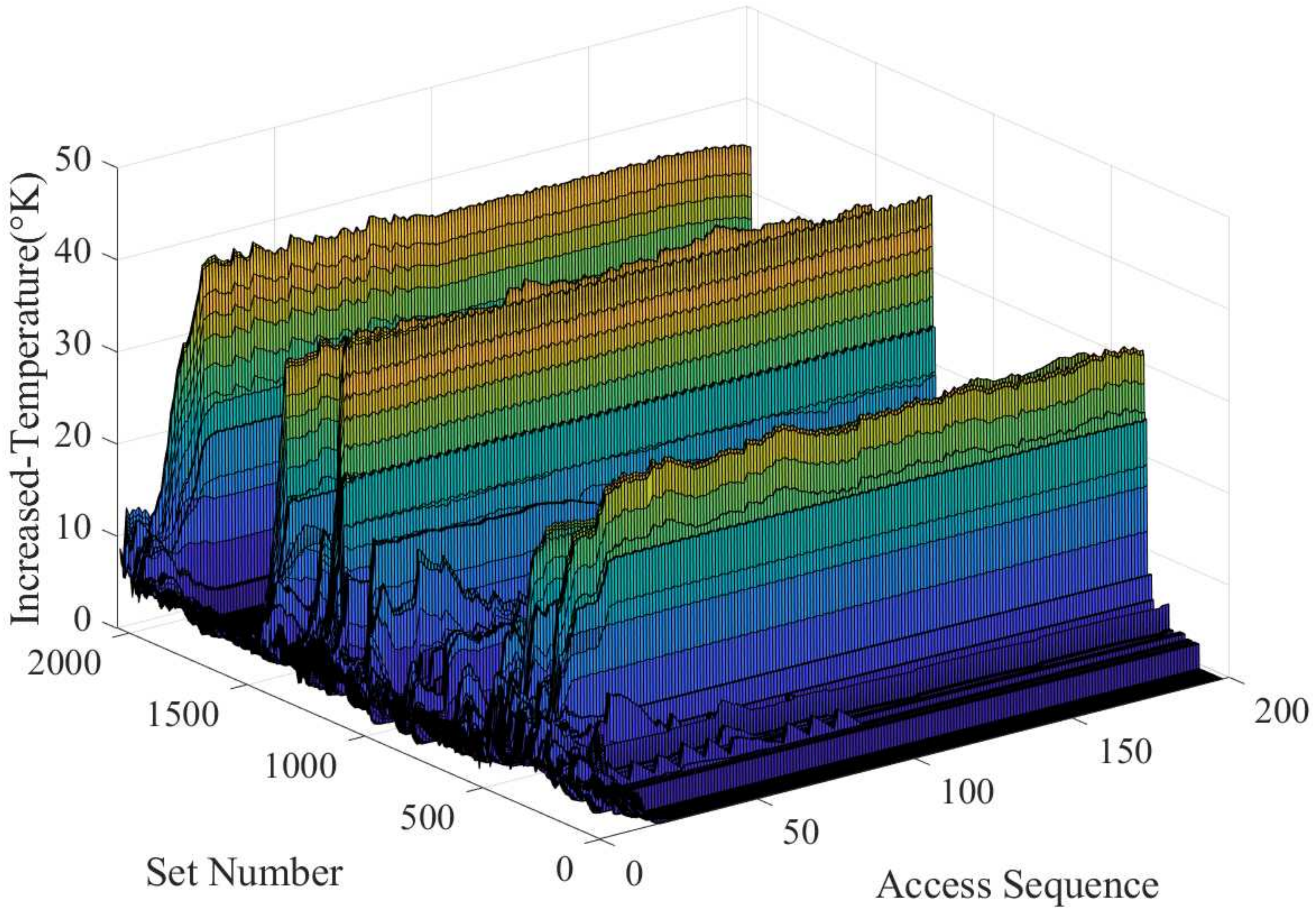}}
				\hfill
				\subfloat[calculix in TA-LRW policy]{\includegraphics[width=0.25\linewidth]{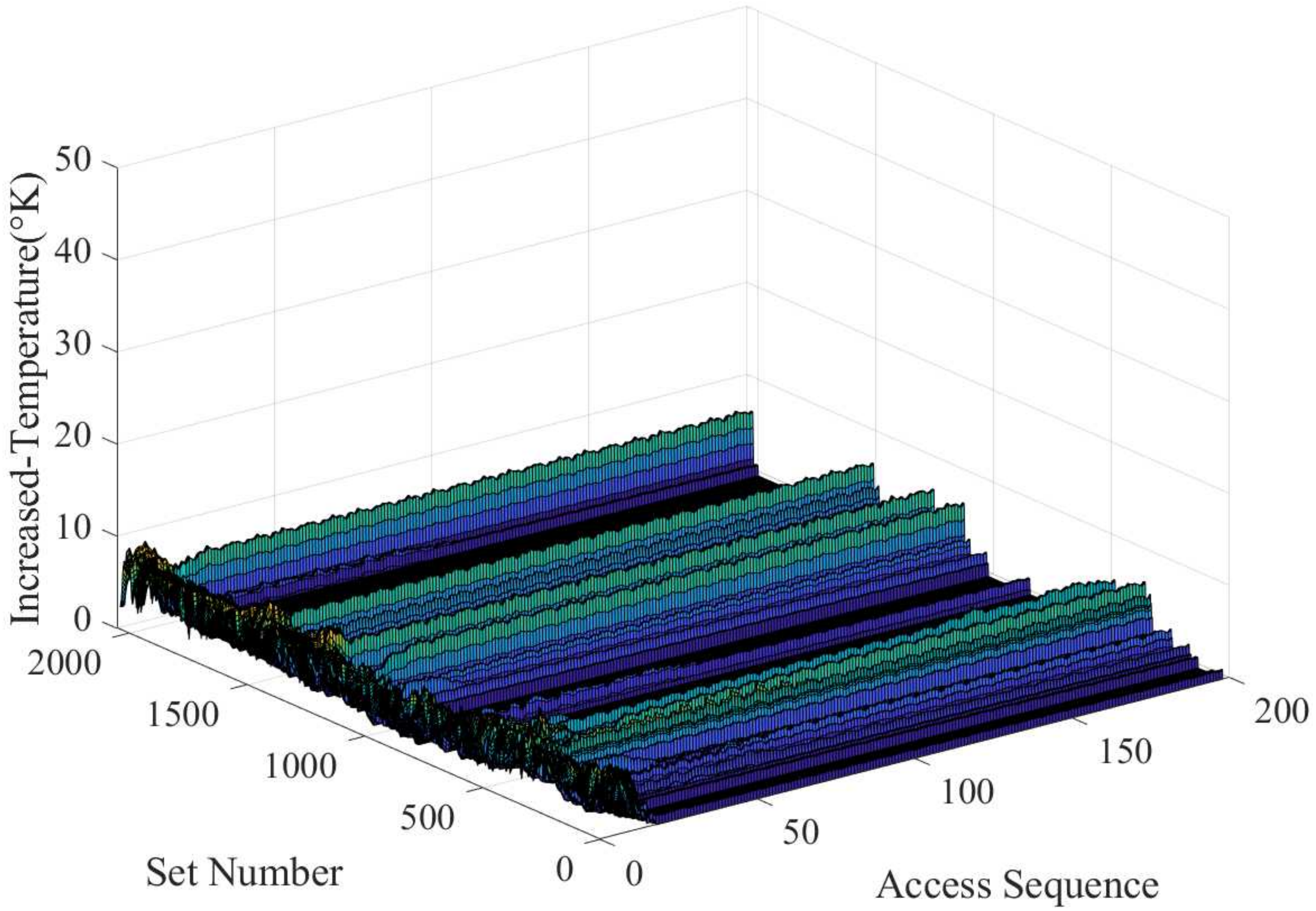}}
				\hfill
				\subfloat[bzip2 in LRU policy]{\includegraphics[width=0.25\linewidth]{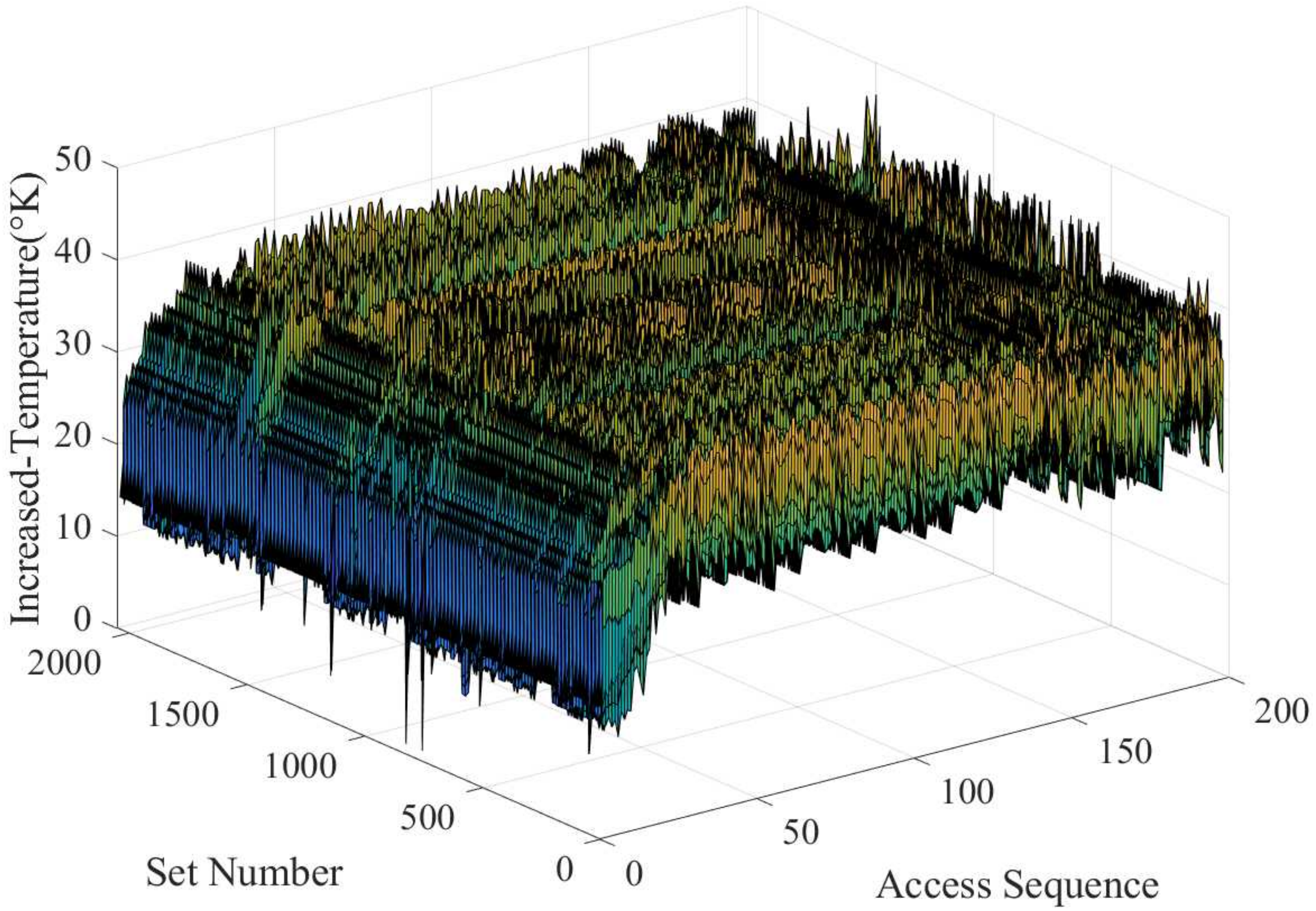}}
				\hfill
				\subfloat[bzip2 in TA-LRW policy]{\includegraphics[width=0.25\linewidth]{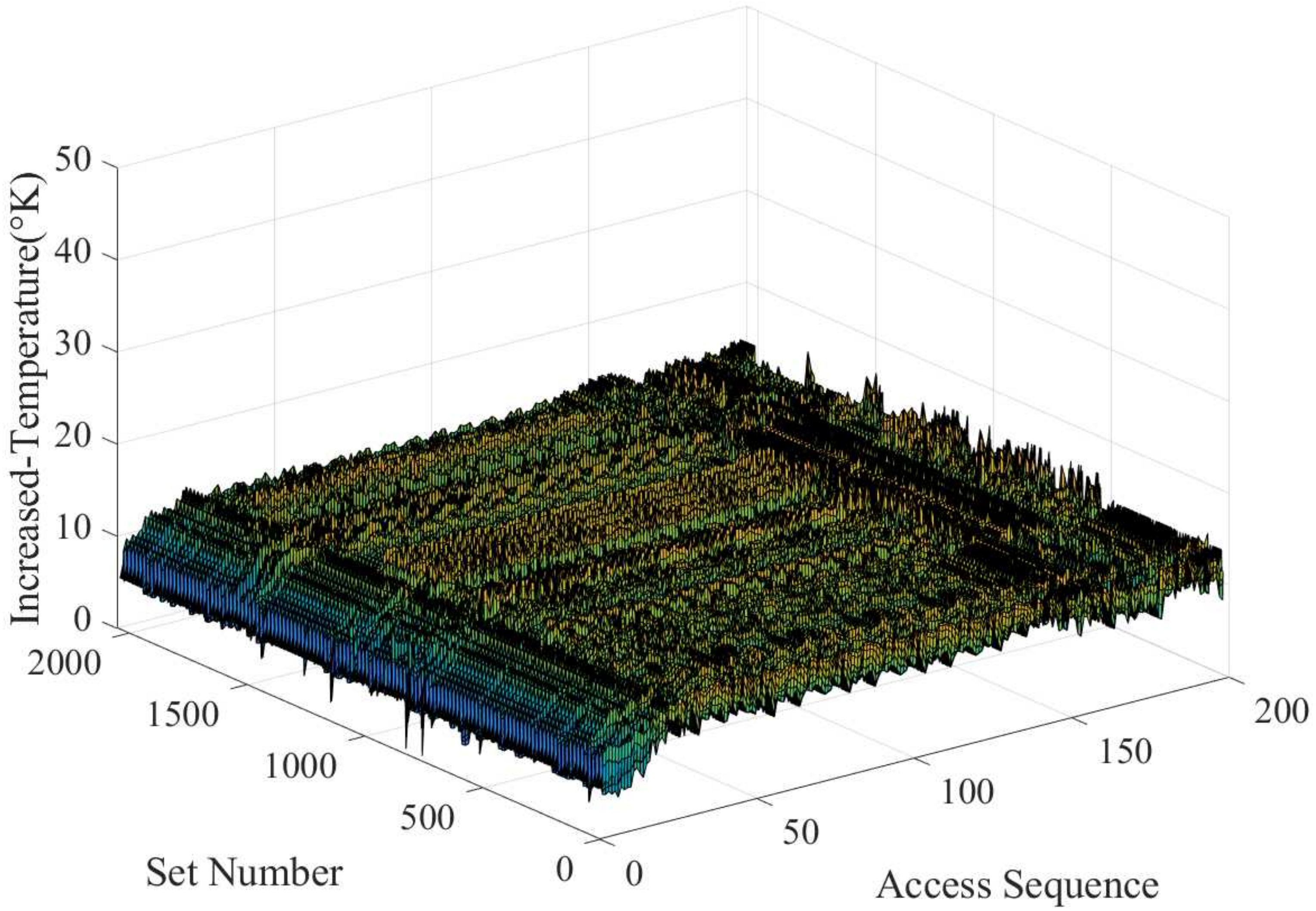}}
				\hfill
				\subfloat[omnetpp in LRU policy]{\includegraphics[width=0.25\linewidth]{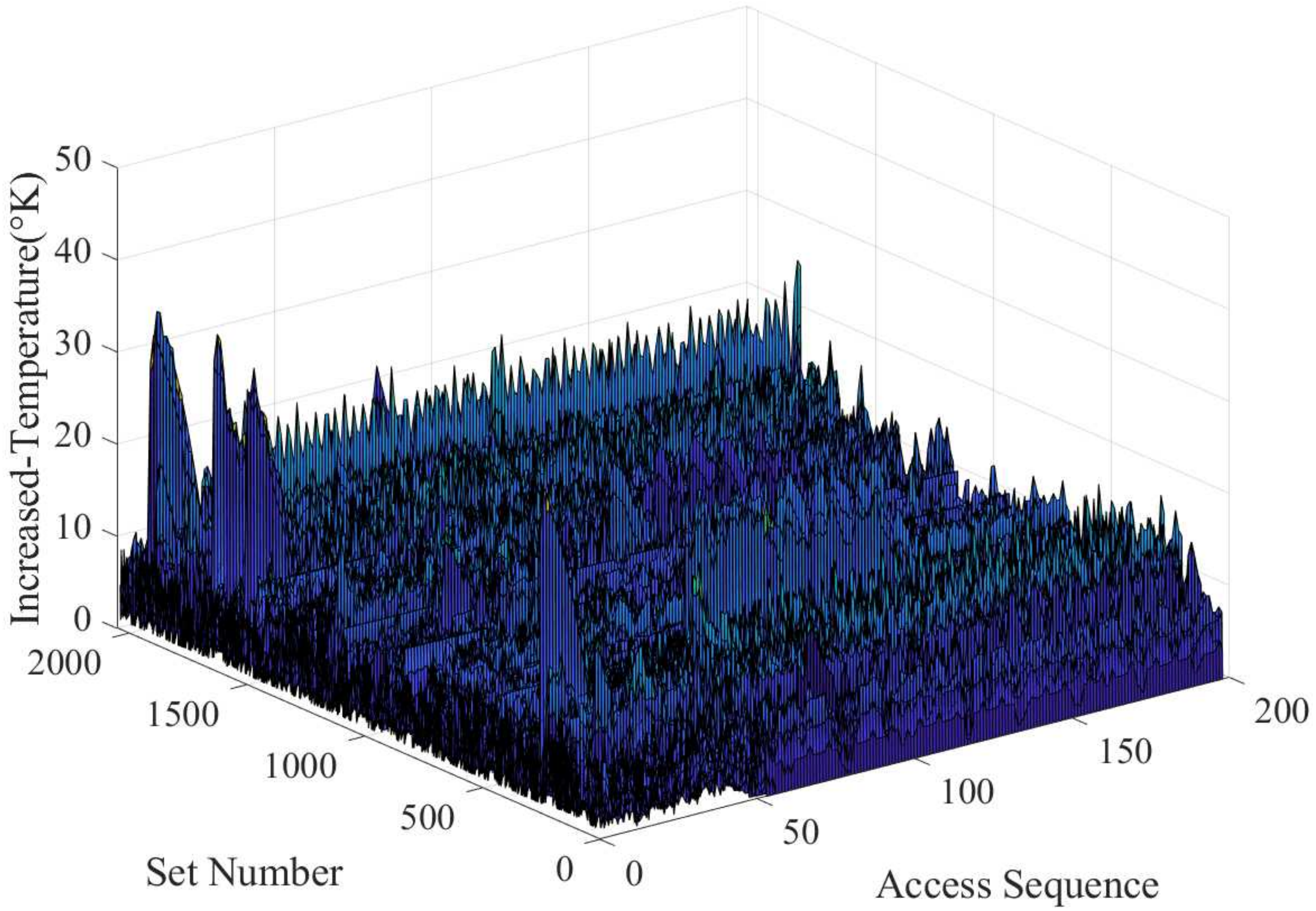}}
				\hfill
				\subfloat[omnetpp in TA-LRW policy]{\includegraphics[width=0.25\linewidth]{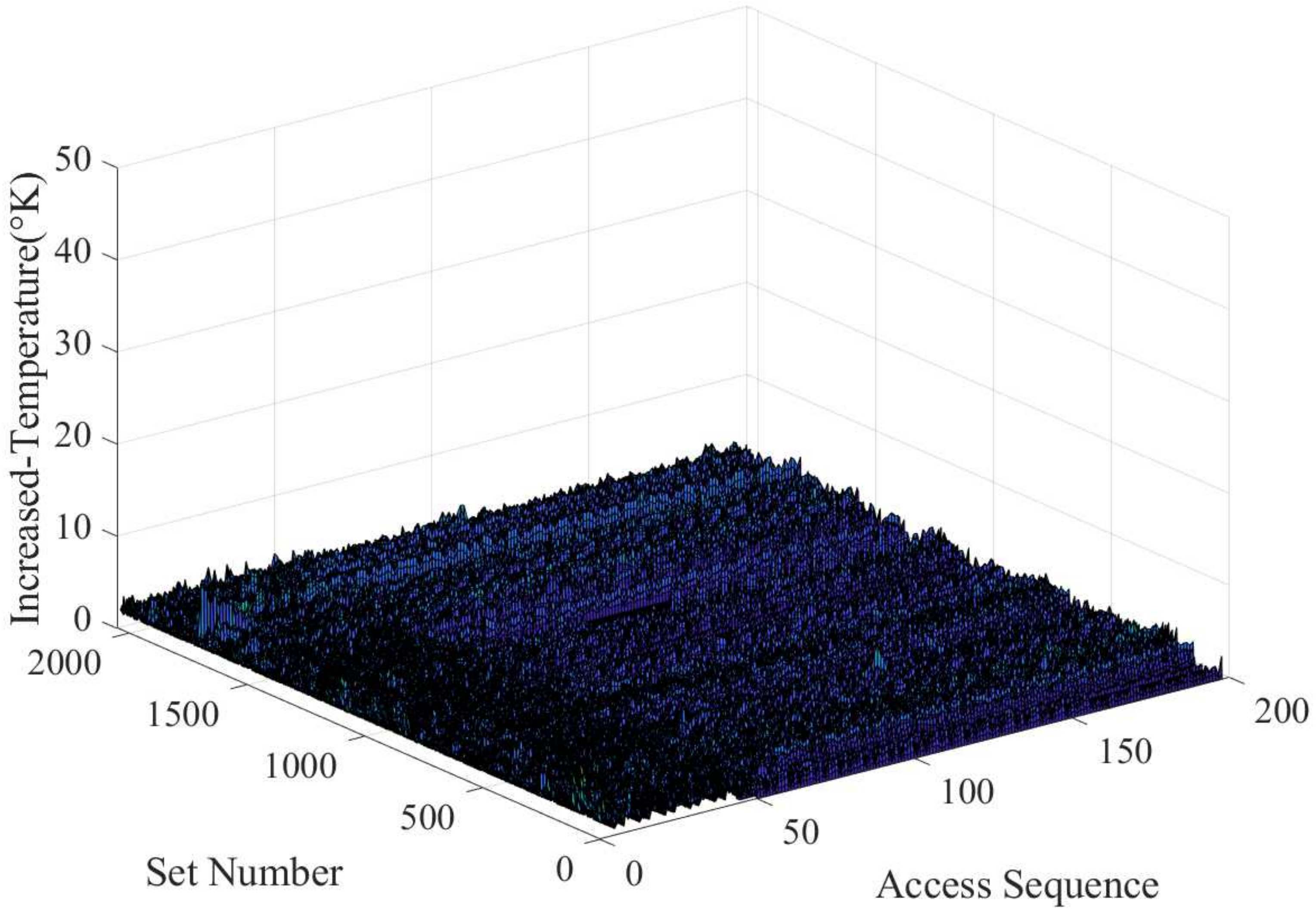}}
				\hfill
				\subfloat[mcf in LRU policy]{\includegraphics[width=0.25\linewidth]{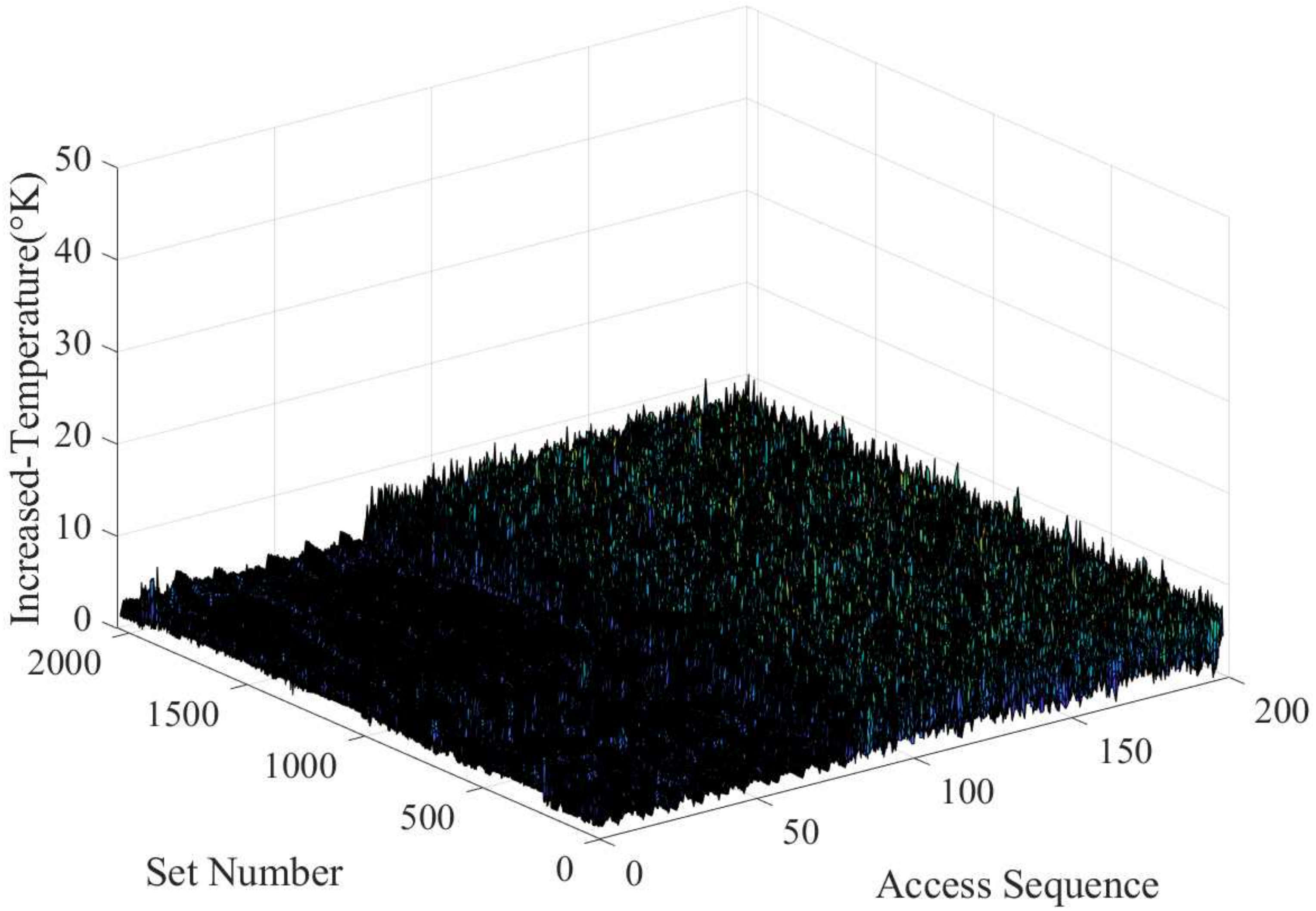}}
				\hfill
				\subfloat[mcf in TA-LRW policy]{\includegraphics[width=0.25\linewidth]{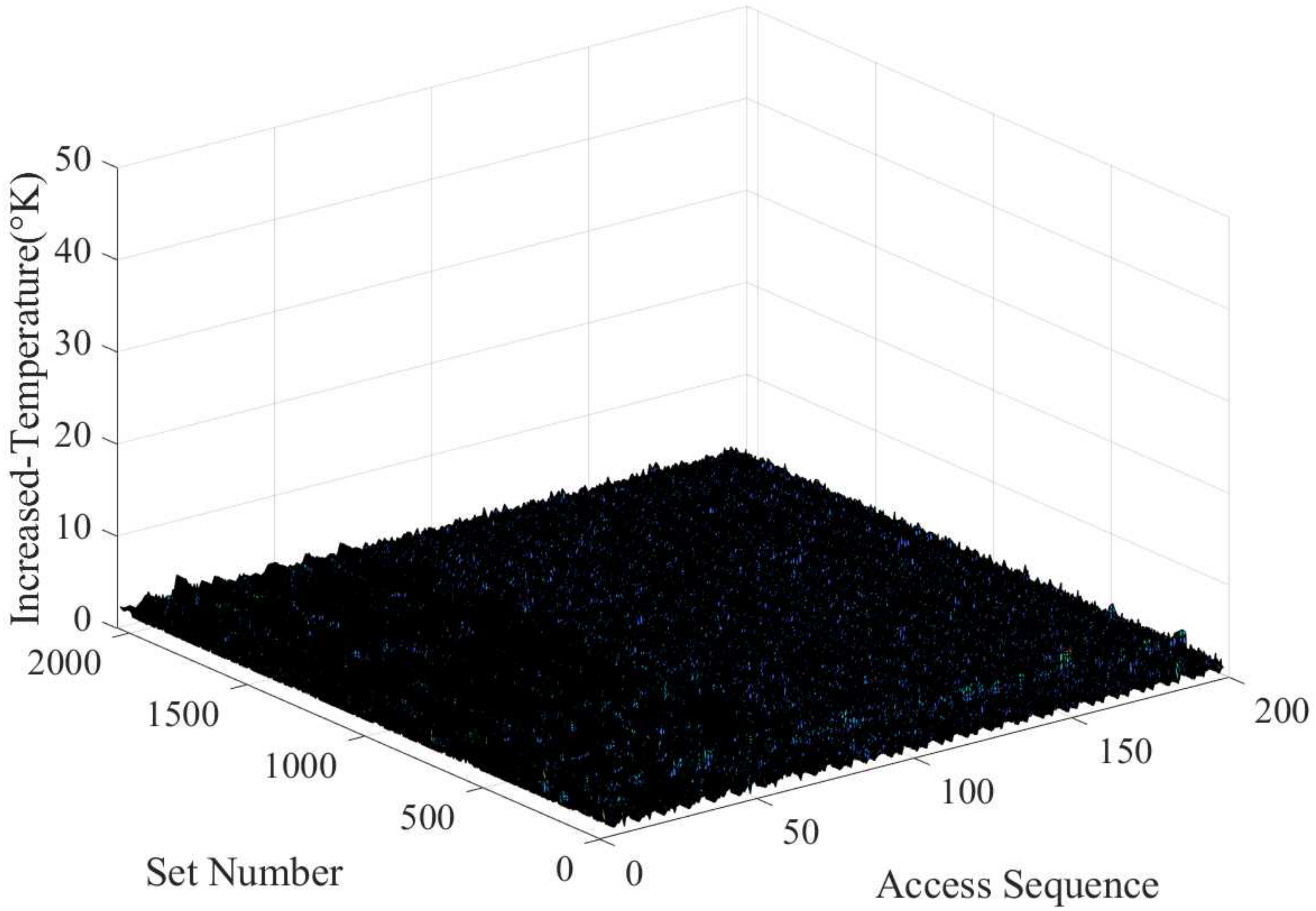}}
				\hfill
				\subfloat[cactusADM in LRU policy]{\includegraphics[width=0.25\linewidth]{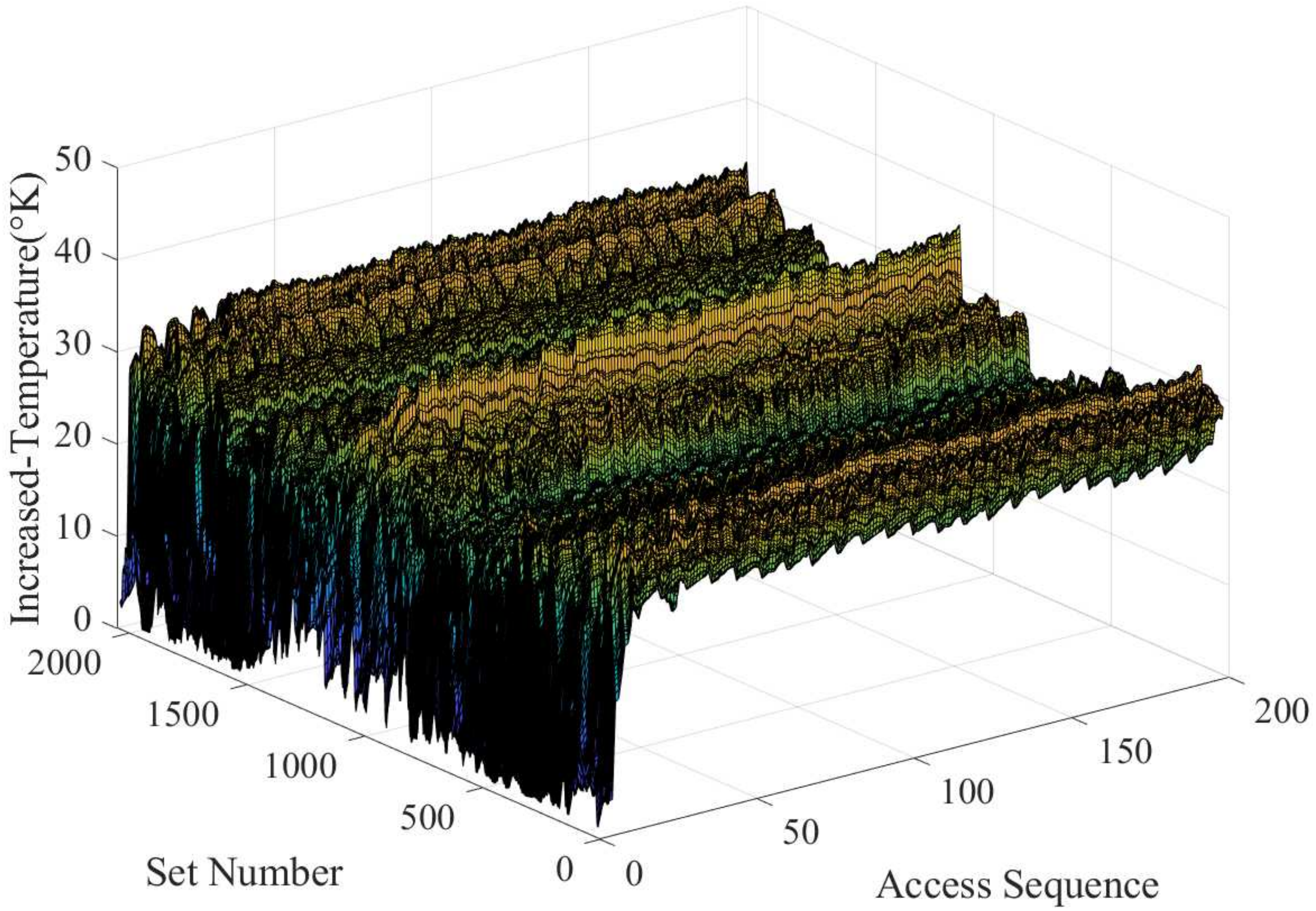}}
				\hfill
				\subfloat[cactusADM in TA-LRW policy]{\includegraphics[width=0.25\linewidth]{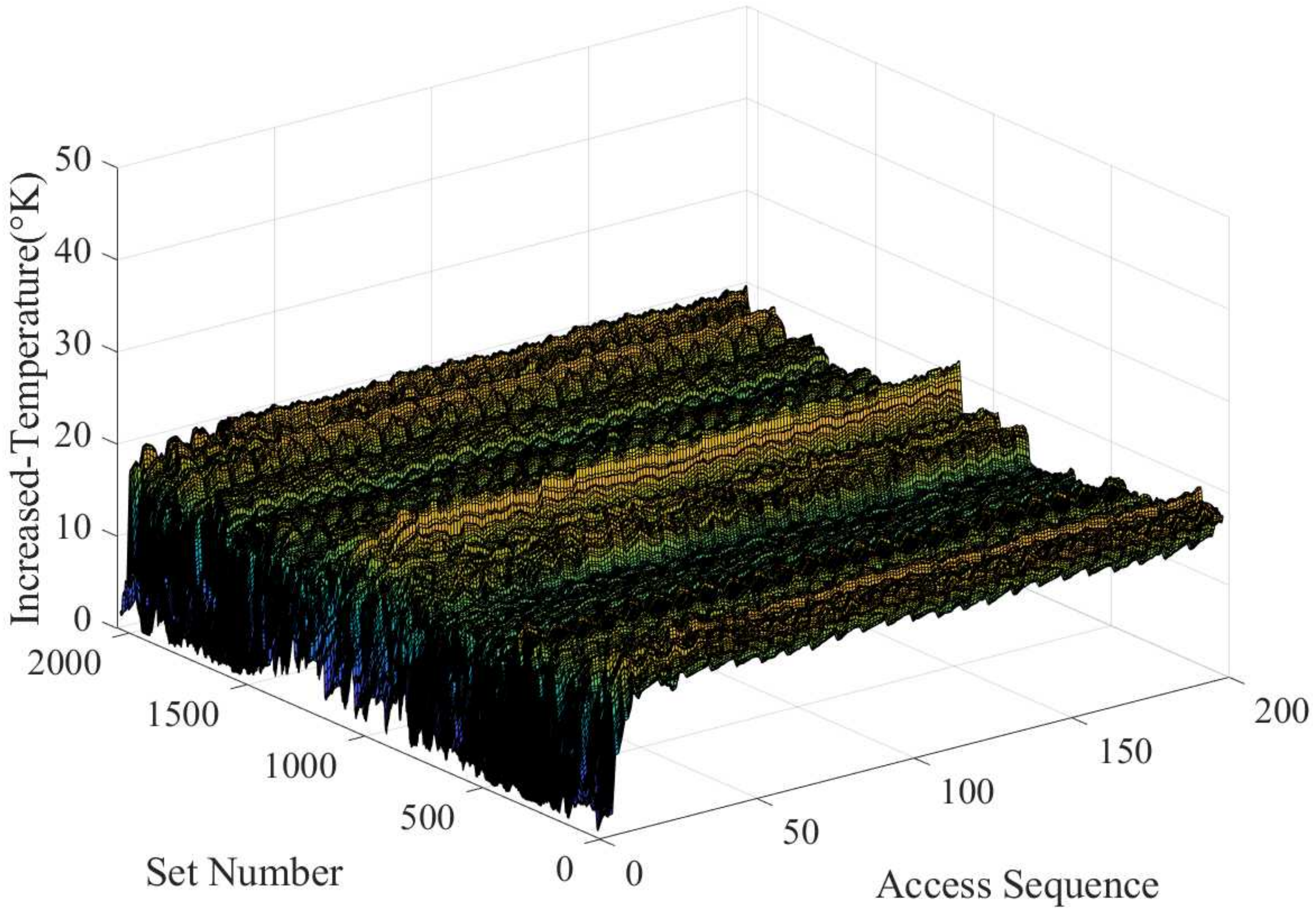}}
				\hfill
				\subfloat[gcc in LRU policy]{\includegraphics[width=0.25\linewidth]{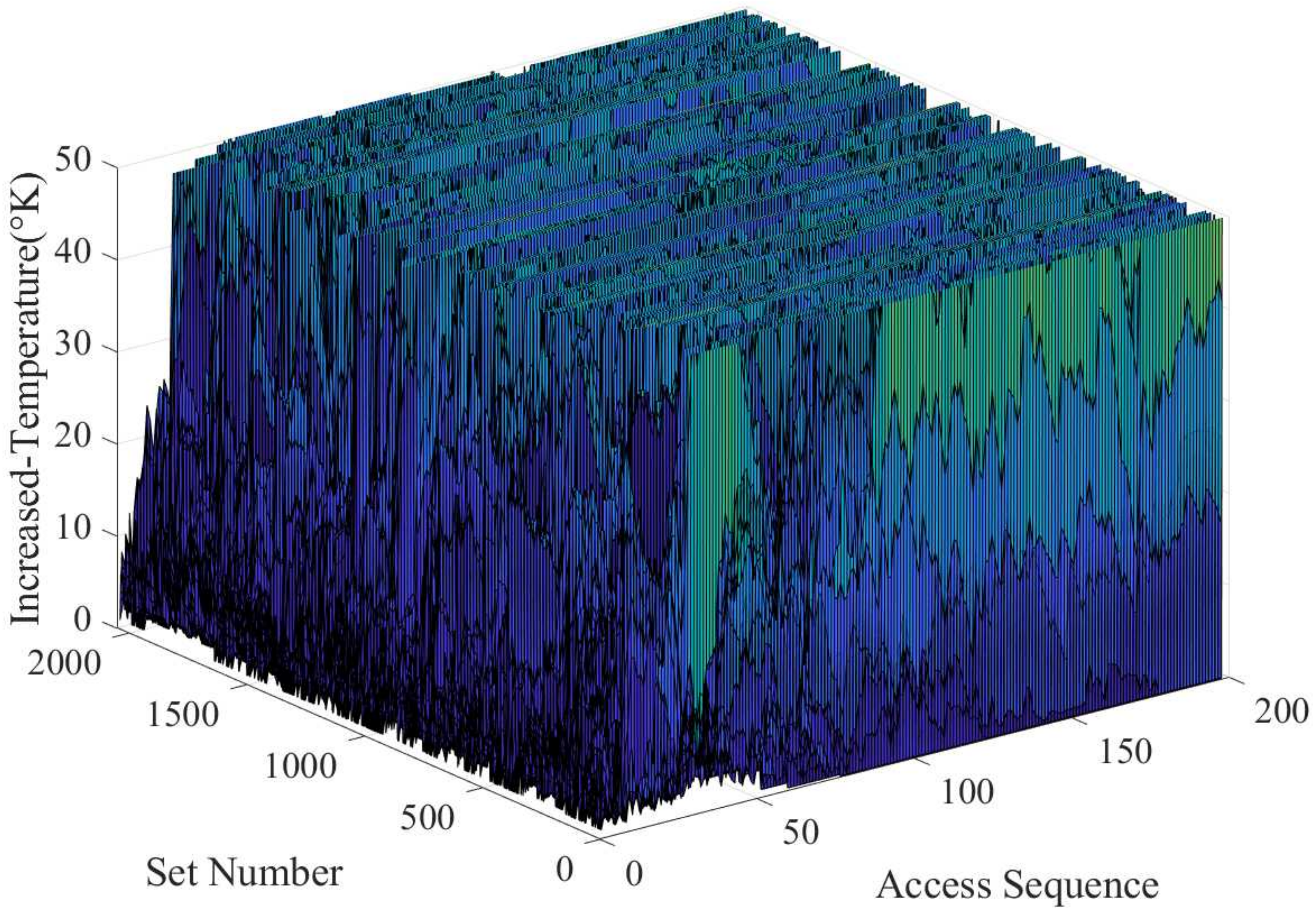}}
				\hfill
				\subfloat[gcc in TA-LRW policy]{\includegraphics[width=0.25\linewidth]{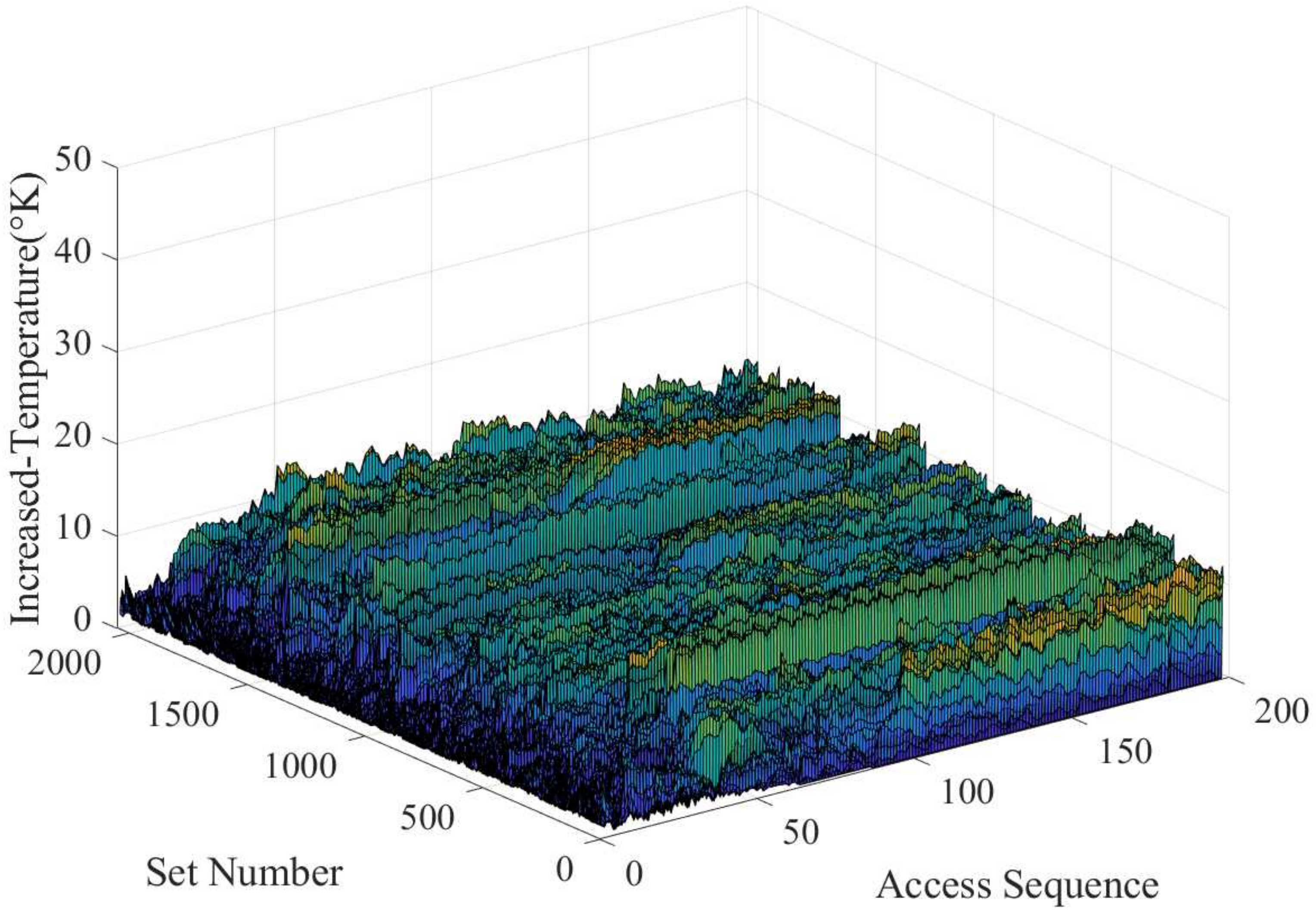}}
				\hfill
				\subfloat[sjeng in LRU policy]{\includegraphics[width=0.25\linewidth]{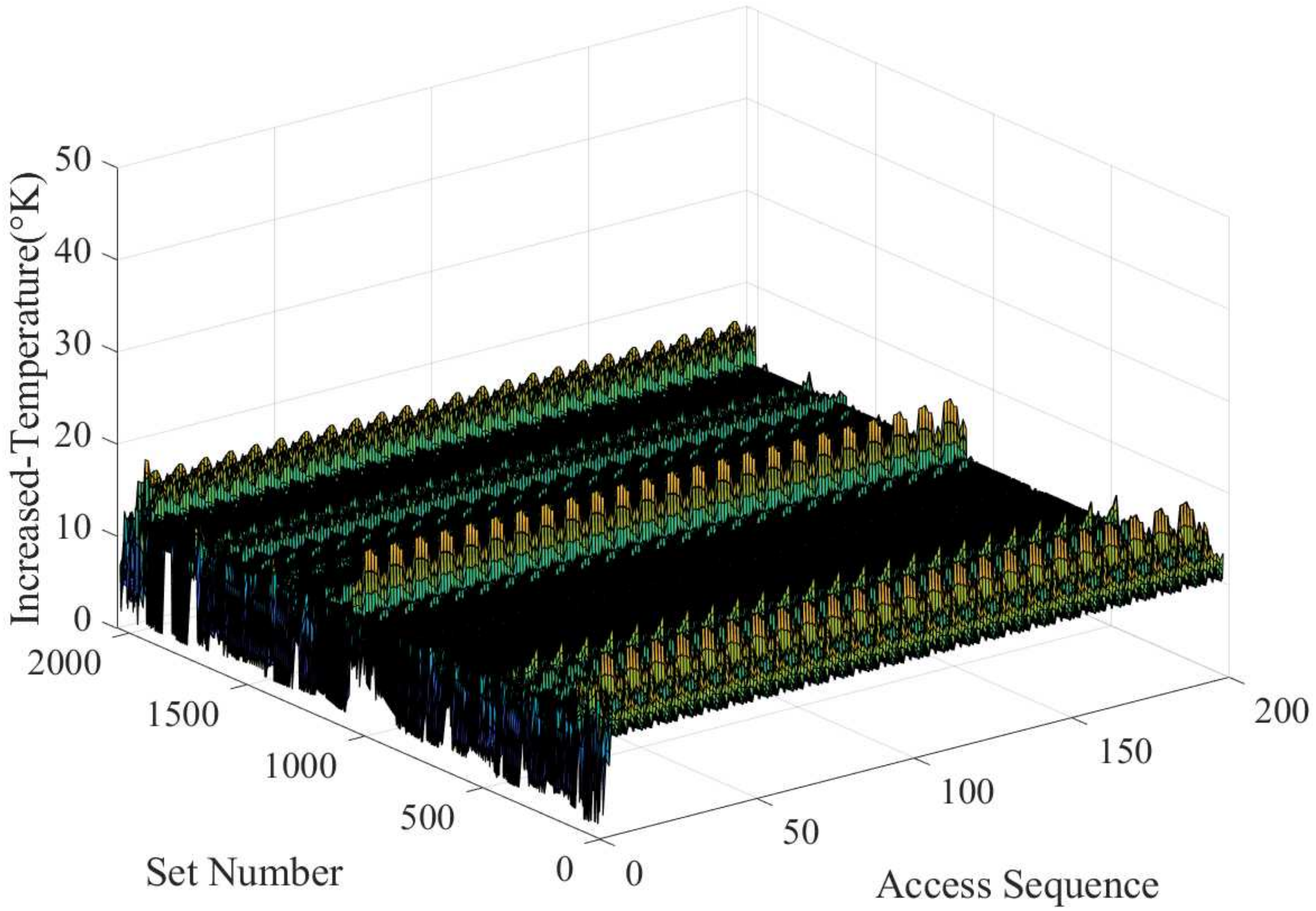}}
				\hfill
				\subfloat[sjeng in TA-LRW policy]{\includegraphics[width=0.25\linewidth]{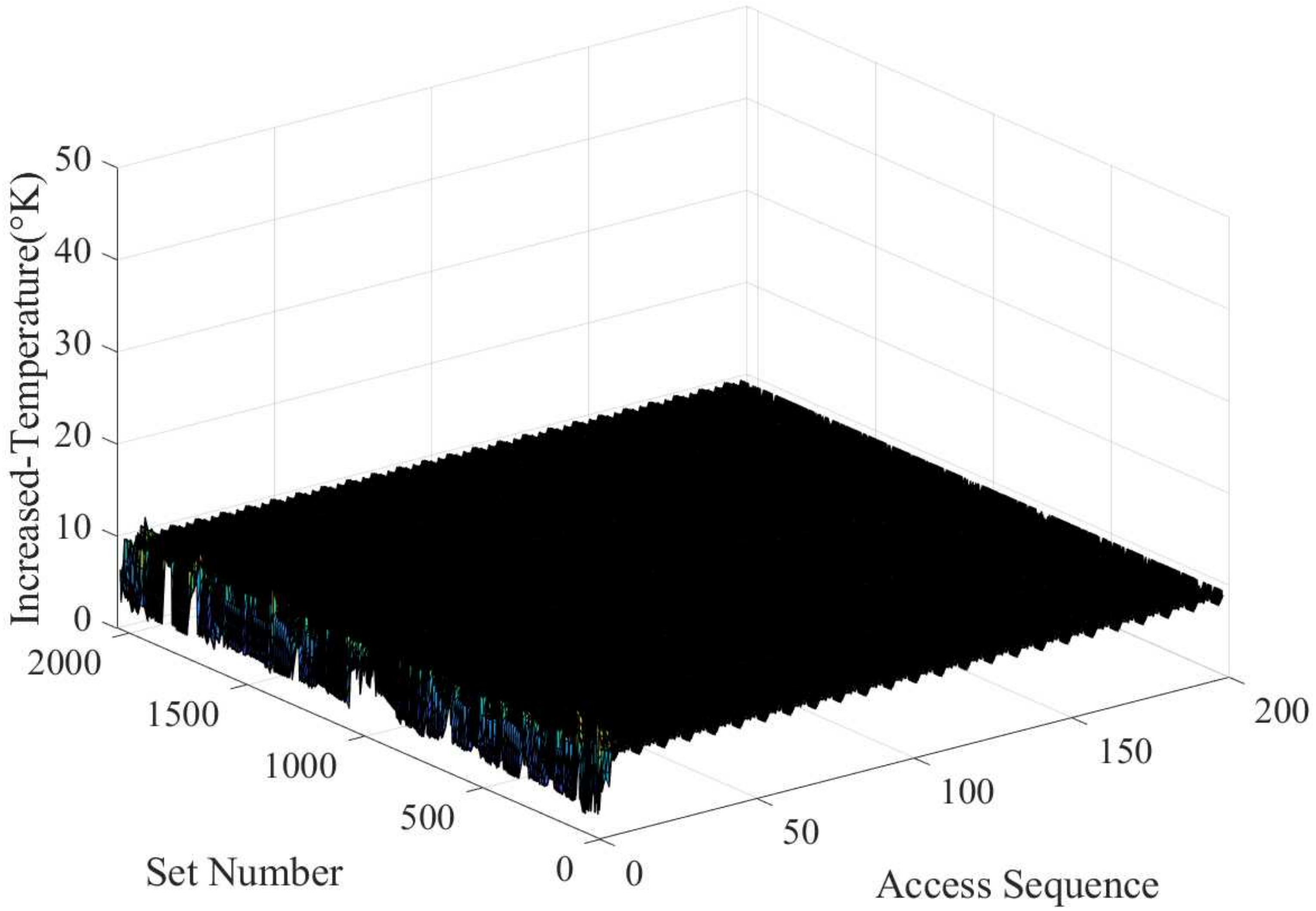}}
				\hfill
				\subfloat[lbm in LRU policy]{\includegraphics[width=0.25\linewidth]{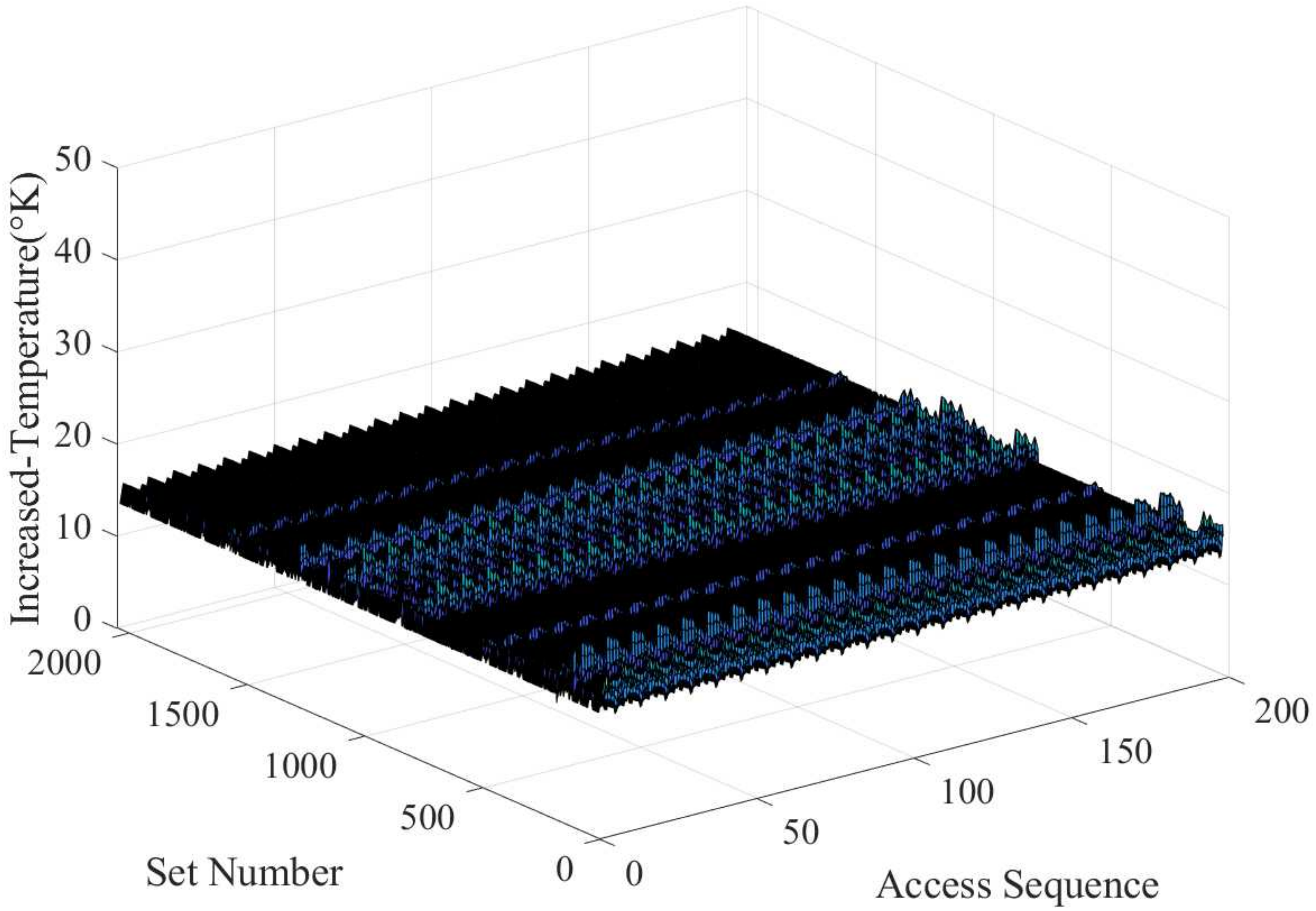}}
				\hfill
				\subfloat[lbm in LRU policy]{\includegraphics[width=0.25\linewidth]{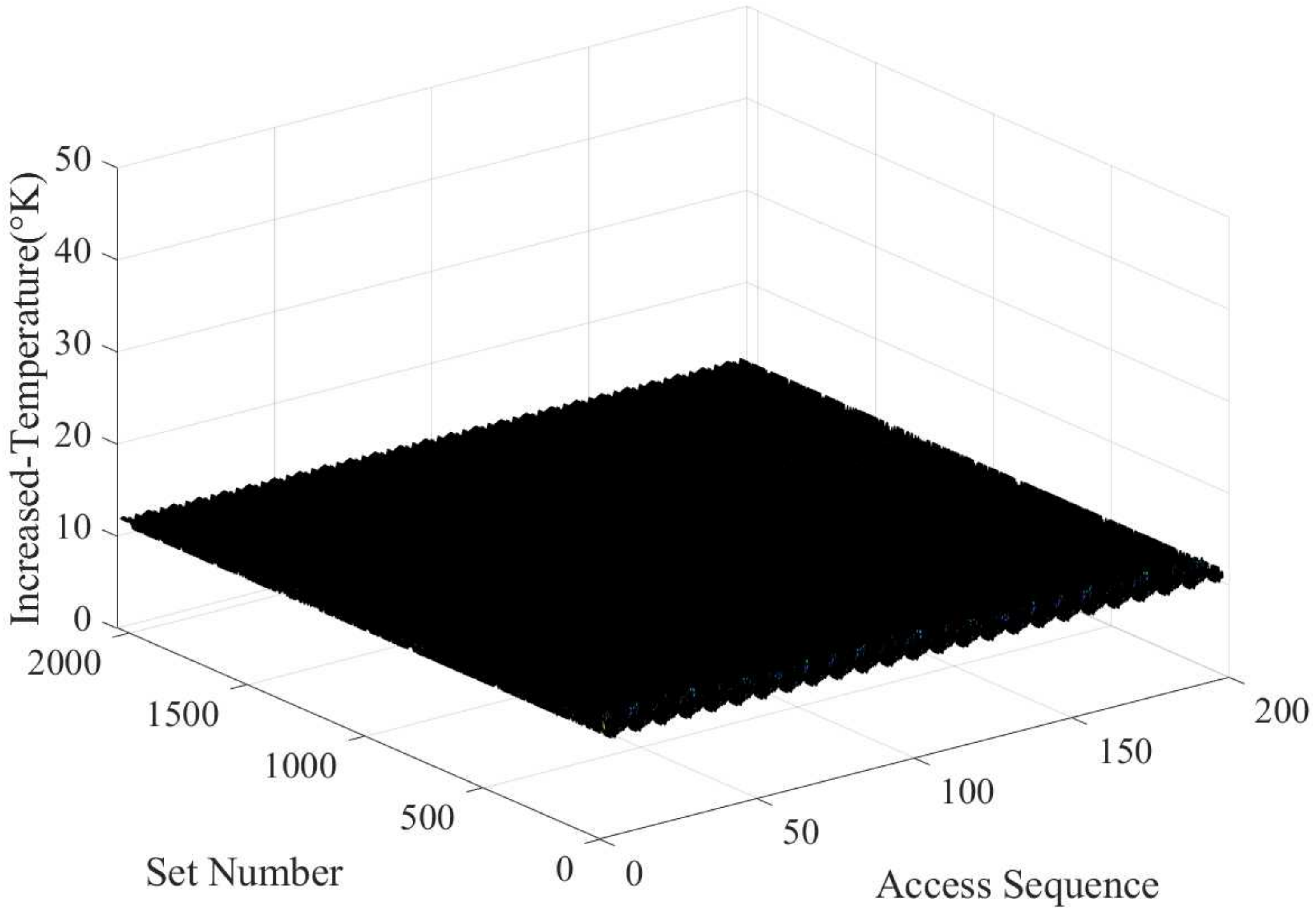}}\hspace{255pt}
				\caption{
				Temperature increase in blocks of STT-MRAM cache with LRU and TA-LRW policies for (a, b) bwaves, (c, d) bzip2, (e, f) calculix, (g, h) omnetpp, (i, j) mcf, (k, l) cactusADM, (m, n) gcc, (o, p) sjeng, and (q, r) lbm benchmarks.}
				\vspace{-8pt}
				\label{fig:app2}
			\end{figure*}
			
			\section*{B. Heat Accumulation in LRU Policy}
			


		Fig. \ref{fig:app2} represents the increased temperature in STT-MRAM cache sets operating under LRU and TA-LRW policies. 			
The results for 200 consecutive write accesses per cache set demonstrate a large increase and high diversity in temperature of cache sets in LRU policy for different workloads. 
			Fig. \ref{fig:app2}(a) shows that the increased temperature for \textit{bwaves} workload in LRU policy varies from 30$^{\circ}$K to 50$^{\circ}$K with the average of about 40$^{\circ}$K. 
			Using TA-LRW policy, the increased temperature is about 20$^{\circ}$K with significantly lower variations compared with LRU, as shown in Fig. \ref{fig:app2}(b).
			According to Fig. \ref{fig:app2}(c), which depicts the LRU policy behavior in \textit{calculix} workload, the temperature is increased by 10$^{\circ}$K in  majority of cache sets. This increase is higher than 30$^{\circ}$K for some sets in all their access sequences. 
			TA-LRW reduces the increased temperature to lower than 7$^{\circ}$K for all sets in all access sequences (Fig. \ref{fig:app2}(d)).

			Using LRU policy, the temperature increase in \textit{bzip2} workload is less than 20$^{\circ}$K for initial accesses and rapidly reaches above 30$^{\circ}$K for all remaining accesses in all sets (Fig. \ref{fig:app2}(e)). 
			TA-LRW limits this increase to about 10$^{\circ}$K (Fig. \ref{fig:app2}(f)).  
			A large variation is observed in Fig. \ref{fig:app2}(g) for the increased temperature of LRU policy in \textit{omnetpp} workload, which not only is efficiently smoothed, but also is reduced to less than 5$^{\circ}$K by employing TA-LRW policy (Fig. \ref{fig:app2}(h)).

			The temperature increase of LRU policy is not considerable in initial accesses of \textit{mcf} workload, as shown in Fig. \ref{fig:app2}(i), and is raised to about 10$^{\circ}$K afterward. 
			Fig. \ref{fig:app2}(j) shows that TA-LRW prevents the temperature to increase more than 3$^{\circ}$K.
			Based on Fig. \ref{fig:app2}(k) and Fig. \ref{fig:app2}(l), the temperature behavior of \textit{cactusADM} workload in LRU and TA-LRW policies, respectively is similar to that in \textit{bzip2} workload.
			The largest temperature variation for LRU policy is observed in \textit{gcc} workload (Fig. \ref{fig:app2}(m)), which indicates 50$^{\circ}$K increase for several sets in a considerable number of accesses. 
			Fig. \ref{fig:app2}(n) depicts that TA-LRW reduces the upper bound of increased temperature to 13$^{\circ}$K where the majority values are less than 10$^{\circ}$K.
			Considering \textit{sjeng} workload in Fig. \ref{fig:app2}(o), the cache temperature is increased by 15$^{\circ}$K in LRU policy and TA-LRW reduces this value to about 10$^{\circ}$K, as shown in  Fig. \ref{fig:app2}(p). 
			The same behavior is observed for \textit{lbm} workload in LRU and TA-LRW policies, as depicted in Fig. \ref{fig:app2}(q) and Fig. \ref{fig:app2}(r), respectively.

			\end{document}